\pdfoutput=1

\documentclass[aps,prd,reprint,groupedaddress, longbibliography]{revtex4-2}

\usepackage{longtable}
\usepackage{graphicx}
\usepackage{fancyvrb}
\usepackage{verbatimbox}
\usepackage{listings}
\usepackage{array}
\usepackage{amsmath}
\usepackage{siunitx}
\usepackage{amssymb}
\usepackage{mathrsfs}
\usepackage[section]{placeins}
\usepackage{enumitem}
\usepackage[title]{appendix}
\usepackage{nicefrac,xfrac}
\begin{document}

\title{Light and charge yields from nuclear recoils in noble liquids: From sub-keV energies to 1 MeV}

\author{Y. Sarkis}
\author{Juan Carlos D'Olivo}
\author{Alexis Aguilar-Arevalo}

\affiliation{Instituto de Ciencias Nucleares, Universidad Nacional Aut\'onoma de M\'exico,  04510 CDMX, Mexico}

\date{\today}

\begin{abstract}
We present a first-principles study of total quanta and charge recombination in noble liquids, based on a solution of Lindhard's integro-differential equation for atomic motion with electronic straggling and detailed modeling of inter-atomic potentials. We introduce the electronic scaling length $\xi_e$ as a fundamental parameter related to a physical observable of the system. We study the energy and electric-field dependence of the exciton-to-ion ratio for nuclear recoils based on the Bates-Griffing process and semiclassical atomic interactions. The resulting light and charge yields are expressed through the recombination probability in the Thomas-Imel box model using the calculated parameter $\xi_e$ and one fitted constant. We compare our model with recent measurements of the electric-field dependence of the box size in LAr and LXe and provide a physical explanation for this behavior. Our model extends to sub-keV energies and provides a consistent description of available data from a few keV to 1 MeV, yielding better overall agreement with existing LAr data than NEST and offering predictions for LXe.
\end{abstract}

\maketitle

\section{Introduction}\label{Intro}

Over the last decade, dual-phase time projection chambers (TPCs) based on liquid argon (LAr) and liquid xenon (LXe) have emerged as the leading technology for the direct detection of dark matter (DM) \cite{LZ_2023, DarkSide20k_2024, XENONnT_2024} and the observation of coherent elastic neutrino-nucleus scattering (CE$\nu$NS) \cite{COHERENT_2021, NUCLEUS_2025,cevnsXe}. The exceptional scalability, high target density, and superior electron-recoil discrimination of these media make them ideal for probing physics beyond the Standard Model (BSM) via nuclear recoil channels.

The observation of CE$\nu$NS \cite{Akimov:2020pdx} has opened a new window for precision phenomenology. As a low-threshold energy process, CE$\nu$NS serves as a powerful tool for measuring the electroweak mixing angle at low momentum transfer and providing stringent constraints on non-standard interactions (NSI), neutrino magnetic moments, and sterile neutrinos \cite{Papoulias_2022, Aristizabal_2023,DeRomeri2026}. However, the success of these precision programs and the possibility of distinguishing a DM signal from the so-called ``neutrino fog" depend critically on accurate calibration of the detector response \cite{Feliciano,duque2026shiftingneutrinofogstudying}. Specifically, the conversion of nuclear recoil energy into detectable quanta (photons and electrons) in the sub-keV regime remains the dominant source of systematic uncertainty.

It is essential to have a comprehensive theoretical model that describes the ionization and scintillation yields of low-energy nuclear recoils. This is a difficult task. Standard Lindhard theory \cite{Lindhard_1963}, while foundational, fails to describe the quenching behavior at energies below 10 keV because it neglects the underlying atomic structure and non-linear electronic stopping effects \cite{Sarkis_2020,BEZRUKOV2011119,Ziegler}. For its part, semi-empirical frameworks, such as the Noble Element Simulation Technique (NEST) \cite{NEST_2021}, rely on extensive fitting to experimental data. Although effective, these models often lack a fundamental predictive basis and may require different parameterizations for different experimental setups.

 Nuclear collisions are governed by the competition between electronic and nuclear stopping processes. Atom-atom interactions involve more degrees of freedom than those between electrons and atoms, and the energy deposited by them is not fully converted into electron-equivalent energy. In the following, we outline the main stages of the physical processes for the production of quanta in noble liquids \cite{BEZRUKOV2011119,ScintillationProcess}:

\begin{enumerate}[label=(\roman*)]
    \item The process begins when a non-ionizing particle hits an atom of a noble element and transfers an initial recoil energy. If this  energy is greater than the minimum required to create quanta, further collisions will occur until the ions or valence electrons scatter with an energy below this minimum value. In cascade collisions, the momentum transferred to the nuclei and the momentum transferred to the electrons are proportional to the nuclear and electronic stopping powers, respectively.

    \item Since the recoil velocity of atoms is much lower than the typical Fermi velocity of valence electrons, it is appropriate to describe nuclear collisions as an adiabatic, elastic, or inelastic process, depending on whether the total kinetic energy is conserved or not during the process.

    \item In a binary atomic collision, the system has many degrees of freedom, and hence, the average energy to create an ion during the cascade is generally not constant. We classify  the dielectronic process as follows:
    
   \begin{itemize}
    \item \textit {Passive channel:} During the binary collision, the projectile electron remains in its ground state. Since it is strongly bound, it behaves as a ``heavy'' particle that transfers momentum to the nucleus. A quasi-molecule is formed during the process because of the partial shielding of the nuclear charges by the inner electron cloud. This reduces the target ionization cross section compared to a bare nucleus \footnote{Alternately referred to as the ``electron-nuclear'' interaction,``monoelectronic'' process, or ``singly inelastic'' channel, its relative strength and dominance over the active channel depend dynamically on the impact parameters, the number of bound electrons, and the relative magnitudes of the projectile and target nuclear charges.}. Subsequently, due to the exchange of momentum, the target  electrons can remain in their orbitals, become excited, or ionize. Therefore, depending on the magnitude of the momentum interchanged, either elastic or inelastic processes can occur in this channel. The passive inelastic channel is also known as the ``single-inelastic channel'' because only one of the atoms is excited or ionized.
    
    \item \textit{Active channel:} This is essentially determined by direct Coulomb interactions between target and projectile valence electrons, which are emitted with an average energy proportional to the valence band width. For the energies involved,  the closest approach distance is not short enough for the resulting ions to form a quasi-molecule. Consequently, only inelastic processes can happen in this channel, which is also known as the ``double-inelastic channel''.   
\end{itemize}

The passive and active channels modify the ionization cross section in opposite ways. One or the other can dominate, depending on  the relative magnitudes of the projectile and target nuclear charges, and the projectile's energy.

    \item The probability of elastic scattering between atoms is proportional to the ratio of electronic to nuclear stopping power. At low energies, elastic collisions in the passive channel dominate.
    
    \item  The elastic electron momentum exchange determines  the electronic straggling in atom collisions \cite{WilsonF}. This process is proportional to the ratio of the electronic straggling to the electronic stopping power multiplied by energy. As energy increases, this ratio decreases.

    \item  The total electronic energy, also called the equivalent electronic energy (eVee), corresponds to the sum of the energies of all nuclear recoils in the cascade with electron emission. The remaining energy, associated with nuclear motion, is converted into heat. Electrons, ions, and excitons form a plasma along a path determined by the first recoiling atom.

    \item In the absence of an external electric field, the electron escape probability is negligible. Consequently, all electrons generated by a nuclear recoil recombine with ions, contributing to the scintillation light. 
    
    \item By contrast, in a dual-phase TPC, the applied  electric field displaces electrons into the gas phase (S2), while photons are captured in the liquid phase by an array of photomultiplier tubes (S1). A region is created where partial electron recombination occurs, the size of which decreases as the intensity of the external field increases, with the consequent reduction of the recombination probability.
    
    \item The probability that an ion captures two electrons during recombination is negligible. Thus, we can consider that there is only one electron per ion and one photon per exciton. 
    
    \item In a cascade with high exciton density, there exists a considerable probability of exciton-exciton collisions, producing a single survival excitation. 
    
    \item If the noble liquid is pure enough (less than 10 ppb) to reduce light-quenching agents such as water and oxygen, then electrons and photons travel through the TPC as the primary detection signal.
\end{enumerate}

 Using the second-order integro-differential equation \cite{QF2023}, derived from the Lindhard integral equation \cite{Lindhard_1963} for atomic motion, we incorporate the effects of electronic and nuclear stopping powers, binding energy, and electronic straggling. From the solution to this equation, incorporating realistic inter-atomic potentials, we implement a unified description of the light and charge yields. A distinctive feature of our model is that it characterizes the medium's response across a broad energy range from sub-keV energies to 1 MeV, in terms of only one adjustable parameter, which is  associated with the recombination process. Our method   accounts for the energy and electric field dependence of the recombination process through a modification of the Thomas-Imel box model which  is valid for different noble liquids. The results show a significant improvement over traditional empirical models, providing the theoretical precision required by the next generation of CE$\nu$NS and DM experiments.
 
The paper is organized as follows. In Sec.~\ref{TotQuanta}, we describe the production of quanta in noble liquids and introduce the total quanta model for noble liquid detectors, covering energy conservation, the passive elastic and inelastic channels derived from the second-order integro-differential equation, the electronic scaling parameter $\xi_e$, and the active-inelastic channel. Section~\ref{BatesGriffingSection} presents the active Bates-Griffing electron scattering model, describing the energy dependence of the exciton-to-ion ratio and incorporating spin exchange effects in binary atomic collisions. Section~\ref{RecombinationModel} describes the recombination process within an extended Thomas-Imel box model, where the box size is derived from electrostatic equilibrium, along with a polarizability model for the electric field dependence of $N_{\text{ex}}/N_i$ and an analytical calculation of Birks' constant. In Sec.~\ref{Recombination FORNOBLE}, we show the unified recombination predictions for light and charge yields across different noble liquids, comparing our results with experimental data and NEST predictions. Finally, Sec.~\ref{Conclusions} summarizes our main conclusions and highlights the implications for future dark matter and CE$\nu$NS experiments. Detailed mathematical derivations and atomic calculations are provided in the Appendices.

\section{Total quanta model for noble liquid detectors}\label{TotQuanta}
\subsection{Energy distribution}\label{SubEnergyDistribution}
 
In dual-phase noble liquid  detectors, the nuclear recoil energy $E_R$ that a particle (e.g., DM or neutrino) transfers to the target nucleus is partitioned into three channels: excitation of atoms $N^R_{ex}$ (excitons), formation of electron-ion pairs $N^R_{i}$ (ions), and atomic motion $E_{\nu}$. 
Consequently, the conservation of energy is expressed by the Platzman equation \cite{PLATZMAN1961116} 
\begin{equation}\label{Eq:platzman}
E_{R}=N^R_{ex} E^R_{ex} + N^R_{i}E^R_{i} + E_{\nu},    
\end{equation}
where $E^R_{ex}$ and $E^R_{i}$ are the average nuclear recoil energies used to create excited states and electron-ion pairs, respectively. 

For electron recoils, the transferred energy $E_e$ is divided between the production of excitons and ions; therefore, the total quanta is  $N^e_q = N^e_{ex} + N^e_{i}$. The average energy to create an exciton or ion can be defined as $W\equiv E_e/N^e_q$ and from this the mean electron ionization energy is given by $W^e_i=W(1+N^e_{ex}/N^e_i)$, where $N^e_{ex}/N^e_i$ is called the (electronic) exciton-to-ion ratio. For LAr, $W=19.5 \pm 1.0$ eV  and $(W^e_{i})_{\rm{LAr}}=23.6 \pm 0.3$ eV and $(W^e_{i})_{\rm{LXe}}= 15.6 \pm 0.3$ \cite{Doke2002}. For LXe, we use the average value from Dahl \cite{Dahl} ($W=13.7\pm 0.2$ eV) and the more recent value from Zurich II \cite{WnewLXe} ($W=11.5\pm 0.25$ eV), $W = 12.84 \pm 0.2$ eV, 
this value is closer to that calculated with the electron-free gas model, see Appendix \ref{AW}. This yields electronic exciton-to-ion ratios of $\left(N^e_{ex}/N^e_i\right)_{LAr}=0.21$, and $\left(N^e_{ex}/N^e_i\right)_{LXe}=0.06$.

In this work, we study the light and charge yields of the photons and electrons originating from nuclear recoils. In general, the exciton to ion ratio is not equal to the value obtained from  electron interactions. The binary collision includes  a pair of nuclei with screened charges; consequently, more excitation levels are available and $N^R_{ex}\cong N^R_i$. The \textit{nuclear quanta efficiency} $f_n(E_R)$ (also known as the Lindhard factor) is defined as 
\begin{equation}\label{Eq:f_ndef}
    f_n(E_R)\equiv \frac{E_{ee}}{E_R},
\end{equation}
 where  $E_{ee}=N^R_qW$ is the electron energy equivalent. Here, $N^R_q=N^R_i +N^R_{ex}$ is the total nuclear quanta and $E_R=E+U$, with $E$ the kinetic energy of the moving ion and $U$ the energy absorbed by atomic electrons. From this equation, one obtains the relation $f_n(E_R)E_R=W^R_iN^R_i$, where 
 \begin{equation}\label{Eq:W_idef}
 W^R_i= (1+\beta)W
 \end{equation}
is the mean ionization energy, with $\beta=N^R_{ex}/N^R_i$.  

 The basic parameters needed to determine  \( f_n(E_R) \) in noble liquids are the electronic stopping power, \( S_e \), the nuclear stopping power, \( S_n \), the electronic straggling, \( \mathcal{W} \), the atomic binding energy, the spin exchange energy, \( E_{ex} \), and the atomic number, $Z$. Two contributions to \( f_n(E_R) \) can be identified: passive and active. The passive component $f^{\mathrm{PI}}_n(E_R)$ is determined by the competition between \( S_e \), \( \mathcal{W} \), and \( S_n \), during electronic momentum transfer in the quasi-molecule. In contrast, the active component $f^{\mathrm{AI}}_n(E_R)$ is governed by the direct electron-electron interaction, which induces ionization without the formation of the quasi-molecule, provided the projectile energy is greater than $W^{R}_i$.  In the rest of the article we will restrict ourselves to the nuclear recoil process and omit the index $R$, except for the recoil energy $E_R$ of the nucleus that initiates the cascade.

\subsection{Passive elastic and inelastic channels}

 We determine \( f^{\mathrm{PI}}_n(E_R) \) from the solution to the second-order integro-differential equation derived from the Lindhard integral equation \cite{QF2023}:    
\begin{equation}\label{Ec:MainintegroDifStraggling}
\begin{aligned}
-\frac{1}{2} & \varepsilon {S_e} (\varepsilon)\left(1 +\frac{\mathcal{W}(\varepsilon)}{S_e (\varepsilon) \varepsilon} \right) \bar{\nu}^{\prime \prime}(\varepsilon)+S_e (\varepsilon)\bar{\nu}^{\prime}(\varepsilon)=\\
& \hspace{-0.55cm}\int_{\varepsilon u(\varepsilon)}^{\varepsilon^{2}} dt \frac{f\left(t^{1 / 2}\right)}{2 t^{3 / 2}}\times[\bar{\nu}(\varepsilon-t / \varepsilon)+\bar{\nu}(t / \varepsilon-u(\varepsilon))-\bar{\nu}(\varepsilon)],
\end{aligned}
\end{equation}
where $\varepsilon\equiv\mathcal{Q}E$, $\bar\nu(\varepsilon)=\mathcal{Q}E_\nu$, and  $u(\varepsilon)=\mathcal{Q}\,U$, with $\mathcal{Q}=16.26\,[Z_1Z_2(Z_2^{0.23} +Z_1^{0.23})\, \rm{keV}]^{-1}$ \footnote{$E_{CM}=E/2$ for identical atoms}. The integration variable is $t=\varepsilon^{2}\sin^{2}(\theta/2)$, with $\theta$ the scattering angle in the center of mass frame.
It is pertinent to point out that the original Lindhard formula for $\bar\nu (\varepsilon)$ \cite{lindhard:1963} is an approximate solution to the first-order integro-differential equation, with $u=0$ and an electronic stopping power valid for energies above 10 keV (bare ions). This precludes its applicability to lower energies. 
\begin{figure}
    \centering
    \includegraphics[width=0.99\linewidth]{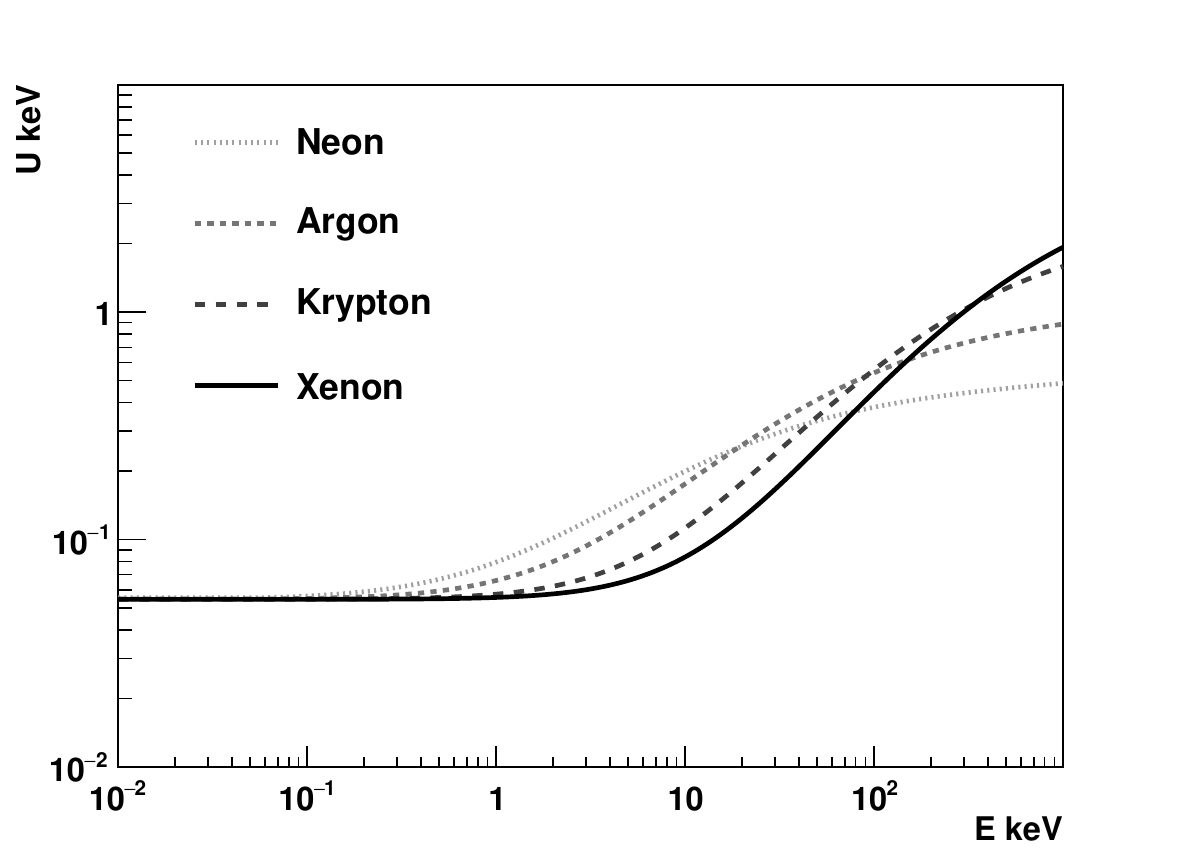}
    \caption{Energy absorbed by the electronic degrees of freedom for different noble elements, $U(E)=U_c +U_{TF}(E)$ as a function of the kinetic energy of the ion in keV, the constant term $U_c$ is the energy required to create an exciton.}
    \label{fig:MultiU}
\end{figure}

The solution to the second order equation in Eq. \eqref{Ec:MainintegroDifStraggling} that we derive here incorporates the effects of atomic binding energy, electronic straggling in nuclear collisions, and electronic stopping. 
For the nuclear stopping power, we use LSS theory \cite{lindhard:1963}, with ion-ion differential cross-section parametrized according to Ziegler  $f(t^{1/2})$ \cite{Ziegler}. The effect of the binding energy is modeled  as the sum  of an $\varepsilon$-dependent term, $u_{TF}(\varepsilon) = \mathcal{Q}\,U_{TF}(\varepsilon)$, plus a constant term, $u_c$, fixed by an energy threshold proportional to the average energy needed to create an exciton in a noble liquid (see Appendix \ref{A00}). $U_{TF}(\epsilon)$ is given by the average energy to excite electrons in the atomic collision, calculated from  the Thomas-Fermi (TF) formalism, and, as
shown in Fig. \ref{fig:MultiU}, it depends on the atomic number of the noble element. 

We adopt the model of Tilinin \cite{Tilinin}, in which, for impact parameters small compared with $a$, the electronic stopping power is governed by the mean free path of the electron, $\lambda_T$, during the formation of the quasi-molecule, and the energy transferred elastically by the electron per collision,
$\Delta E_T = (m\bigl(v+v_F\bigr)^{2} - m v_F^{2})/{2} \simeq m v v_F$, where $m$ is the electron mass, $v$ is the ion velocity and $v_F$ is the Fermi velocity of the electron, with $v \ll v_F$ in the quasi-molecular system. $S_e$ is obtained by integrating the volume of the quasi-molecule and multiplying the result by the atomic density, $N_b$, namely,
\begin{equation}\label{TilininFormulaP1}
S_e = N_b \int_{V} \frac{\Delta E_T}{\lambda_T}\, dV,
\end{equation}
with  $\lambda_T^{-1} = n_{eT}\,\sigma_{tr}$, where $n_{eT}$ is the electronic density in the quasi-molecule and $\sigma_{tr}$ is the momentum transfer cross section (MTCS) of the electron. The electronic density determines the Coulomb repulsion effects \cite{QF2023,BEZRUKOV2011119} and can be approximated by the sum of the contributions of the projectile, $n^{P}_{eT}$, and the target, $n^{T}_{eT}$. For atoms with the same mass and atomic number, we have $n_{eT}=2n^{P}_{eT}$. 

\begin{figure}[h]
    \centering
    \includegraphics[scale=0.46]{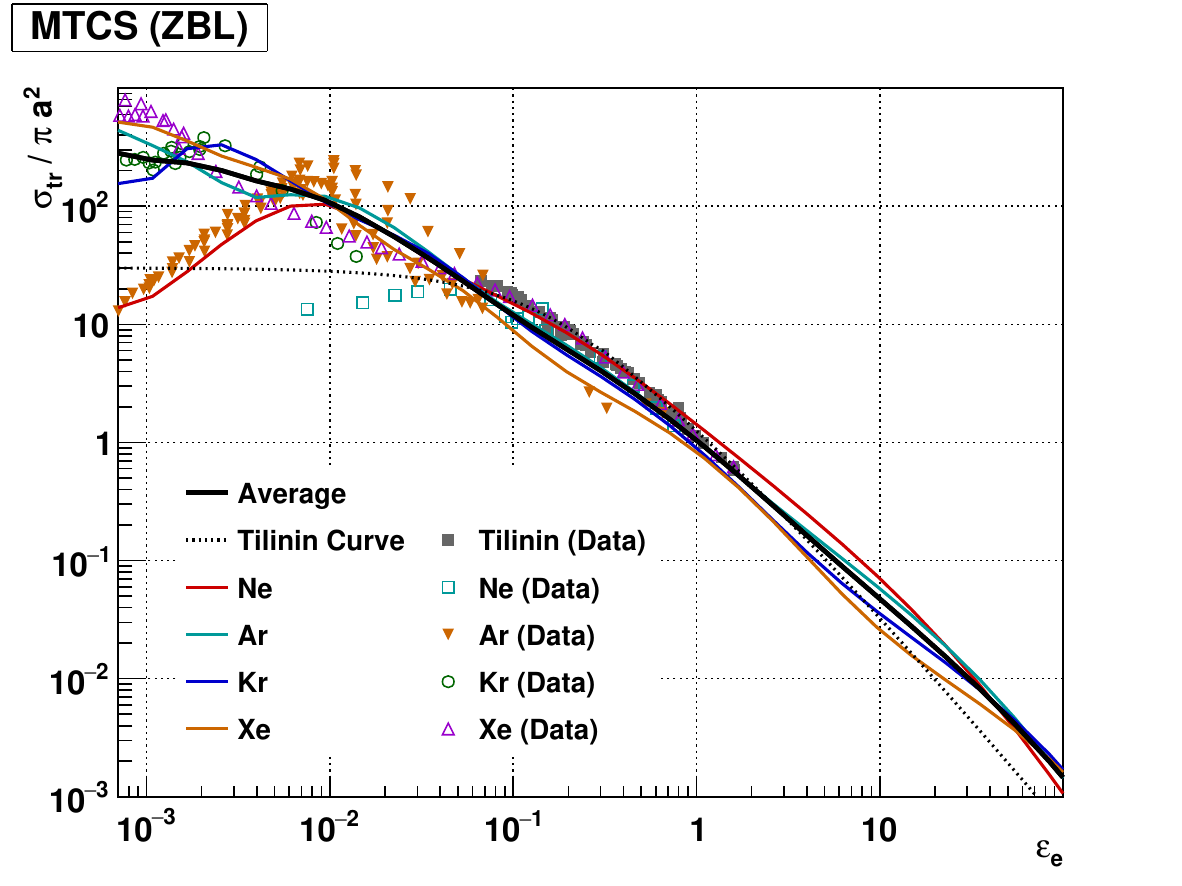}
    \caption{Momentum-transfer cross sections (MTCSs) for electron--atom scattering in Ne, Ar, Kr, and Xe. The black curve denotes the mean cross section across these noble-gas targets. Colored markers indicate measurements from distinct experimental datasets; the Tilinin dataset additionally includes $\rm{O_2}$, $\rm{N_2}$, $\rm{Ar}$, and $\rm{Ne}$. Solid lines show MTCS predictions obtained using the semiclassical Tilinin formalism (dotted) and,  partial-wave calculation employing the ZBL interaction potential.}
    \label{fig:MTCS}
\end{figure}

This model better describes the low energy region than the one in \cite{lindhard:1963} and  works  well for silicon \cite{Chavarria,QF2023}. However, it uses a semiclassical momentum transfer cross section for electron-atom scattering, and its applicability is limited to energies higher than a threshold proportional to the atomic number ($1.4$ keV for Ne and $70$ keV for LXe). To examine the low-energy regime, we implemented a    solution for the MTCS using a partial-wave analysis with the ZBL potential for the atomic numbers of Ne, Ar, Kr, and Xe.   In Fig.~\ref{fig:MTCS}, we compared our computed values of $\sigma_{tr}$ with available experimental data for these noble-gas elements \cite{MTCS,MTCSKr,MTCSLXe,MTCSNe,Hayashi1983DeterminationOE}. The comparison is presented as a function of the reduced energy, $\varepsilon_e = 32.52\, E_e\,\text{keV}\, Z^{-4/3}$. In addition, we report the mean trend obtained by averaging across all species, which highlights the near-universal behavior of the MTCS at electron energies exceeding 10~eV. We have  used the standard Fermi scale, $a=0.885\,a_0\,Z^{-1/3}$, where $a_0$ is the Bohr radius, and add the data for other elements reported in \cite{tilinin1988elastic}.  In this way, the validity of our model 
extends from 
$\varepsilon_e = 0.1$ to 
$\varepsilon_e = 10^{-3}$. For $E_R = 2\sqrt{2}ZE_e$, this interval corresponds approximately to nuclear recoil energies from $10\,\text{keV}$ to $0.1\,\text{keV}$, for LXe. As shown in the Figure, this energy region corresponds to the peak of the cross section for both LAr and LXe. Since existing data are for noble gases, extending the applicability for liquid phase to lower energies would require considering the liquid structure and polarization potentials \cite{MTCSLAr,MTCSdat2}.

For electronic straggling \(\mathcal{W}\), in a first-order approximation, we assume that the electronic and nuclear straggling contributions are equal. Consequently, we adopt the expression provided by \cite{WilsonF}, in which \(\mathcal{W}\) is evaluated on the basis of the nuclear straggling. Given that \(\mathcal{W}\) appears only through the second-order derivative in Eq.~\ref{Ec:MainintegroDifStraggling}, rescaling the electronic straggling by a factor \(M_0\) induces only a small modification of \(f_n(E_R)\), with the effect being more pronounced at low recoil energies.

\subsection{Electronic scaling parameter}\label{SubScallingPar}
As  discussed in \cite{QF2023}, the electronic stopping power must be corrected with the fermion factor $\xi_e$, which accounts for the energy dependence of the electronic excitations and depends on the media. Then, the scaling length becomes $a=0.885\,a_0\,Z^{-1/3}\,\xi^{2/3}_e$, and the electronic stopping power can be written as $S_e(\epsilon, \xi_e)=\xi_e S_e(\epsilon)$. Lindhard et al \cite{lindhard1954properties} derived the expression $S_e(\varepsilon,\xi_e)=\mathcal{K}\,\varepsilon^{1/2}$ for bare ions, where $\mathcal{K}=0.133\,(Z/A)^{1/2}\xi_e$, with $\xi_e\cong Z^{1/6}$.

 For high-energy collisions, all electrons  up to the Fermi level can be  effectively excited (Lindhard assumption) and $\xi_e \to 1$, whereas for low energies, only electrons close to the Fermi level can be excited and $\xi_e\rightarrow (5/3)^{3/2} \cong 2.15$.   
Instead of the fit for $\xi_e$ given in \cite{QF2023}, here we define the scaling factor  from Thomas-Fermi (TF) theory, using the plane-wave approximation for electrons:  
 \begin{equation}\label{ScalingLenght}
     \xi^{2/3}_e \equiv \frac{5}{3}\frac{\int^{k_F}_{k_F-\Delta k}E(k)k^2dk}{ E(k_F)\int^{k_F}_{k_F-\Delta k}k^2dk},
 \end{equation}
where $E(k)={\hbar^{2} k^2}/{2m}$, $k_F$ is the average Fermi wave-number of electrons, $\Delta k$ is the maximum electron dispersion in the collision, which depends on the projectile energy, and $E(k_F)=U_0/\xi_e^{2/3}$, with $U_0=24.04\,Z^{1.23}$ eV for a two atom system and $U_0=20.92\, Z^{4/3}$ eV for a single atom. 

Associated with $\Delta k$ there is an average energy available for excitation below the Fermi level, which we denote as the average Fermi energy of skin penetration  $\Delta E_F$ \footnote{$<\frac{(\bar k_F-\Delta \bar k)^2}{2m}>=E_F +\Delta E_F$, since for the free electron gas distribution, the term $<\bar k_F\cdot\Delta\bar k>=0$.}. Another interpretation of $\Delta k$ is as follows. Suppose that electrons scatter elastically, then $\Delta k=k_F\sin(\theta_c/2)$, where $\theta_c$ is the scattering angle in the CM frame, as shown in Appendix \ref{A0}. For  $\theta_c=\pi$, we have $\Delta k=k_F$, and only radial collisions are allowed \footnote{For $\theta_c<\pi$ we have electron collisions described  by a  radial contribution potential with a non central term (e.g. electron-electron, spin coupling).}. For angles above $\theta_c$ we have damping  due to energy transfer from electrons to inner shells, then only electrons up to the  energy $\Delta E_F$ have to be taken into account. 

For noble elements, the $\xi_e$ factor has to decrease as the atomic number increases, because the  TF model has been shown to be exact in the limit $Z\rightarrow\infty$ \cite{NARNHOFER1981128}. For the cascade process, the value of the momentum transferred is related to the average kinetic energy of the electrons in the atom, and consequently the electronic stopping power depends on this energy. Thus, the upper limit for $\xi_e$ can be estimated (in the gas phase) using the expected TF   average kinetic energy for electrons in an atom,  
 \begin{equation}\label{Eq:KineticEnergy}
 T_F/Z \equiv <T_F>= 20.92\,\left(Z^{2}/\xi_e\right)^{2/3}\,\text{eV}.
 \end{equation}
 For $\xi_e =1$  the average kinetic energy is overestimated, because Eq. \eqref{Eq:KineticEnergy} ignores the exchange effect and sudden variations in electron density near the nuclei and in the  valence orbital region.  The maximum value for $\xi_e$ can be calculated using the formula  \eqref{Eq:KineticEnergy} for the hydrogen atom  and $<T_F>=<T_e>=13.6057$ eV, which gives us $\xi^{max}_e=1.91$.

For multielectron atoms we can determine $<T_F>$ using  atomic orbital calculations based on local density approximations \cite{OrbitalComp,NIST}, see Table \ref{tab1}. It can be shown that the values for $\Delta E_F$ follow the approximate relation
 $\Delta E_F\cong 0.29\,Z^2 \,\rm{eV}$.

\begin{table}[h!]
\begin{tabular}{ | c | c | c | c | } 
  \hline
  Element & $<T_F>_{RLDA}$ eV & $\xi_e$ &  $\Delta E_F$ eV \\ 
  \hline
  He & 37.7 & 1.66 &  1.2\\ 
  \hline
  Ne & 348.4 & 1.47 & 26.2 \\ 
  \hline
  Ar & 797.1 & 1.38 &  85.3\\ 
  \hline
  Kr & 2114.4 &1.27 &  347.0\\ 
  \hline
    Xe & 3793.9 &  1.19 & 855.6 \\ 
  \hline
\end{tabular}
 \caption{Z orbital average kinetic energies for different noble elements calculated in refs.\cite{OrbitalComp} and \cite{NIST}, using the average value among  the local-density, relativistic local-density, local-spin-density  and
scalar-relativistic local-density approximations. The $\xi_e$  factor has been computed by using Eq. \eqref{Eq:KineticEnergy} and  $<T_F>_{RLDA}$. In the last column, we tabulate the corresponding average skin Fermi penetration energy $\Delta E_F$.}
    \label{tab1}
\end{table}

The electron-electron interaction can  be significant for low $Z$ elements, resulting in an ideal fermion system with lower $\Delta E_F$. According to the TF approach and Eq. \eqref{ScalingLenght} electrons with kinetic orbital energy up to $\Delta E_F$ behave as free particles. Therefore, for light elements, only the outer electrons fulfill this assumption. In the limit $Z\rightarrow \infty$, we have $\Delta E_F\rightarrow E_F$ as expected.

The estimation of $\xi_e$ for noble liquid elements  can be done by adding to 
 $\Delta E_F$ the gap energy  to create a pair of electrons, $E_g$. Then, we can compute the 
 $\xi_e$ factor for the liquid phase by means of  Eq. \eqref{ScalingLenght}, see Table \ref{tab2}. This addition only affects those elements with an atomic number lower than $Z=36$ and, in particular, for LAr the correction is only of $3\%$.  In the case of He, the MIMAC experiment \cite{HeQF}  indicates a lower value than the Lindhard model ($\xi_e=1.12$), in better agreement with the present model.
\begin{table}[h!]
\centering
\begin{tabular}{ | c | c| c | } 
  \hline
  Element & ($\Delta E_F + E_g$)\,eV & $\xi_e$\,(liquid phase)  \\ 
  \hline
  He &  25.8 &  1.03 \\ 
  \hline
  Ne & 47.7 & 1.33 \\ 
  \hline
  Ar & 99.6 &  1.34\\ 
  \hline
  Kr & 358.7 &  1.26\\ 
  \hline
    Xe &  864.9 & 1.19 \\ 
  \hline
\end{tabular}
 \caption{Effective Fermi energy of skin penetration for noble liquid elements calculated using the energy gap values from \cite{GapEnergy,GapNe,HeBandG}. The last column shows the solutions of Eq. (\ref{ScalingLenght}) for the tabulated energies.}
    \label{tab2}
\end{table}

\subsection{Active-Inelastic channel}
\begin{figure}
    \centering
    \includegraphics[width=0.99\linewidth]{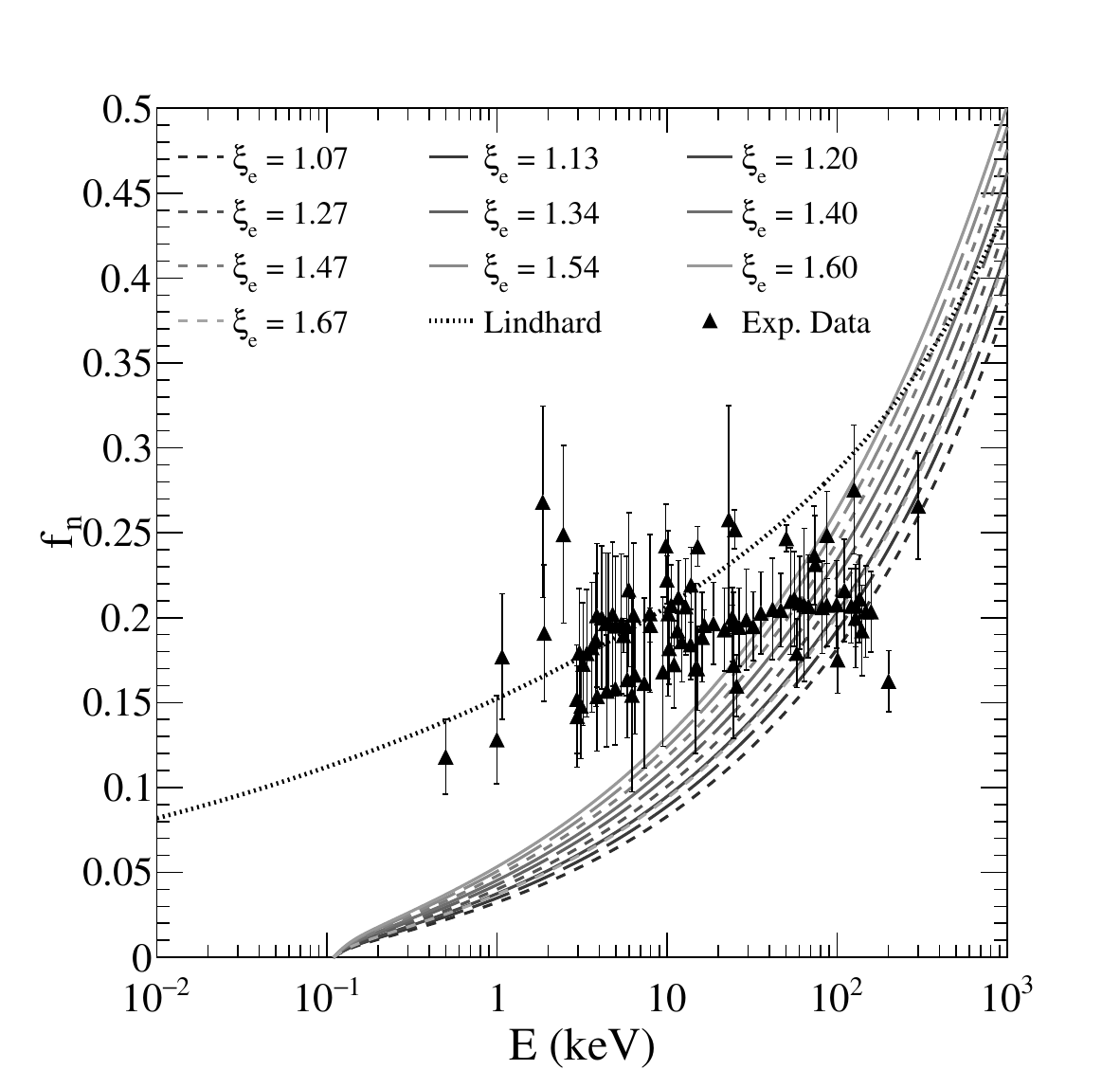}
    \caption{ (gray lines) Passive total quanta efficiency using  Tilinin electronic stopping power for LXe considering several $\xi_e$, (dotted) Lindhard model and data from \cite{NESTModel}. }
    \label{fig:TilininvsXe}
\end{figure}

For elements with a high atomic number such as LXe, the Tilinin model, used to account for the contribution of the passive channel, yields a low value of $f_n(E_R)$. The discrepancies between the predictions of the model and the observed data are illustrated in Fig. \ref{fig:TilininvsXe}. This indicates that other electron emission processes, particularly inelastic electron emission, should be considered.

When an electron associated with the projectile atom interacts with the inner-shell electrons of the target atom, an electronic interaction potential energy $u^e_{P}$ arises. This quantity is taken to correspond to the atomic electronic binding energy, $u_{TF}(\varepsilon)$, as described within the Thomas–Fermi  framework. By contrast, direct interactions between the valence electrons of the projectile and those of the target predominantly drive inelastic electron emission during the cascade process. These electrons are emitted with an energy $u^e_{A}=\bar\varepsilon_D$, where $\bar\varepsilon_D$ is the average electron kinetic energy associated with the valence-orbital energy $\varepsilon^{0}_D$. From \cite{FanoDoke} we can estimate $\varepsilon^0_D$ using two models; Shockley $\varepsilon^0_D=3E_{ex}/5$ and Alkhazov's $\varepsilon^0_D=E_{ex}/2$, where $E_{ex}$ is the exciton energy, see Table \ref{tab:kinetic_energy_noble}. 
 
 \begin{table}[htb]
\centering
\caption{\label{tab:kinetic_energy_noble}First exciton transition energy and estimated average electron kinetic energy  across noble liquid elements.}
\begin{ruledtabular}
\begin{tabular}{lccc}
{Element} & {$E_{\text{ex1}}$ (eV)} & { $(\varepsilon^0_D)_{\text{sub}}$ (eV)} & { $(\varepsilon^0_D)_{\text{Shockley}}$ (eV)} \\
\hline
{ He} & 19.80 & 9.90 & 11.88 \\
{ Ne}   & 17.50 & 8.75 &  10.50\\
{ Ar}  & 12.06 & 6.03 &  7.24 \\
{ Kr} & 10.17 & 5.09 &  6.10 \\
{ Xe}  & 8.37  & 4.19 &   5.02\\
\end{tabular}
\end{ruledtabular}
\end{table}
 
 Consequently, the total energy satisfies the following
\begin{eqnarray}
    \label{InelasticBalance}
    \varepsilon&=&(\bar\eta(\varepsilon) +u^e_{A} -u^{e}_{P}) +(\bar\nu(\varepsilon) -u^\nu_{P})\\\nonumber
    &=&\bar\eta (\varepsilon) + \bar \varepsilon_D + \bar \nu (\varepsilon)-u,
\end{eqnarray}
where $u^\nu_{P}$ is the nuclear recoil energy associated with the first process, and we define the total atomic binding energy as $u=u^e_P + u^\nu_P=u_{TF}(\varepsilon) +u_c$. The resulting total-quanta efficiency is
\begin{equation}
    \label{eq:totQuantaInelastic}
    f_n(\varepsilon _R)=\frac{\bar\eta(\varepsilon) +\bar\varepsilon_D}{\varepsilon_R},
\end{equation}
 where we identify $\bar\varepsilon_D/\varepsilon_R=f^{\rm{AI}}_n(\varepsilon_R)$ as the active channel contribution. Inelastic electron emission is energetically allowed only when the projectile energy exceeds the threshold $W_i$, which is equivalently expressed as requiring an interaction distance smaller than $r_M$. At $r_E=r_M$, the emission probability $p(r_M)$ attains its maximum. For $E>W_i$ (i.e., $r_E<r_M$), electron emission may still occur, but with a reduced probability. The expected number of emitted electrons is given by $p(r_E)E/W_i$, and the corresponding mean electron energy is $\bar\varepsilon_D=p(r_E){\varepsilon^{0}_D}E/W_i$. Incorporating this inelastic energy-loss channel into the quenching factor yields:
\begin{equation}
    \label{EqfnDielectronic}
    f_n(\varepsilon_R)=\frac{E_{ee}}{E_R}=\frac{\bar \eta(\varepsilon)}{\varepsilon_R} + \frac{p(r_E)\epsilon^0_{D}E}{W_iE_R}. 
\end{equation}
Given that the process is governed by Coulomb interactions between electrons, the scaling relation \(E/Z^{2} \cong 1/r_{E}\) follows. Consequently, \(p(r_{E})\) is expected to increase proportionally to \(Z^{2}\).
The IDES formulation for $f_n(E_R)$ is determined by the underlying interatomic interaction potential, the effective scaling length $\xi_e$, and the atomic number and atomic mass of the target medium. In the present work, the effective atom–atom interaction is modeled using the Ziegler potential \cite{Ziegler}.

\section{Active Bates-Griffing electron scattering}
\label{BatesGriffingSection}
\subsection{Energy-dependent $\beta$ model}\label{SubBatesGriffing}
\begin{figure}
    \centering
    \includegraphics[width=0.92\linewidth]{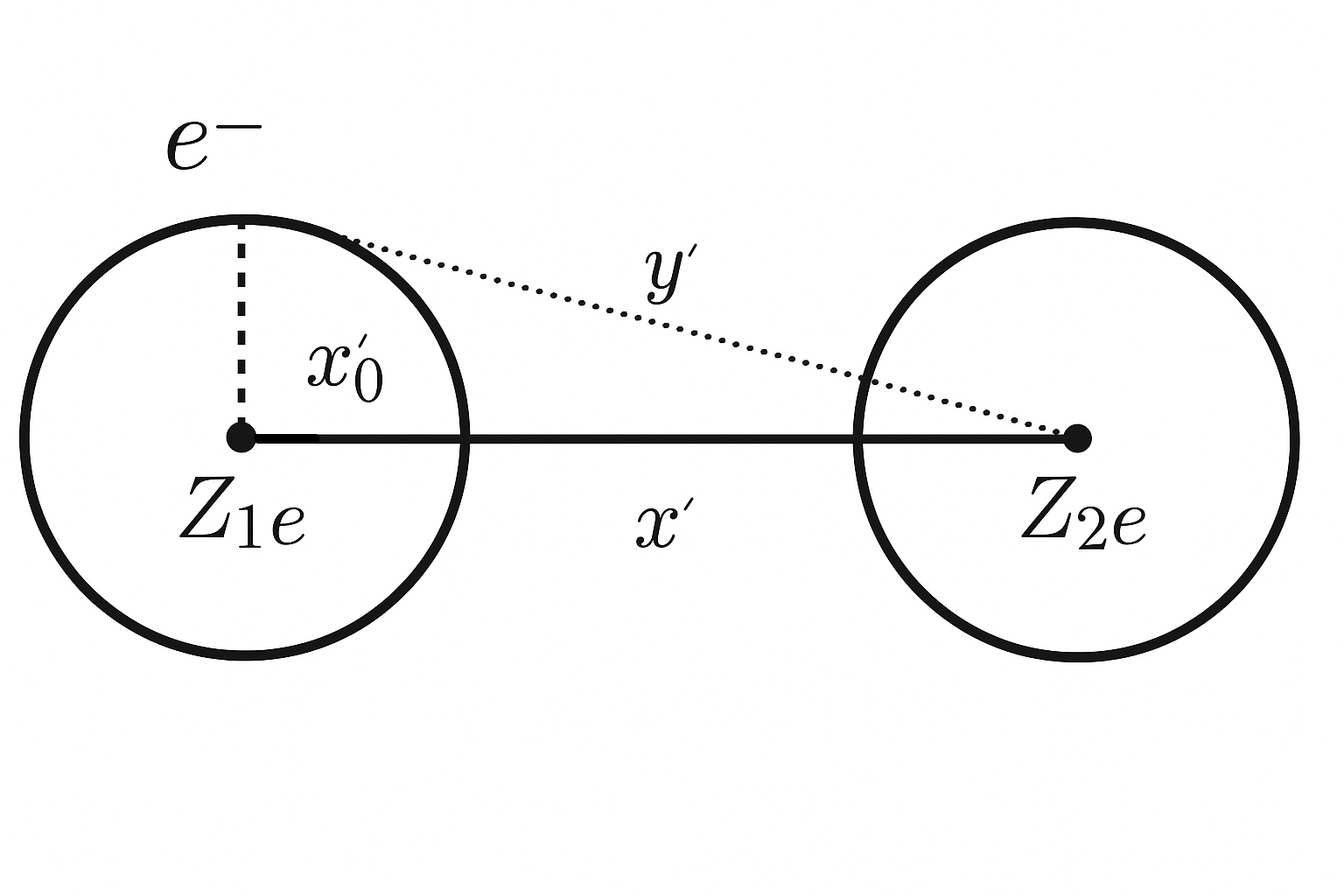}
    \caption{Diagrammatic  description of the semi-classical BG model. The electrons in the target are centered around the nuclei position ($\cong 0$) with orbital radius $x'_0$. These electrons interact with Coulomb potential by a separation $x'$ and with the screen potential by the distance $y'$.}
    \label{fig:Bates}
\end{figure}

In the active channel of collisions between pairs of identical atoms, one electron from each atom is released from its orbital through its energy-dependent electrostatic interactions with the electrons and nucleus of the atom it collides with. This process is modeled in terms of the Bates-Griffing (BG) scattering adapted to the present case for energies of the order of keV. We use a semiclassical approach, assuming that only valence electrons participate in the process. These are located at the radii of their valence orbitals. Since the electron clouds are not expected to change significantly during the collision, the average positions of the electrons can be considered nucleus-centered. When the two atoms are separated by a distance $x'$, ignoring spin effects, a valence electron from one atom interacts through the Coulomb potential of the valence electrons and the shielded nucleus of the other atom, see Fig. \ref{fig:Bates}.

The atom-atom system is described by a multi-electron wave function $\psi(r_e,R)$, where $R$ is the nucleus separation and $r_e$ denotes the set of vector positions of electrons in both atoms. In the Born-Oppenheimer approximation 
$\psi(r_e,R)\cong\psi_e(r_e)\psi_\mathcal{N} (R)$ \cite{Born-Opi}. We define  $x=R/a_Z$, in terms of the Ziegler units $a_Z=0.885\,q\,\xi^{2/3}_e a_0/(2 Z^{0.23})$, with $Z_1=Z_2=Z$. Here, $q$ is a correction factor for atom resizing, which, ignoring relativistic corrections, can be determined by the Hartree-Fock model \cite{qFactor}. The respective values for LNe, LAr, LKr, and LXe are 0.667, 0.92, 0.95, and 1.08.  For LXe, a relativistic contraction correction is needed, giving $q=1.05$ \cite{relativistiqCorrection, atomicqCorr2}, while for LAr, using data from Ref. \cite{WilsonF}, we get $q=0.942$. 

We use the semi-empirical potential energy \cite{Ziegler},
\begin{equation}\label{PotentialDefinition2}
    e\phi_Z(x,\xi_e)=\left(\frac{Z_1Z_2e^2}{a_Z x}\right)\chi_{Z}(x),
\end{equation}
where $\chi_{Z}(x)$ is the Ziegler screening function:
\begin{multline*}
     \chi_{Z}(x)=0.18175\,e^{-3.1998\,x} + 0.50986\,e^{-0.94229\,x} + \\
      0.28022\,e^{-0.40290\,x} + 0.02817\,e^{-0.20162\,x}.
\end{multline*}
Consider an incoming atom with kinetic energy $E$, then the minimum approach distance, $x_m$, is obtained from the equality $E={e\phi_Z(x_m,\xi_e)}$ .

At the end of the cascade process, for a given incoming energy $E$, the atoms involved in the collisions may have an effective number of electrons $N<Z$  determined  by the  Bohr stripping criteria  \cite{BohrStrpC,GumusH}, see Appendix \ref{A1}. Assuming that \(N\) remains constant during the recombination process, the potential describing a neutral atom (\(Z = N\)) must be replaced by a potential characterized by an effective charge \(Z - N\). This modification can be implemented using the Amaldi model. Within this framework, the electron–ion potential energy can be written as
\begin{equation}
    e\phi^e_Z(x',\xi_e,N)=-\frac{Ze^2}{bx'}\left[\left(\frac{N}{Z}\right)\chi_Z (x') +\frac{(Z-N)}{Z} \right],
\end{equation}
with $x'=r/b$ and $b=q\,0.885\,\xi^{2/3}_e\,a_0N^{-1/3}$. For details on the validity of this potential, see Appendix \ref{A2}. 

  The orbital radius of the valence electron with $l\neq 0$  is determined by condition $e\phi^e_Z(x'_0,\xi_e,N)/2=-\,U_e$, where $U_e$ is the average electron binding energy of the valence electrons: $21.6$ eV, $15.82$ eV, $14.1$ eV, and $12.57$ eV for Ne, Ar, Kr, and Xe, respectively \cite{ValenceUe,IonizationEnergies_Rossler1970}. 

\begin{figure}
    \centering
    \includegraphics[width=0.95\linewidth]{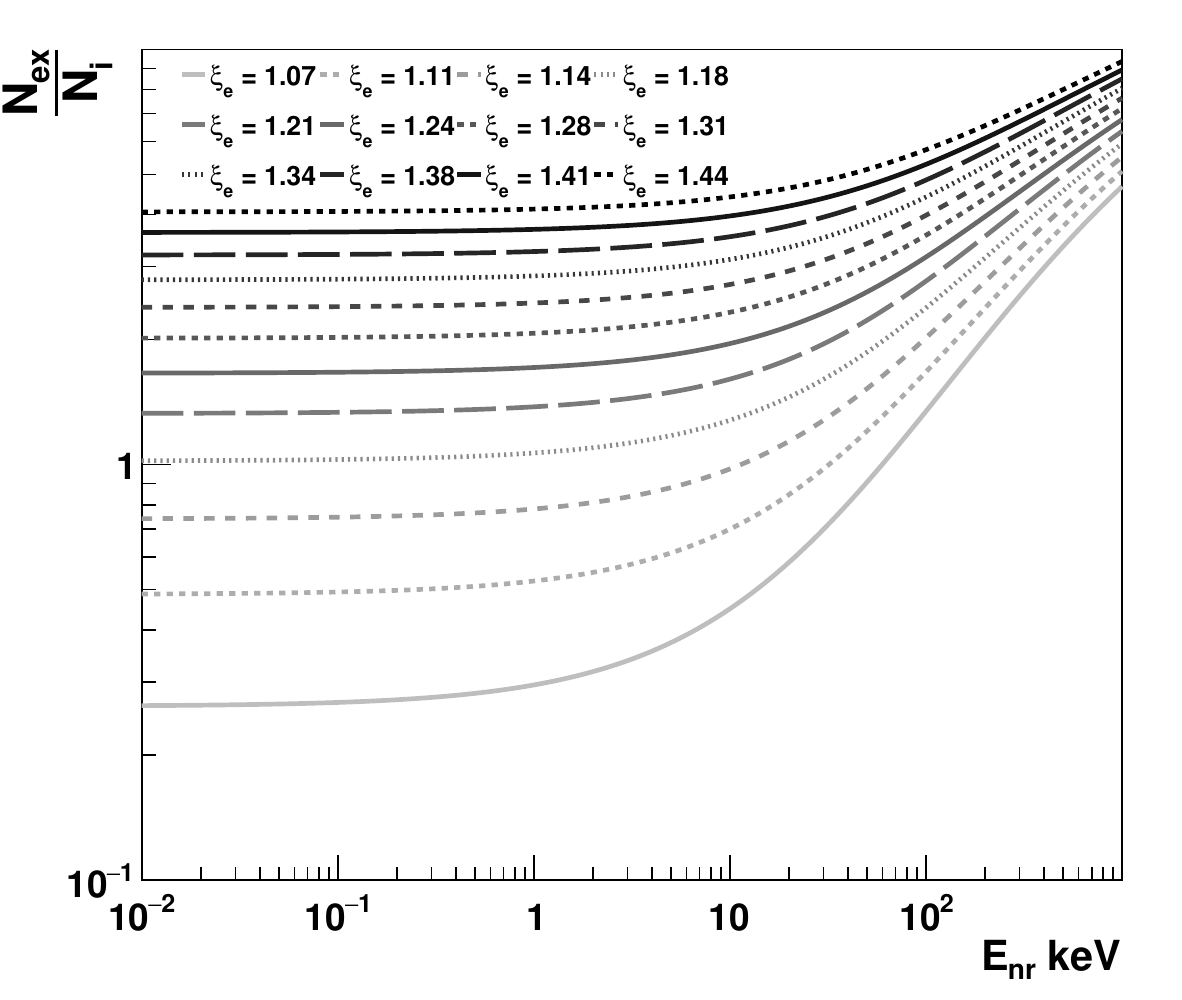}
    \caption{Exciton to ion ratio for LNe computed with Eq \eqref{Eq Ratio Nex/Ni} for several scaling lengths.}
    \label{fig:RatioLNe}
\end{figure}

\begin{figure}
    \centering
    \includegraphics[width=0.95\linewidth]{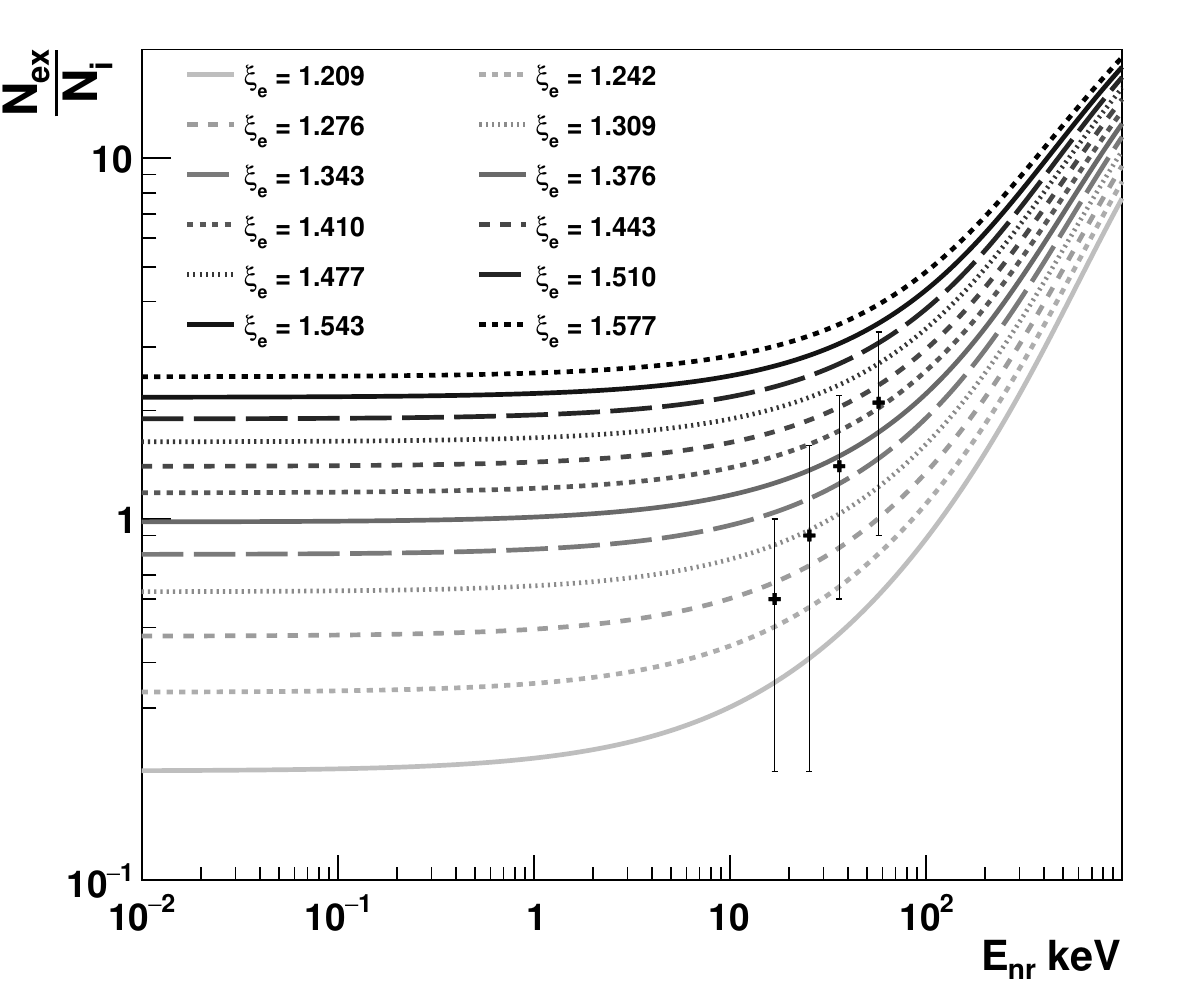}
    \caption{Exciton to ion ratio for LAr computed with Eq \eqref{Eq Ratio Nex/Ni} for several scaling lengths. We also show SCENE \cite{SCENELAr} data, which agree with $\xi_e=1.34$ with $\chi^2/\mathrm{ndf}=1.6/4$.}
    \label{fig:RatioLAr}
\end{figure}

\begin{figure}
    \centering
    \includegraphics[width=0.95\linewidth]{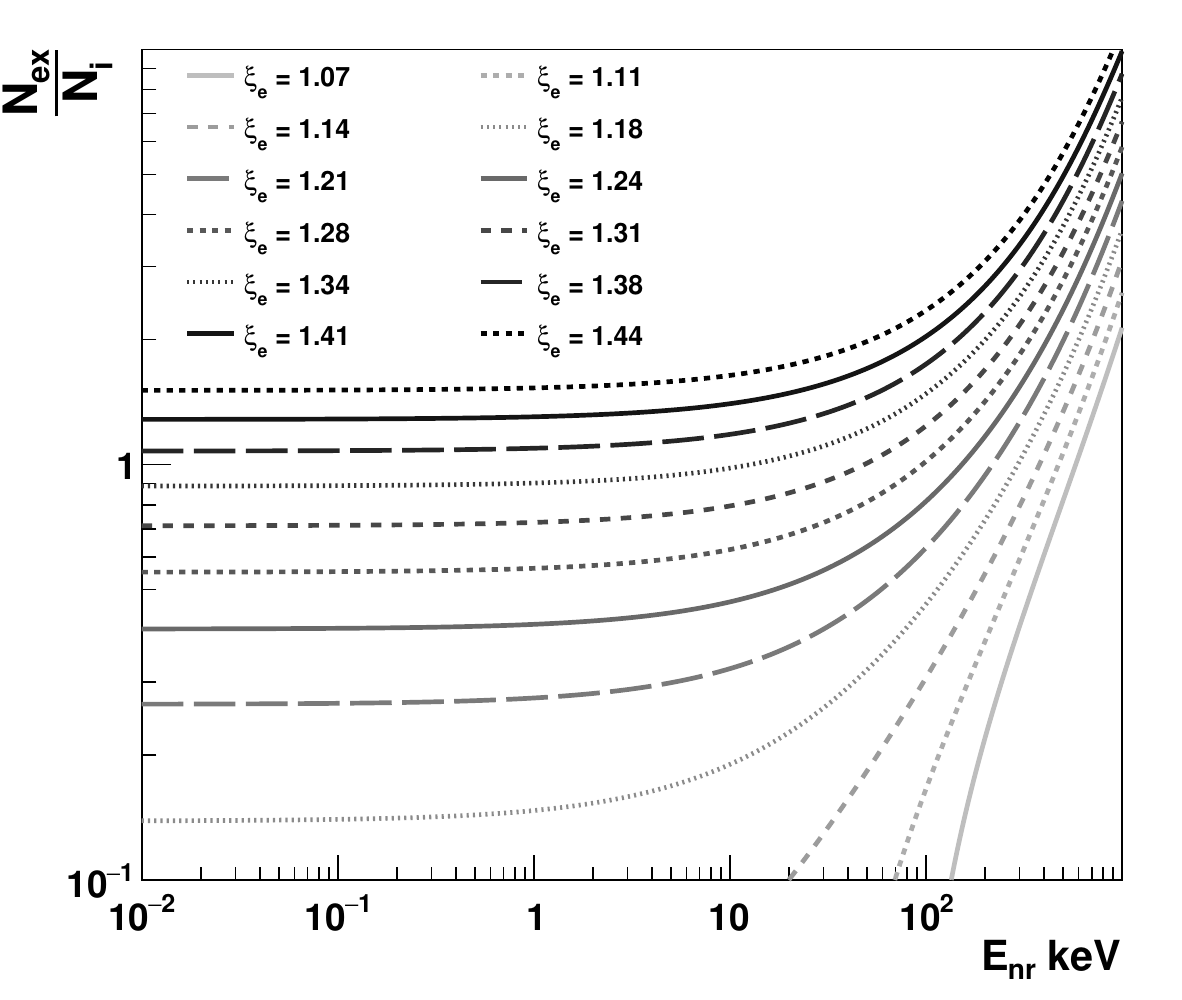}
    \caption{Exciton to ion ratio for LKr computed with Eq  \eqref{Eq Ratio Nex/Ni} for several scaling lengths.}
    \label{fig:RatioLKr}
\end{figure}

\begin{figure}
    \centering
    \includegraphics[width=0.95\linewidth]{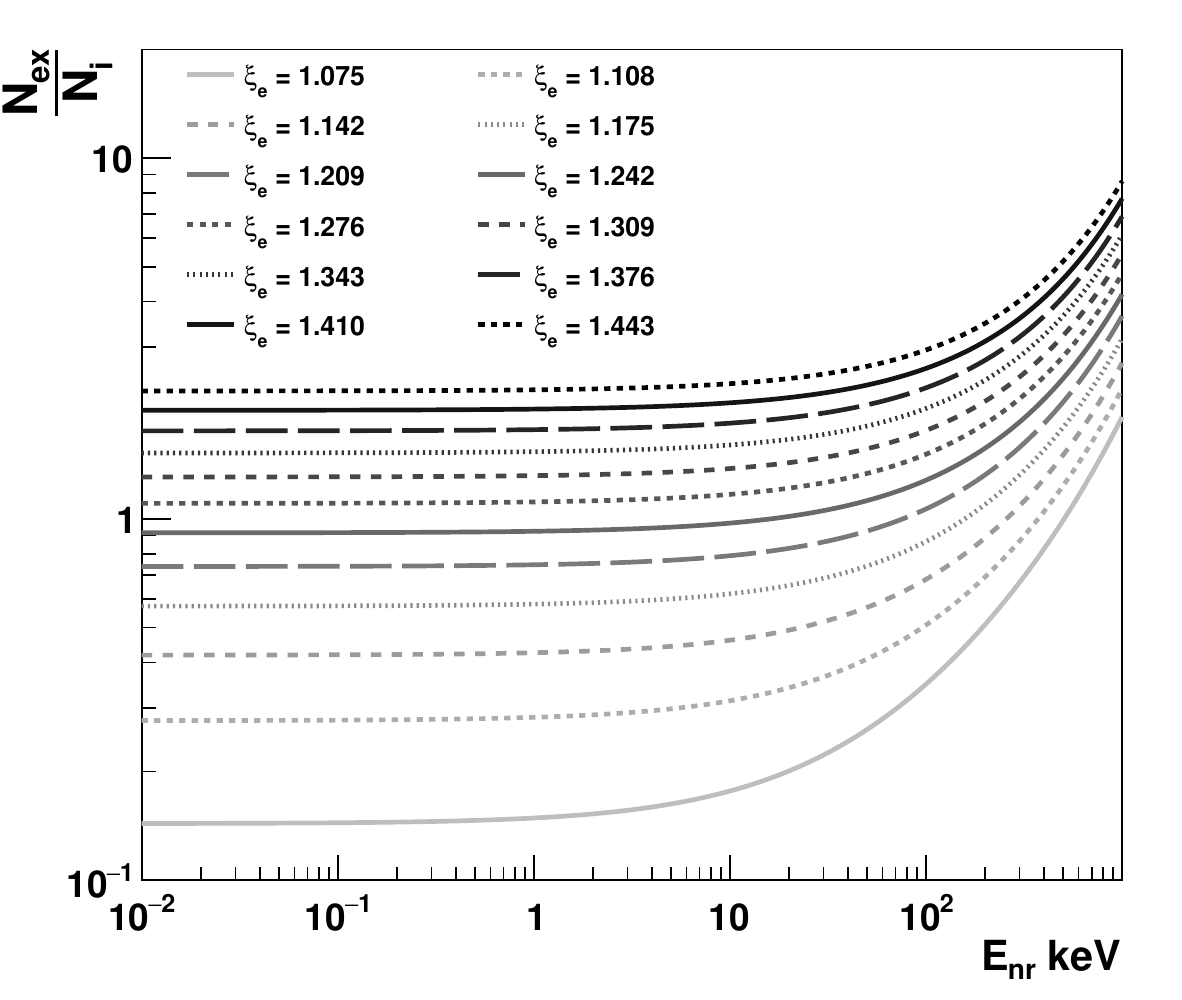}
    \caption{Exciton to ion ratio for LXe computed with Eq  \eqref{Eq Ratio Nex/Ni} for several scaling lengths.}
    \label{fig:RatioLXe}
\end{figure}

The total energy of the system is given by the sum of negative projectile binding $\mathcal{E}_U=-\,U_e/2$, negative projectile-target binding $\mathcal{E}_y=e\phi^e_Z(y',\xi_e,N)/2$, with $y'=\sqrt{x'^2 + x'^2_0}$, and positive Coulomb electron-electron repulsion ($T_C=0$) $\mathcal{E}_C={e^2}/{(bx')}$. For an electron to be emitted from the projectile, the total energy must be zero, which gives the transcendental equation,
\begin{equation}\label{MainPotentialEquation}
     \frac{U_e}{2}=\frac{e\phi^e_Z(y',\xi_e,N)}{2}+ \frac{e^2}{ bx'}.
\end{equation}
We can determine $x'$ by solving Eq. \eqref{MainPotentialEquation}  for a given  $N$. Subsequently, the  associated ion-ion potential energy, denoted as $W_i$, required to ionize an electron from one of the ions can be calculated. 
\begin{equation}\label{WiEquation}
    \frac{W_i(E_R,\xi_e)}{2}=\frac{e\phi_Z(x,\xi_e)}{2},
\end{equation}
with $x=x'b/a_Z$.  Using the definition \eqref{WiEquation},
\begin{equation}\label{Eq Ratio Nex/Ni}
\frac{N_{ex}}{N_i}(E_R,\xi_e)=\left(\frac{W_i(E_R,\xi_e)}{W}-1\right) ,
\end{equation}
we can obtain the exciton to ion ratio. Since $N$ is expected to decrease with $E_R$ the screening in the potential $\phi^e_Z$ will decrease, and therefore the right-hand side of \eqref{WiEquation} will increase with energy; see Figures \ref{fig:RatioLNe}, \ref{fig:RatioLAr}, \ref{fig:RatioLKr} and \ref{fig:RatioLXe}. Consequently, we infer that $N_{ex}/N_i$ has to be an increasing function of energy, with minimum  value at the limit $E_R\rightarrow 0$ where $N\rightarrow Z$. $N_{ex}/N_i$ is expected to also be sensitive to the scale parameter in the allowed interval of values $1 \leq \xi_e\leq 2.15$.  For values near $\xi_e = 2.15$ the electrons below the Fermi energy cannot be ionized and only excited by the exchange mechanism, hence it is expected that $N_{ex}/N_i$ be greater than for the opposite case.  For electron recoils in LAr and LXe, $(N_{ex}/N_i)_{min}=0.23$ and $(N_{ex}/N_i)_{min}=0.06$ \cite{ArisLAr,XeEXP1} may be considered the lowest limit value for the respective nuclear recoil ratios. 
\subsubsection{Spin effects}\label{Sec3Spin}
  Because of permutation symmetry between the electrons, for a multi-electronic wave function with spin there is an extra interaction energy contribution to the total Hamiltonian of nonclassical origin (exchange interaction) \cite{Exchange}. Consequently, a more precise description of the BG process can be given by considering the spin wave function in the binary process.  

  For two approaching atoms, at separation distance $x\gg 1$, the total potential energy of an electron in each atom is given by the sum of the two individual potentials: $V_1(x'_1) +V_2(x'_2)$, where $x'_1$ and $x'_2$ are the corresponding electron coordinates. Taking into account the spin, the orbital wave function of the electrons becomes 
$\psi(x'_1,x'_2)=(\psi_1(x'_1)\psi_2(x'_2) \pm \psi_1(x'_2)\psi_2(x'_1))/\sqrt{2}$, where we can define the average Coulomb potential $<\psi\mid V_C\mid \psi>$ for $x\gg x'_i$ ($i=1,2$). At closer distance we have to include the perturbation potential $V_{nn} +V' +V'_{ee}$,
where $V_{nn}$ is the nuclei-nuclei potential, the electron potential in atom $1$ with the nuclei in  atom $2$, $V'$, and the electron-electron interaction between both atoms, $V'_{ee}$. Since 
$\mid V'+V'_{ee}\mid/V_{nn}\cong x'_i/Zx \ll 1$, the potential $V_{nn}$ makes the dominant contribution.

   Consequently, the perturbation potential is positive, and  the orbital energy level shifts to higher values. To gain more stability, electron orbitals can lower their energies through exchange interaction, which, according to the generalized Hund rule \cite{HundRulesTwoAtoms}, flip the spins to point in the same direction. For noble element quasi-molecules, we can  have  three electron spins down for the $p_{1/2}$ (shifted) orbitals and  three spins up for $p_{3/2}$. In this way, the total spin is conserved ($S=0$) and the multiplicity is maximized.

   For the singlet, the spatial wave function has to be symmetric and the Coulomb  potential energy  will have two terms, the direct term $e^2/(bx)$, considered previously, and the exchange term $V^{ex}$; $V_{C}=e^2/(bx') +V^{ex}_{ee}$. In this situation, since the overlap of the wave function is not too strong, for distances ($x\cong 2x'_0$) the active BG process is favored, where Eq. \eqref{Eq Ratio Nex/Ni} is satisfied ($x\cong 2x'_0$). Thus, $e^2/(b2x'_0) \gg V^{ex}_{ee}$ and Eq. \eqref{MainPotentialEquation} remains valid as a good (classical) approximation.

  For the triplet case, the spatial wave function is antisymmetric, leading to a potential energy reduction  $V_{C}=e^2/(bx') - V^{ex}_{ee}$. For distances $x<2x'_0$ the overlap of the wave function becomes significant and disfavors the active BG process.  Therefore, this case may lead to a passive BG process. 
In principle, the active channel can be computed from the interaction between two atoms and Eq. \eqref{Ec:MainintegroDifStraggling}. This calculation requires the emission cross section, which depends on the atomic orbitals of both atoms, as well as on the energy and spin degrees of freedom. A rigorous treatment of this process would require a spinor wave function obtained from the Dirac equation \cite{relativistiqCorrection}. Such a relativistic analysis lies beyond the scope of the present work; therefore, we adopt a semiclassical approach.

  To examine the probability of electron spin exchange, we account for the Coulomb energy of the quasi-molecule and the exchange energy. We assume that the Coulomb interaction is dominant.  We define the average electron exchange interaction distance $x_\lambda=(a_Z\sigma_t N_e)^{-1}$ and, assuming a Poisson process, determine the probability of an odd number of flip-flops during the collision,  $N_{ff}=x/x_\lambda \geq1$:
\begin{equation} \label{fnDielectronic}
    p(x)\cong\frac{P_s}{2}(1-e^{-2N_{ff}})\Theta(E-W_i),
\end{equation}
where $P_s$ is the singlet probability ($\cong 1/4$),  $\sigma_t$ is the total cross section, $\sum\sigma_i+\sigma_{el}$,  $\sigma_i$ is the ionization cross section for each sub-level $i$ and $\sigma_{el}$ is the elastic cross section,  $N_e$ the electron density of the quasi-molecule that contributes to the dielectronic process (by exchange interaction) and $x$ the minimum separation distance of the atoms; for details, see Appendix \ref{AppendixE}. 

   From Eq. \eqref{fnDielectronic} it can be argued that for low $Z$ elements $N_e$ is expected to be low, then  $x\ll x_\lambda$, so the emission probability is suppressed $p\propto x/x_\lambda\ll1$. For Ne and Ar, the BG effect is negligible compared with that observed for heavier elements such as Kr and Xe.



\section{Recombination process: Thomas-Imel box model}\label{RecombinationModel}
The recombination process is relevant only for liquid-phase detectors because the probability of its occurrence in solids is negligible.
 The partition between ionization and scintillation produced in a noble liquid by the initial recoil of an atom is governed by the recombination probability $p_r$ between the ions and excitons that were generated \cite{BEZRUKOV2011119,ThomasImel}. 
 The Thomas-Imel box model \cite{ThomasImel} assumes that the recombination mechanism is described by a classical process of ions and electrons recombining in a box of size $a_B$ in the presence of a uniform electric field $F_0$, given in units of $\rm{V/cm}$. 
 
 It is generally assumed that each ionization produces a single electron and that each excited atom and each recombination emit a scintillation photon \cite{Doke2002}. Then, for  $N_i$ ion-electron pairs generated by the cascade process in the box, the number $N_e$ of electrons escaping recombination predicted by Thomas-Imel is 
\begin{equation}\label{ThomasImel}
     \begin{aligned}
\frac{N_e}{N_i} & =1-p_r=\frac{1}{\xi} \ln (1+\xi), 
\end{aligned}
 \end{equation}
 where
 \begin{equation}\label{ThomasImel2}
     \xi =\frac{N_i  \alpha}{4 a_B^2 v_e},
 \end{equation}
  where $\mu_e$  is the electron mobility, $v_e=\mu_eF_0$ is the electron drift velocity and  $\alpha$ is a constant recombination parameter  with dimensions of
 $\rm{m^3 s^{-1}}$. Scintillation is equal to exciton quanta plus the fraction of recombined ions, that is, $N_{\gamma}= N_{ex} + p_rN_i$. The remaining $(1-p_r)N_i$ electrons escape recombination. Therefore, $N_{ex}+N_i=N_e + N_\gamma$.
 
\subsection*{Biexcitonic process}\label{SubBiexcitonic}

Excitons can self-trap at a frequency similar to that of the electron-hole recombination process. Before trapping, free excitons can diffuse and undergo biexcitonic quenching, \textit{i.e.}, $Ar^*+Ar^* \rightarrow Ar+Ar+e$. This mechanism is proportional to the energy deposited and produces light quenching in noble liquids \cite{Hitachi2}. Consequently, the total quanta is also quenched \cite{flFactor}:
\begin{equation}\label{Eq:Bioxitonic}
N_{{ex}}+N_{{i}}=\frac{N_\gamma}{  f_l(E_R)} +N_{{e}}=N_q,
\end{equation}
where $f_l(E_R)$ is the biexcitonic quenching factor, which is related to the Penning effect and can only be present in nuclear recoil ionization. Some authors \cite{MEI200812,BEZRUKOV2011119} have parameterized this process in terms of Birks' saturation law,
\begin{equation}\label{BirksLaw}
    f_l(E_R)=\frac{1}{1+k_BS_e(E_R)},
\end{equation}
where $k_B$ is called Birks' constant. For this model to work, the timescale of exciton self-trapping must be shorter than the recombination time at this recoil energy \cite{FOXE201524}. 

Let us consider the total quanta efficiency measured in a dual phase detector $f^*_n(E_R) = (N_{\gamma} +  N_{e})W/E_R$; from Eqs. \eqref{Eq:f_ndef} and \eqref{Eq:Bioxitonic}, it can be expressed as: 
\begin{equation}\label{eq: f_nMeasured}
    f^*_n(E_R)=f_n(E_R)f_l(E_R) +(1-f_l(E_R))\frac{N_eW}{E_R}.
    \end{equation}
Some observations are in order. First, $f_n(E_R)$ is an upper limit for the measured efficiency, and second, $f^*_n(E_R)$ depends slightly on the electric field through $N_e$. Because $N_e/E_R$ is much smaller than the light yield for $E_R$ greater than $50$ keV and $f_l(E_R) \cong 1$ for recoil energies lower than this value, we can make $f^{*}_n(E_R)\cong f_n(E_R)f_l(E_R)$. 

  Using $N_i=(E_R/W)f_n(E_R)/(1+\beta)$, and taking into account Eqs.~\eqref{Eq:Bioxitonic}, \eqref{ThomasImel}, and \eqref{Eq:f_ndef}, we obtain the following expressions for the charge and light yields:
 \begin{eqnarray}\label{LyandQyEq}
\frac{N_e}{E_R}&=& \frac{f_n(E_R)\ln (1+\xi)}{W(1+\beta)\xi},\\
\frac{N_\gamma}{E_R} &=&\frac{f_n(E_R)f_l(E_R)}{W}\left(1 - \frac{\ln (1+\xi)}{(1+\beta)\xi}\right) .
 \end{eqnarray}

 In recent works \cite{NESTModel,SCENELAr,ArisLAr}, the recombination parameters $\beta$, $\xi$, and $\alpha$ are determined from direct measurements. In some studies, the model depends on up to five parameters for each noble liquid \cite{kimura} and, particularly for LAr, the consistency with charge and light yields is not entirely satisfactory. Next, we present a comprehensive model that explains the recombination parameters and the total quanta of all noble elements as functions of a single experimental parameter. In a subsequent publication, we will provide the explicit formulas and software to implement the model.  

 For recoil energies below $\cong 1$ MeV ($v\ll Z^{2/3}v_0$ and $Z\geq10$) the recombination is not expected to drastically alter electronic orbitals, so that when  the cascade process ends and exciton-ion recombination begins, $\Delta k$ should not change significantly. It is important to note that the previous equations strongly depend on $f_n(E_R)$, the exciton-ion ratio $\beta$ and implicitly on the length of the electronic scaling $\xi_e$. Considering the interpretation of Dahl \cite{Dahl}, we will assume that the box size is determined by the electrostatic length scale, which implies a non-hyperbolic dependence of $\xi$ on $F_0$.


\subsection{Box size description in terms of the electrostatic electron-ion interaction}\label{SubBoxSize}

In this part of the study we present a model for the box size parameter $a_B$, valid for any noble liquid  detector, which  explains the electric field behavior of the parameter $\xi$ in Eq. \eqref{ThomasImel2}. The model depends on the microscopic properties of the noble liquid, the scale length $\xi_e$, and the electrostatic potential. 

During recombination, electrons and ions coexist until thermalization begins to contribute significantly. In the  Thomas-Imel  model this is characterized by the box size parameter $a_B$, see Eq. (\ref{ThomasImel2}), with the box assumed to be isolated and the ions and electrons uniformly distributed. Typical values of $a_B$ are of the order of micrometers \cite{szydagis2023reviewnestmodelsapplication,Dahl}. 

In many studies \cite{MOZUMDER1995143,BURNS1987381}, for the fit function of $\xi$, $a_B$ is taken as an average quantity and the dependence on the electric field is a power law: $\xi \propto 1/F^{\gamma}_0$, where $\gamma \neq 1$ is a nonuniversal parameter that depends on the medium. Instead, we will follow Dahl's ideas \cite{Dahl} and consider that $a_B$ is determined by the electrostatic equilibrium between electron-ion pairs and the applied electric field, namely, $\phi_{ext}(z,\xi_e)=-bzF_0$, where $z$ is the average  electron-ion separation. To estimate the lower electric field for which this criterion can be applied, we use the Coulomb potential for the electron-ion interaction and assume that the electron's time inside the box $t=2a_B/v$ is less than the thermalization time ($\cong 1$ ns), which gives us $F_0> 10 ~\rm{V/cm}$. These assumptions are supported by several observations \cite{Dahl, ArisLAr,flFactor, kimura, NESTModel}.

 Because the box size is expected to be much larger than the atomic scale, screening effects due to the induced charge and atom polarization have to be included. Since we are not in a relativistic regime,  we can use the linear Thomas-Fermi screening theory \cite{LiniarTFscreening}, where a semiclassical approximation for a slowly varying potential is assumed. Additionally, for low energy ion tracks ($< 1$ MeV), only  the  valence electrons need to be considered inside the box, thus the electron-ion potential 
becomes
\begin{equation}
    (\phi^e_Z(z,\xi_e,N))_{\rm{sc}} \cong \phi^e_Z(z,\xi_e,N)e^{-kz}, 
\end{equation}
where $k^{2}= 6\pi e^2 n_e/\Delta E_F$ is the Thomas-Fermi wave vector, with $n_e$ denoting the electron density in the box and, according to Eq. \eqref{ScalingLenght}, $\Delta E_F$ is determined by the value of the scaling length $\xi_e$. In this model, the number $N$ of effective electrons for each atom inside the box is given by the average ion energy, $E_i=2Z(\Delta E_F /2)$.

We define the electron density $n^0_e$ in thermal equilibrium by $N_{v} n$, where $n$ is the number density of atoms in the medium at a certain equilibrium temperature and $N_{v}$ is the number of valence electrons. The electron density is expected to vary from $n_e$ to $n^0_e$. To establish an approximate relation between $n_e$ and $n^0_e$, we make the following specific assumptions:

The electron density is proportional to the mean free path $\lambda_e$ of electron-ion interactions, $n_e\propto 1/\lambda^3_e$. Here, $\lambda_e =1/n_{i}\sigma$, where $n_{i}$ is the density of ions within the box and $\sigma$ is the elastic cross section for the electron-ion dispersion, which, for $\Delta E_F\gtrsim 10$ eV, is proportional to $Z/\Delta E_F$ \cite{Tilinin}.

The density of ions is proportional to the mean separation $x_i$ of the ions inside the box, determined by the ion-ion potential $\phi_Z(x_i,\xi_e)$ in the weakly coupled dense plasma approximation \cite{PlasmaRef2}. This, for the interval from $x_{min}=2x_0$ to the thermal equilibrium distance $x_{th}=2r_W/a$, where $r_W$ is the Van der Waals radius. 

\begin{figure}
    \centering
    \includegraphics[width=0.95\linewidth]{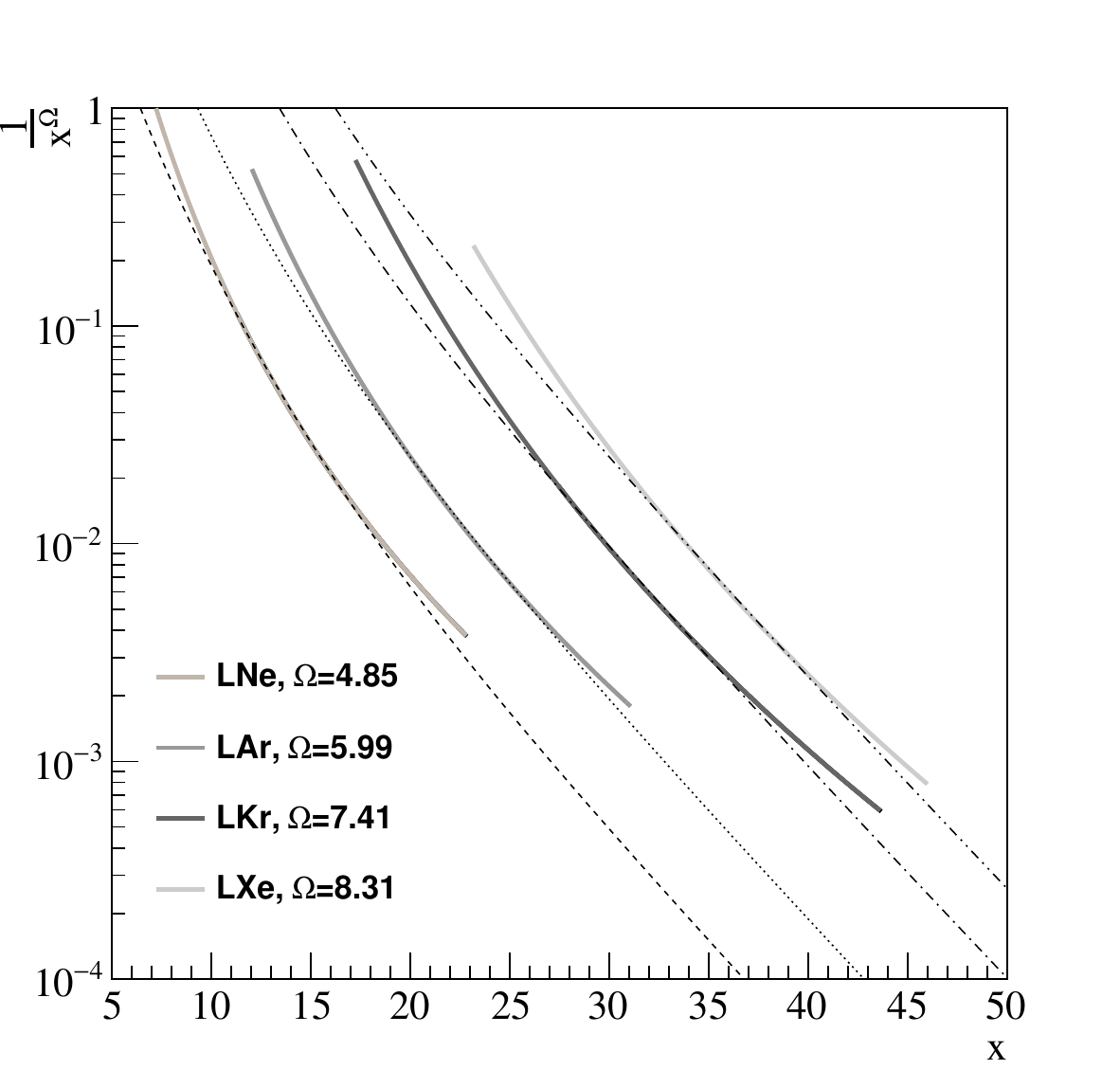}
    \caption{Power-law approximations to Ziegler atom-atom potentials for several noble liquids in the range between $x_{th}$ and $x_{min}$ discussed in the text.}
    \label{fig:PowerLawOmega}
\end{figure}

For simplicity, we approximate the ion-ion potential  by a power law, $\phi_Z(r,\xi_e)\propto Z^{2}/r^{\Omega}$, where $\Omega$ is a noble liquid  dependent parameter, with $\Omega > 1$ \footnote{It is expected that for $\Omega\rightarrow 1$ the ion has lost all electrons}, see Figure \ref{fig:PowerLawOmega}.  Then, we can establish  an explicit proportionality relationship for the average ion energy for the entire range of $x_i$: $\Delta E_F\cong\phi_Z(x_i,\xi_e)/Z\propto Z/x^{\Omega}_i$. This relation is  valid only if   $q^{\Omega} \ll 1$, with $q=(x_{th} -x_{min})/(x_{th} +x_{min})$. Then $n_{i}\propto (1/x_i)^{3} \propto (\Delta E_F)^{3/\Omega}Z^{-3/\Omega}$.

From the previous results, for the mean electron-ion scattering length, we get 
\begin{equation}
\label{lambdae}
\lambda_e = \lambda_0(\Delta E_F/10\,\rm{eV})^{1-3/\Omega}Z^{3/\Omega-1},
\end{equation}
 where $\lambda_0$ is a proportionality constant with units of length. Then, the electron density becomes
\begin{equation}\label{Ne equation}
    n_e=n^{0}_e\left(\frac{\Delta E_F}{(\Delta E_F)_0}\right)^{(9/\Omega -3 )}\left(\frac{N}{Z}\right)^{(3-9/\Omega)},
\end{equation}
with $(\Delta E_F)_0$ being the average electron energy for thermal electrons, which can be estimated from the gap energy, $E_g$. Thermal equilibrium is reached when $\Delta E_F\rightarrow (\Delta E_F)_0 $ for $\Omega>3$ \footnote{In this case the box size is on the order of 10 nm and the mean free path of the electrons is about 50 nm, verifying $\lambda^3_en_e\gg1$}. On the other hand, condition $q^{\Omega} \ll 1$ limits the applicability of Eq. (\ref{Ne equation}) for high energies ($\Omega < 3$), in which the electron density, $n_e$, increases with energy.  

To determine the size of the box, $a_B$, we equal the screened electron-ion potential to the external field potential
\begin{equation}\label{BoxSizeEquation}
     -bzF_0 = (\phi^e_Z(z,\xi_e,N))_{\rm{sc}},
\end{equation}
and solve for $z$ as a function of the electric field $F_0$. Then, the size of the box is determined by doing $a_B=z(F_0)$. For an electric field  of about $10\;\rm{V/cm}$ we obtain $a \cong 10 $ nm for LAr and $a \cong 80$ nm for LXe. As expected, in both cases, the box size is smaller than the Onsager radius, which is about $100$ nm. From Eq. \eqref{ThomasImel2} we see that now $\xi\propto 1/(a_B^2 F_0)$ and its dependence becomes more complicated than the hyperbolic one. In general, $a_B(F_0)$ is expected  to be a monotonic decreasing function of the electric field, and its derivative with respect to $F_0$ is sensitive to $\xi_e$. The minimal value of $F_0$ for Eq. (\ref{BoxSizeEquation}) to be valid can be estimated by equating  the box size to the Onsager radius, with is related to the average thermal energy $E_T$, with the formula $F_3=(E_T/E_g)(W_i/E_g-1)$, where $F_3$ is the excitation Fano factor \cite{FanoDoke}. For LAr, $F_3=0.004$ and then  $E_T=9$ meV, which corresponds to an Onsager radius of 10 nm, and a minimal electric field of  10 $\rm{V/cm}$. In the case of LXe, $F_3=0.0006$, $E_T=1$ meV, and the Onsager radius is 67 nm, which gives a minimal  electric field $20\; \rm{V/cm}$.

For simplicity, to have some insight on the solution of Eq. \eqref{BoxSizeEquation},  we can perform a power-law fit to the box size, $a_B=a'_0(F'_0/F_0)^\delta$ around a specific electric field value, where $a'_0$ and $F'_0$ are fit constants. Typical electric fields in liquid TPCs are approximately $100$--$1000$ $\rm{V/cm}$. As discussed above, the slope of the box size depends on the scale parameter $\xi_e$. For noble liquids considered here we obtain different values for the slope $\delta$ at different electric fields, see Table \ref{tabledelta}. The slope $\delta$ depends on $\xi_e$, the atomic number (for noble elements), and the electric field. From Eq.~\eqref{BoxSizeEquation}, we find $\delta_{\max}=1/2$, reached only for a purely Coulomb interaction potential. For elements with large atomic numbers, the valence electrons are located relatively far from the nucleus, and the effective nuclear charge estimated using Slater’s rules is approximately equal to the net charge experienced by the valence shell after shielding by the other valence electrons. Consequently, the screening parameter $\delta$ is expected to increase and approach its maximal  value as the atomic number increases.

For $\delta$ values corresponding to $\xi_e$ in Table \ref{tab2} we get the fit function,
\begin{equation}\label{Eq:Box_aB}
    a_B\cong  13Z^{1.94}(F_0 \,\rm{V^{-1}cm})^{-\delta} \, \rm{pm}.
\end{equation}
This result demonstrates that, to a very good approximation, the recombination parameter exhibits an inverse proportionality to the electric field and obeys a power law dependence, $\xi/N_i\propto 1/F^{(1-2\delta)}_0$. 
\begin{table}[h]
\centering
\begin{tabular}{|c| c |c|} 
 \hline
 Element & $\delta$ ($10^2 \,\rm{V/cm}$)  & $\delta$ ($10^3\, \rm{V/cm}$) \\ [0.5ex] 
 \hline\hline
 Neon & 0.161 & 0.211    \\ 
 \hline
 Argon & 0.203 & 0.270    \\
 \hline
 Krypton & 0.288 & 0.371    \\
 \hline
 Xenon & 0.385 & 0.449    \\
 \hline
\end{tabular}
\caption{Slope value obtained from a fit to the solution of Eq. \eqref{BoxSizeEquation} for LNe, LAr, LKr and LXe. We compare how the slope varies with the electric field.}
\label{tabledelta}
\end{table}
\subsection{Polarizability model for $N_{ex}/N_i$}\label{SubStarkEff}
When an electric field $F_0$  is applied to a noble  liquid , the ions and atoms that recombine are expected to interact with the field. This interaction produces a small displacement $d$ of the outer electron orbitals. For an electric field applied  along   the $z$ direction, assuming a spherical shape of the atom in the absence of the field, to first order, the deformation due to the displacement is $x'=x'_0+x'_d\cos\theta$, where $x'_d=d/b$ and $\theta$ is the angle with respect to the field direction.

The potential of the atom-atom system, $V(r_e,x,\xi_e)$, is given by the sum of the following contributions:  nuclei-nuclei $V_{\mathcal{N}\mathcal{N}}(x)$, electron-electron $V_{ee}(r_e)$, and electron-nuclei $V_{e\,\mathcal{N}}(r_e,x,\xi_e)$. We consider this interaction in the context of a semiclassical theory, where the quantum interference effects for distances lower than $x$ are ignored. The average potential is given by
\begin{multline}\label{PotentialDefinition}
    \bar  V(x,\xi_e)  \cong  \\
     \int_{r_e}\int^{\infty}_x d^3x'' d^3r_e \mid \psi_e(r_e)\mid^2 \mid\psi_\mathcal{N}(x'')\mid^2 V(r_e, x'',\xi_e). \quad
\end{multline}  
For an atom with $Z$ electrons, the electron-atom potential can be defined in a similar manner as in Eq. (\ref{PotentialDefinition}). For a test charge at a distance $x'$ from the nuclei, 
\begin{multline}\label{SphericalPotentialDef}
         \bar  V_{\mathcal{N}e}( x',\xi_e) +\bar V_{ee}(x',\xi_e)  = \\
\phi^e_Z(x',\xi_e,Z) \cong\frac{4\pi b^3}{e\bar Z(x')}\int^{\infty}_{x'} \rho(x'') x''^2 V_Z(x'') dx'',
\end{multline}
 where  $e\bar Z(x')=4\pi b^3\int^{\infty}_{x'}\rho (x'') x''^{2} dx''$ is the effective charge and   $\rho(x')=e\!\mid\!\psi_Z(x')\!\mid^2$. Here,  $\psi_Z(x')$ and $V_Z(x')$ are the wave function and the effective potential, respectively, resulting after the integration of the electronic coordinates.
 
 The correction $\delta \phi^e_Z (x',F_0)$, due to the external field potential $\phi_{ext}(x',\xi_e)$, modifies the total potential for electron-atom and atom-atom systems, 
 \begin{align}\label{ShiftedPotentials}
     (\phi^e_Z(x',\xi_e,Z,F_0))_S=\phi^e_Z(x',\xi_e,Z) +\delta \phi^e_Z (x',F_0),\\
\nonumber\\
        (\phi_Z(x,\xi_e,Z,F_0))_S=\phi_Z(x,\xi_e,Z) +2\delta \phi^e_Z (x,F_0).
\end{align}
 We determine both shifts in terms  of the Stark effect \cite{SchrodingerStark} for a deformed  charge distribution  using nondegenerate first-order perturbation theory \footnote{This gives a similar result to the second order perturbation theory for non deformed charge distribution}, 
 \begin{multline}\label{delta}
     \delta \phi^e_Z (x',F_0) =\\
     \frac{2\pi b^4 F_0}{e\bar Z(x')} \!\int^{\pi}_{0} \!\int^{\infty}_{x'} \!x''^3 \rho(x''+x'_d \cos\theta ) \cos\theta \sin\theta d\theta dx''.
 \end{multline}
     To evaluate the double integral on the right-hand side of this equation, we make the following approximations: (i) Since $x'_d \ll x'$,  $\rho(x'+x'_d \cos\theta)\cong \rho(x') +x'_d \cos\theta \,d\rho'/dx'$, the radial integration is separated and, only the term dependent on $\cos^2\theta$    gives a nonzero result to the integration over $\theta$. (ii) Only interaction with valence electrons is considered and consequently $x'\gg 1$; therefore, we can use the Kohn-Sham asymptotic electron density, $\rho(x')\cong  \rho_0 e^{-\gamma x'}$, with $\gamma=2(2U_{e1}/E_h)^{1/2}q\,\xi^{2/3}_e\,0.885\,Z^{-1/3}$ \cite{Kohnsham}, where $E_{h}$ is the hydrogen binding energy ($\cong 13.6$ eV), $U_{e1}$ is the second ionization energy of the atom, and $\rho_0$ is a constant associated with the electron density in the region close to the valence electrons. (iii) We assume that $x'_d=\alpha_p F_0/eb$ and  $\alpha_p$ is the polarizability constant. Using these approximations we get, 
 \begin{equation}\label{shiftStark}
     \delta \phi^e_Z (x',F_0)\cong -\left(\frac{4\pi\varepsilon_0 \gamma\alpha_p (100F_0)^2}{3e} \right)(x'+\gamma).
 \end{equation}
Since we use first order perturbation theory and assume $x'\gg x'_d$, the validity of Eq. \eqref{shiftStark} is limited to electric fields not very intense.  Since the shift is negative, the exciton to ion ratio $N_{ex}/N_i$ is expected to be a decreasing function of $F_0$. 

\begin{figure}
       \centering
       \includegraphics[width=0.99\linewidth]{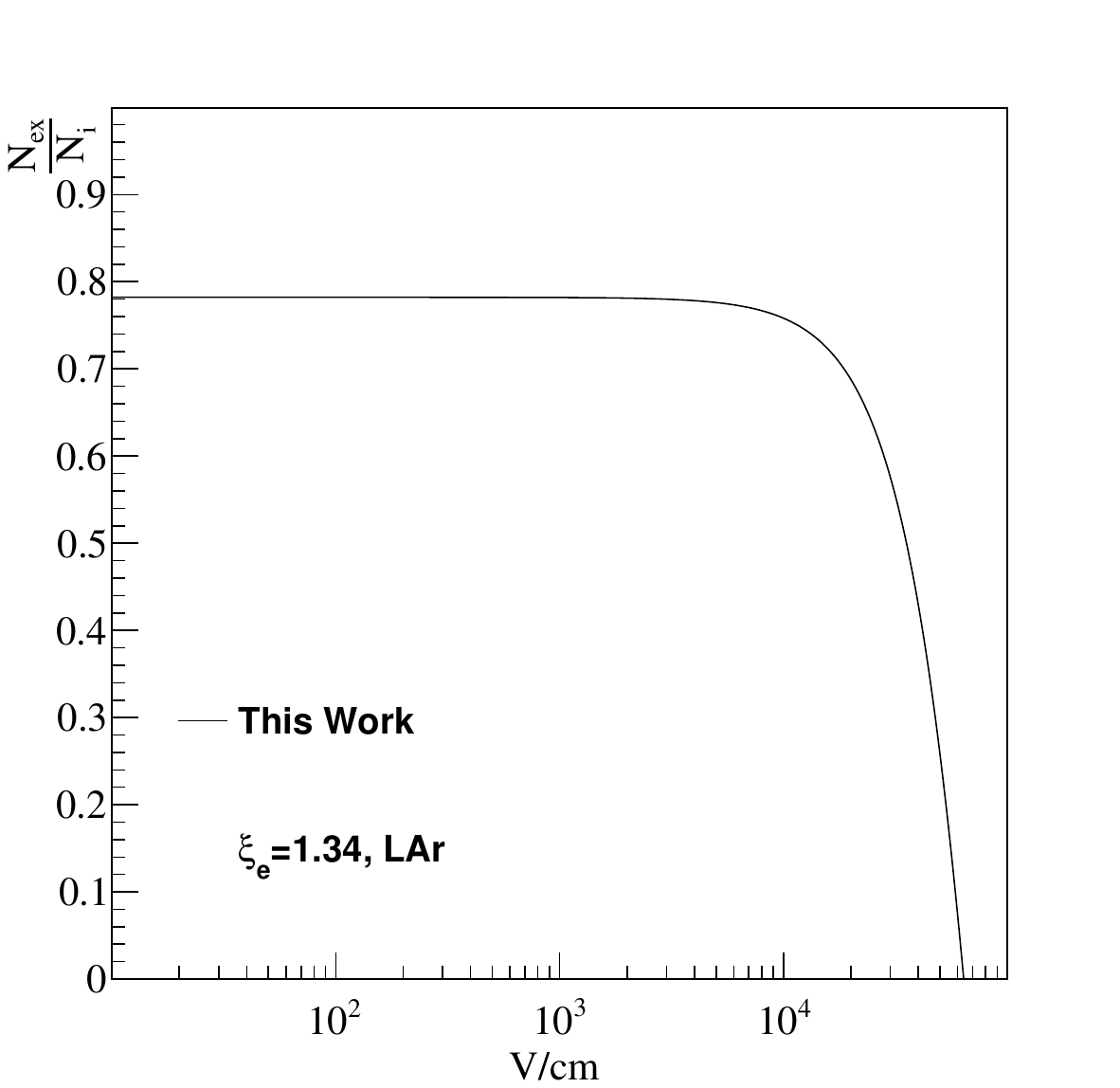}
       \caption{Exciton-to-ion ratio for LAr as a function of the electric field.}
       \label{fig:RatioAr}
   \end{figure}

\begin{figure}
       \centering
       \includegraphics[width=0.99\linewidth]{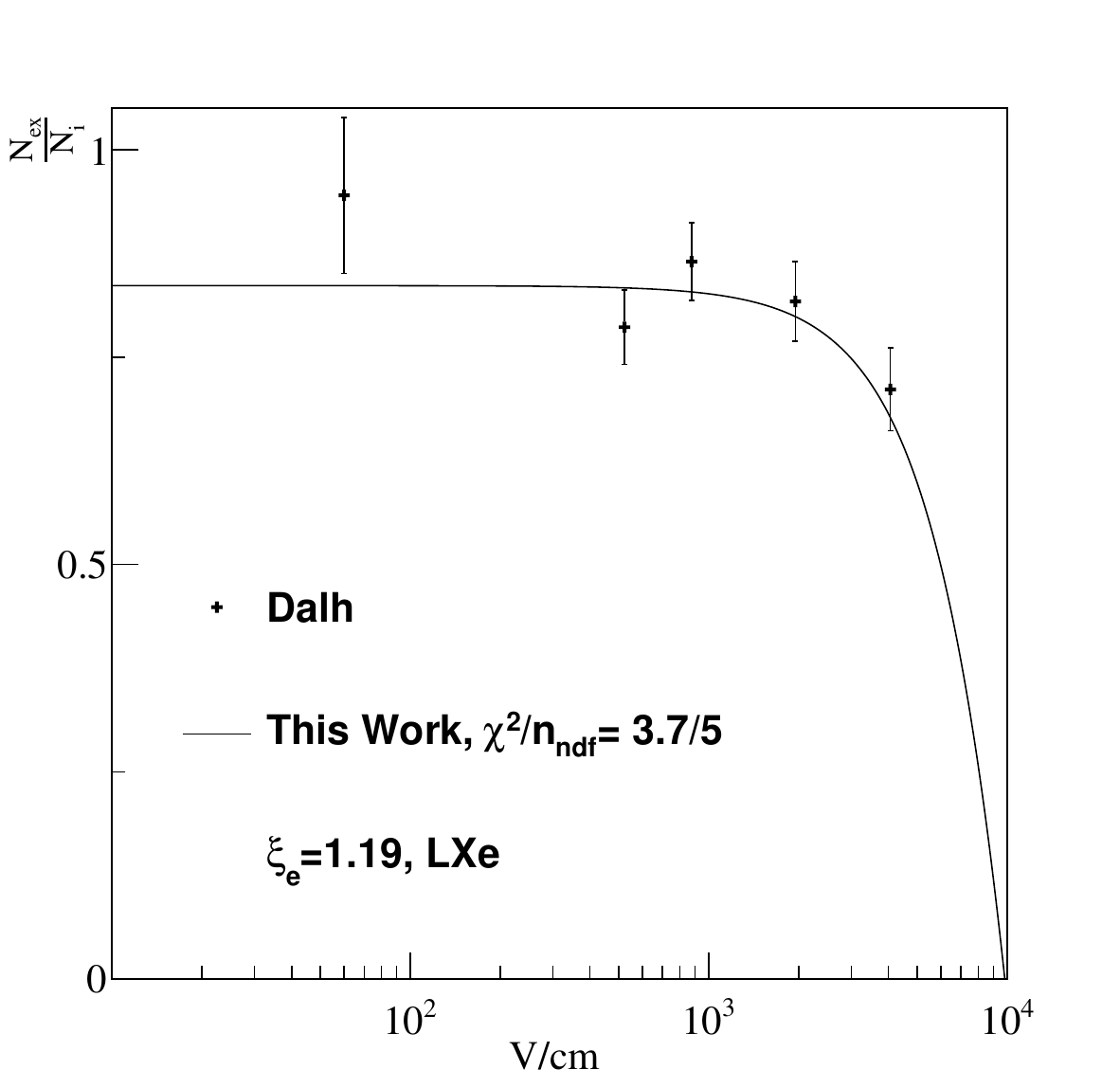}
       \caption{Exciton-to-ion ratio for LXe as a function of the electric field.}
       \label{fig:RatioXe}
   \end{figure}

   Using the same procedure as in Sec. \ref{SubBatesGriffing}, we  compute the ratio $N_{ex}/N_i$ by means of  Eqs. \eqref{MainPotentialEquation} and \eqref{Eq Ratio Nex/Ni} but with the shifted potentials defined in Eq. \eqref{ShiftedPotentials}. The results of this computation for LXe and LAr are shown in Figs. \ref{fig:RatioAr} and \ref{fig:RatioXe}. From these figures, we see that the upper value for the electric field is of the order $\sim10^4\; \rm{V/cm}$, which is sufficiently large for the current noble liquid experiments \cite{szydagis2023reviewnestmodelsapplication}.

   
\subsection{Birks constant}\label{BirksCte}
The coefficient $k_B$ in Birks' law (Eq. \eqref{BirksLaw}) is  related to the probability that two excitons form a bonded system at high energies. Assuming a simple orbital bonding, the symmetric wave function (singlet) is required  and  $k_B\propto k'_B/4$, where $k'_B$ is another constant.  From the Birks law, the constant $k'_B$ has to be proportional to $S_e(\varepsilon_0,\xi_e)^{-1}$, where $\varepsilon_0$ is the energy of the ions such that the effective ion charge, $Z_{eff}$, allows the production of an exciton pair. The effective charge for an exciton must be in the range $0<Z_{eff}<1$, thus, to estimate $\varepsilon_0$ we use the Ziegler effective charge formula for ions, $Z_{eff}=Z(1-e^{-v_i/v_0Z^{2/3}})$, where $v_i$ is ion velocity, such that, $Z_{eff}=1/2$, see Table \ref{tab:zeff_half}.
\begin{table}[t]
\centering
\begin{tabular}{|c|c|c|c|}
\hline
Ion & $Z$ & $A$ & $\varepsilon_0/\mathcal{Q}$ (keV) \\
\hline
Ne & 10 & 20.180  & 27.92 \\
Ar & 18 & 39.95  & 35.96 \\
Kr & 36 & 83.798 & 48.1 \\
Xe & 54 & 131.293 & 56.98 \\
\hline
\end{tabular}
\caption{Energy required for noble gas ions to reach $Z_{{eff}} = 0.5$ using a Ziegler parametrization.}
\label{tab:zeff_half}
\end{table}
 From the energies listed in Table \ref{tab:zeff_half} we can make an analytic estimate using Lindhard's electronic stopping power, including the effect of the correction factor $q$ for the inter-atomic potential, with $S_e(\varepsilon_0,\xi_e)=q^{3/2}0.133\xi_e(Z\varepsilon_0/A)^{1/2}$, then 
\begin{equation}\label{AproxkBequation}
    \xi_ek_B =\frac{1}{4q^{3/2}}\left(\frac{7.52}{\varepsilon^{1/2}_0}\right)\left(\frac{A}{Z}\right)^{1/2},
\end{equation}
where  we can use the approximate trend $\varepsilon_0\cong 121.52/Z^{1.91}$ ($\varepsilon_0/\mathcal{Q}\sim Z^{1/3}$) to give a closed formula for $\xi_ek_B$ for all noble elements, see Table \ref{tab:xiekb_noble}.

\begin{table}[h!]
\centering
\begin{tabular}{|c|c|}
\hline
Element & $q^{3/2}\xi_e k_B$,  Eq.\eqref{AproxkBequation}  \\
\hline
Ne &  2.18  \\
Ar &   3.97\\
Kr &  7.99\\
Xe &  12.01 \\
\hline
\end{tabular}
\caption{Values of $\xi_e k_B$ for noble elements}
\label{tab:xiekb_noble}
\end{table}

 The results obtained for LAr and LXe are in good agreement with the experimental measurements, \cite{BEZRUKOV2011119,Birks2}. This model predicts that for higher $Z$ the biexcitonic quenching is more active.

\section{Recombination for noble liquids }
\label{Recombination FORNOBLE}
We now show that the ratio $\xi/N_i$  can be determined for all noble liquids in terms of a  single fit constant. It is convenient to write  $\alpha=\alpha_0r^2_0v_{max}$, where $\alpha_0$ is a dimensionless constant, $v_{max}\cong v_0$  is the maximum drift velocity of the electrons, and $r_0$ is the minimum box length. 
The electron mobility is $\mu_e\propto\lambda_e/v^0_{e}$,
where $\lambda_e$ is given in Eq. \eqref{lambdae} and $v^0_{e}$ is  the electron velocity within the box, $v^0_{e}=(2\Delta E_F/m)^{1/2}$. From the empirical formula for the atomic radius  $r_0\cong0.18Z^{0.44} \AA$,  Eq. \eqref{ThomasImel2}, and  Eq. \eqref{Eq:Box_aB}, we find 
\begin{equation}\label{ParameterFit}
    \frac{4\xi}{N_i}=\frac{\alpha_0r^2_0v_F}{a^2_B\mu_eF_0}\cong C_0\frac{\Delta E_F^{-1/2+3/\Omega } }{Z^{2.0+3/\Omega} F^{1-2\delta}_0},
\end{equation}
where $C_0=(\alpha_0/\lambda_0)(271v_0m^{1/2}(10\,\rm eV)^{1-3/\Omega}/e)\, $, where $\alpha_0/\lambda_0$ is an unknown constant.

\begin{figure}
    \centering
    \includegraphics[width=0.995\linewidth]{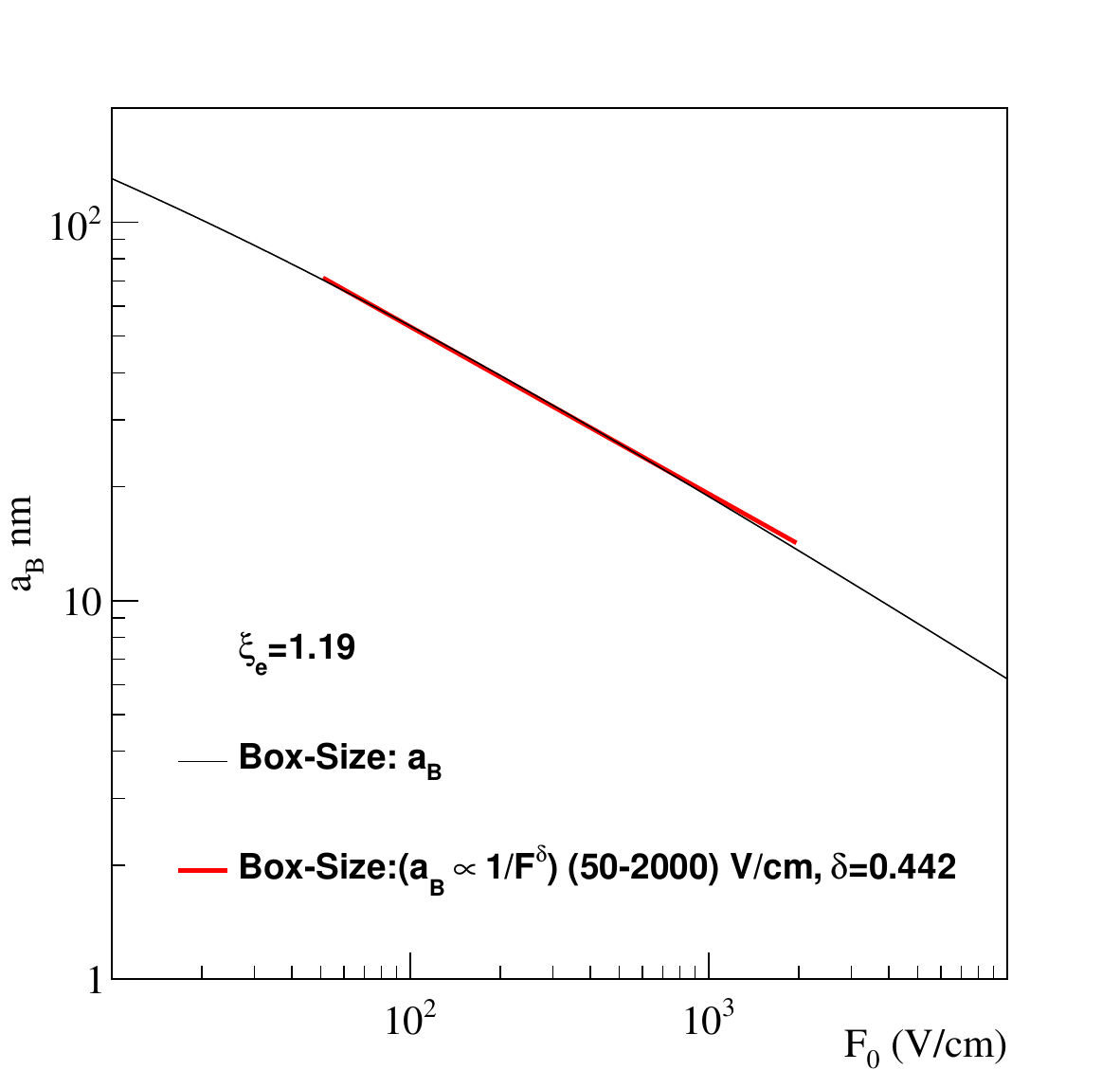}
    \textbf{a)}\\
        \includegraphics[width=0.995\linewidth]{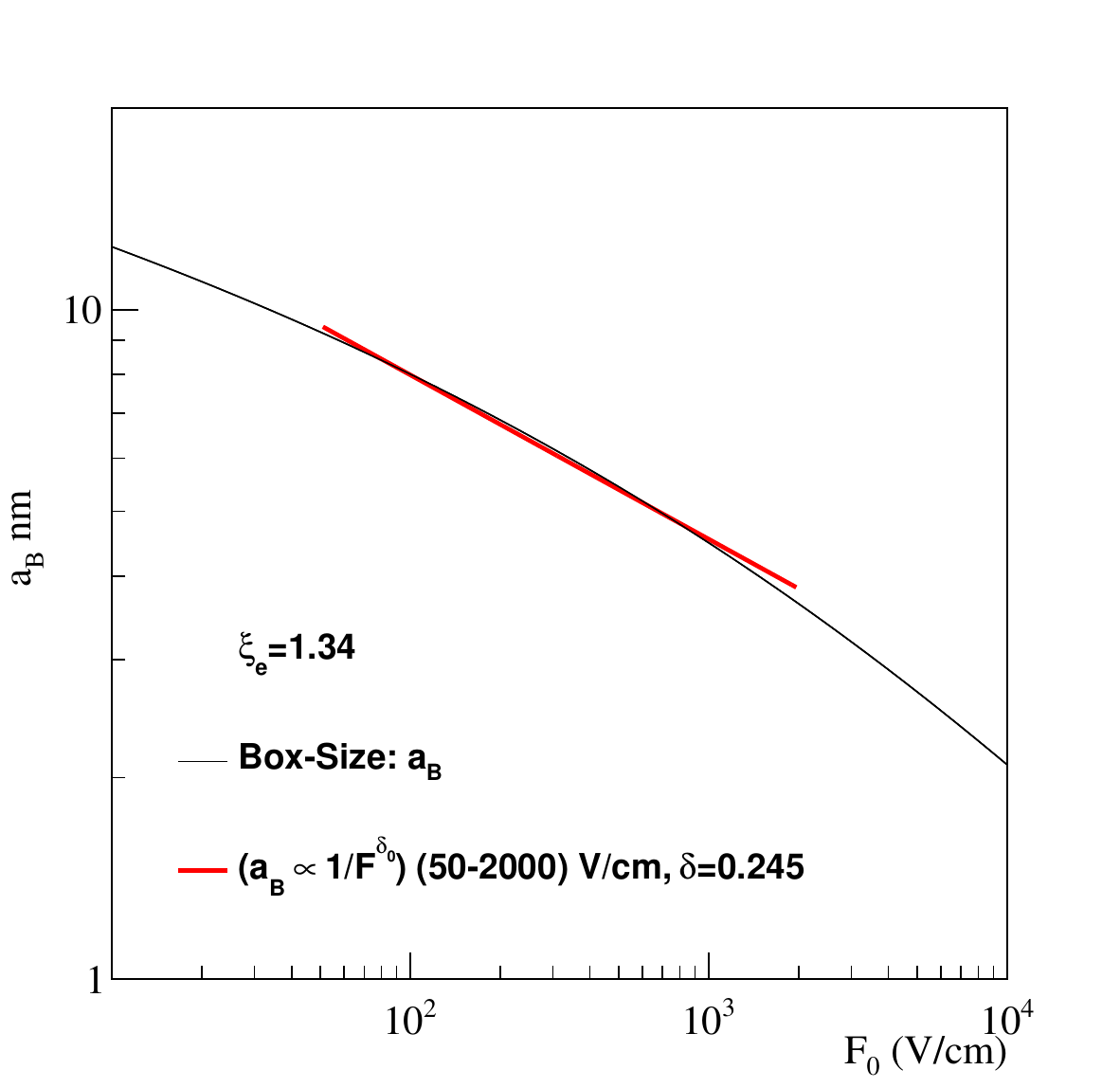}
            \textbf{b)}\\
    \caption{Box size $z=a_B(F_0)$ (black line) as a function of the electric field $F_0$ according to Eq. \eqref{BoxSizeEquation} for (a) LXe and (b) LAr. In both cases, a power-law approximation (red line) is performed over the specified range.}
    \label{fig:BoxSize_aB}
\end{figure}

\begin{figure}
    \centering
    \includegraphics[width=0.99\linewidth]{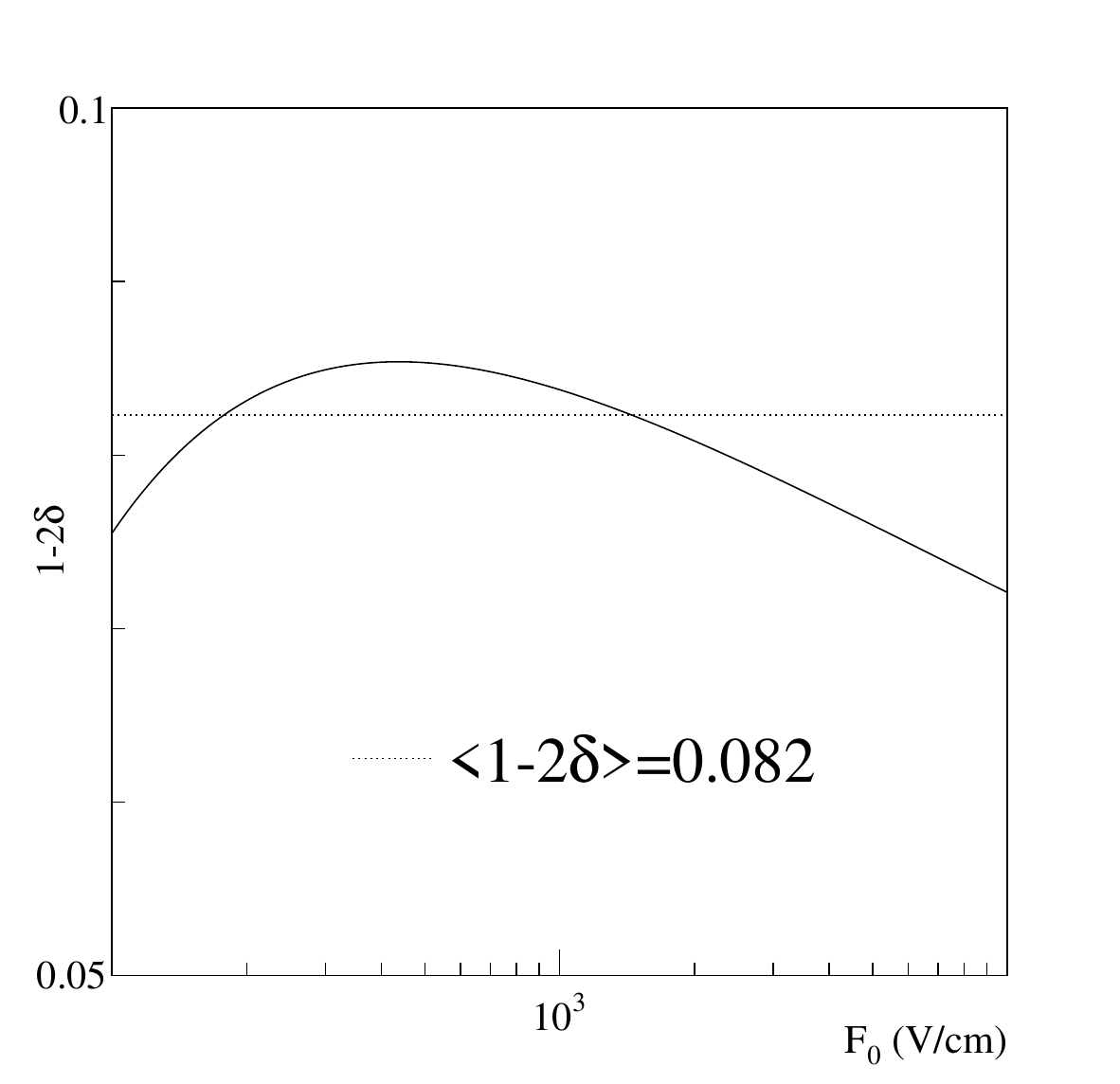}
     \textbf{a)}\\
        \includegraphics[width=0.99\linewidth]{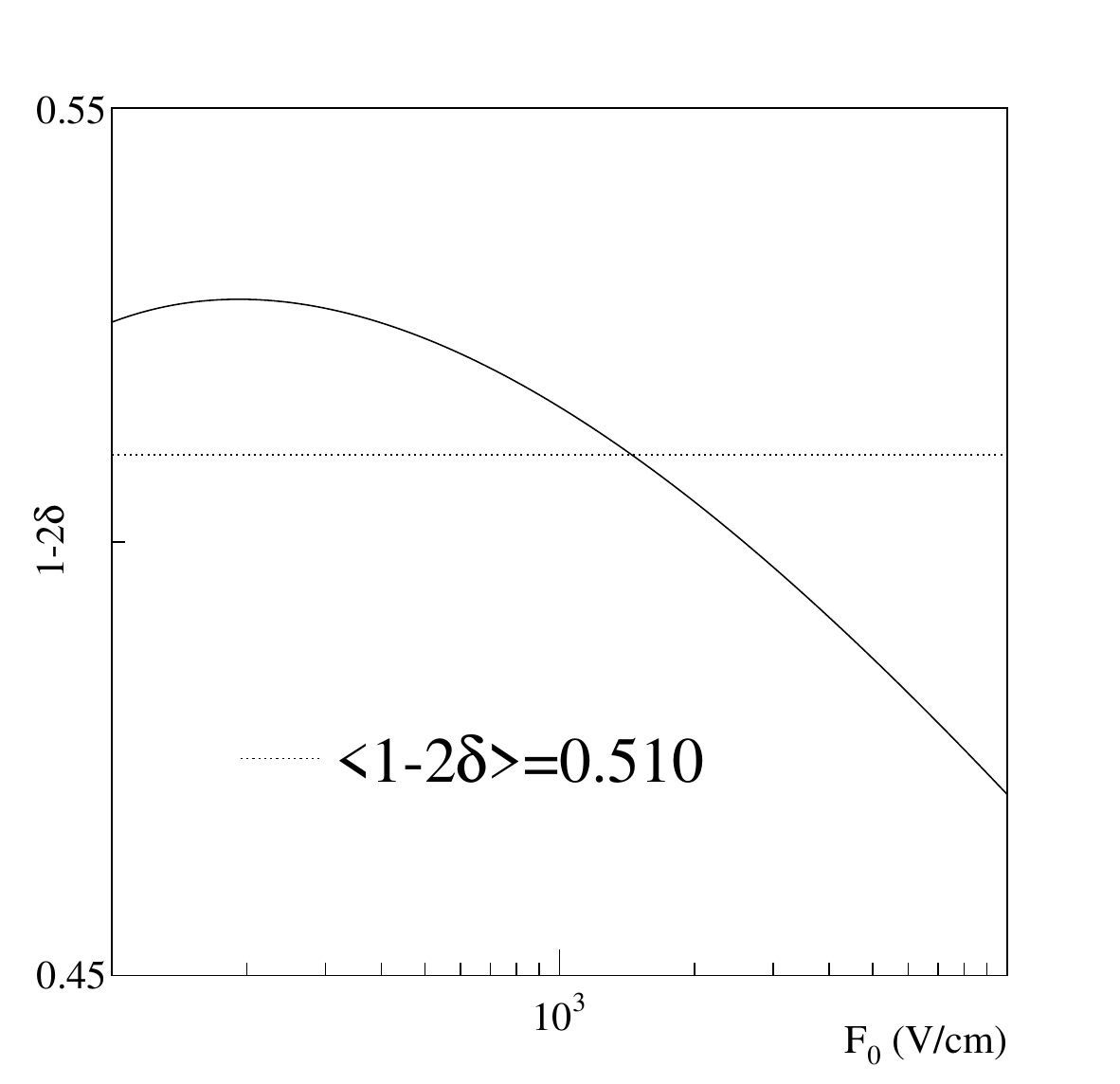}
         \textbf{b)}\\
    \caption{Predicted dependence of $1-2\delta$ for (a) LXe and (b) LAr with the electric field according to Eq. \eqref{BoxSizeEquation}. }
    \label{fig:Etafun}
\end{figure}

To find $C_0$ we have to determine $\delta$. From the fit with the power law of Eq. \eqref{Eq:Box_aB}, in the range $100 \,{\rm{V/cm}} < F_0 <2000\, {\rm{V/cm}}$ for the electric field, we obtain $1-2\delta = 0.106$ for LXe and $1-2\delta = 0.53$ for LAr, see Fig. \ref{fig:BoxSize_aB}. Using the exact solution of Eq. \eqref{BoxSizeEquation} we obtain the curves in Fig. \ref{fig:Etafun} for $1-2\delta$ as a function of the electric field. We also show the mean values, $<1-2\delta>=0.082$ for LXe and $<1-2\delta>=0.51$ for LAr, in the range $100\,{\rm{V/cm}}, < F_0 < 10000\,{\rm{V/cm}}$. We observe that the exponent behavior is nearly constant with the electric field. Except for the cases where the model is not reliable for low electric fields, $F_0<50 \rm{V/cm}$ or near the breakdown electric field, $F_0>10^{4} \rm{V/cm}$.  

\begin{figure}
    \centering
    \includegraphics[width=0.99\linewidth]{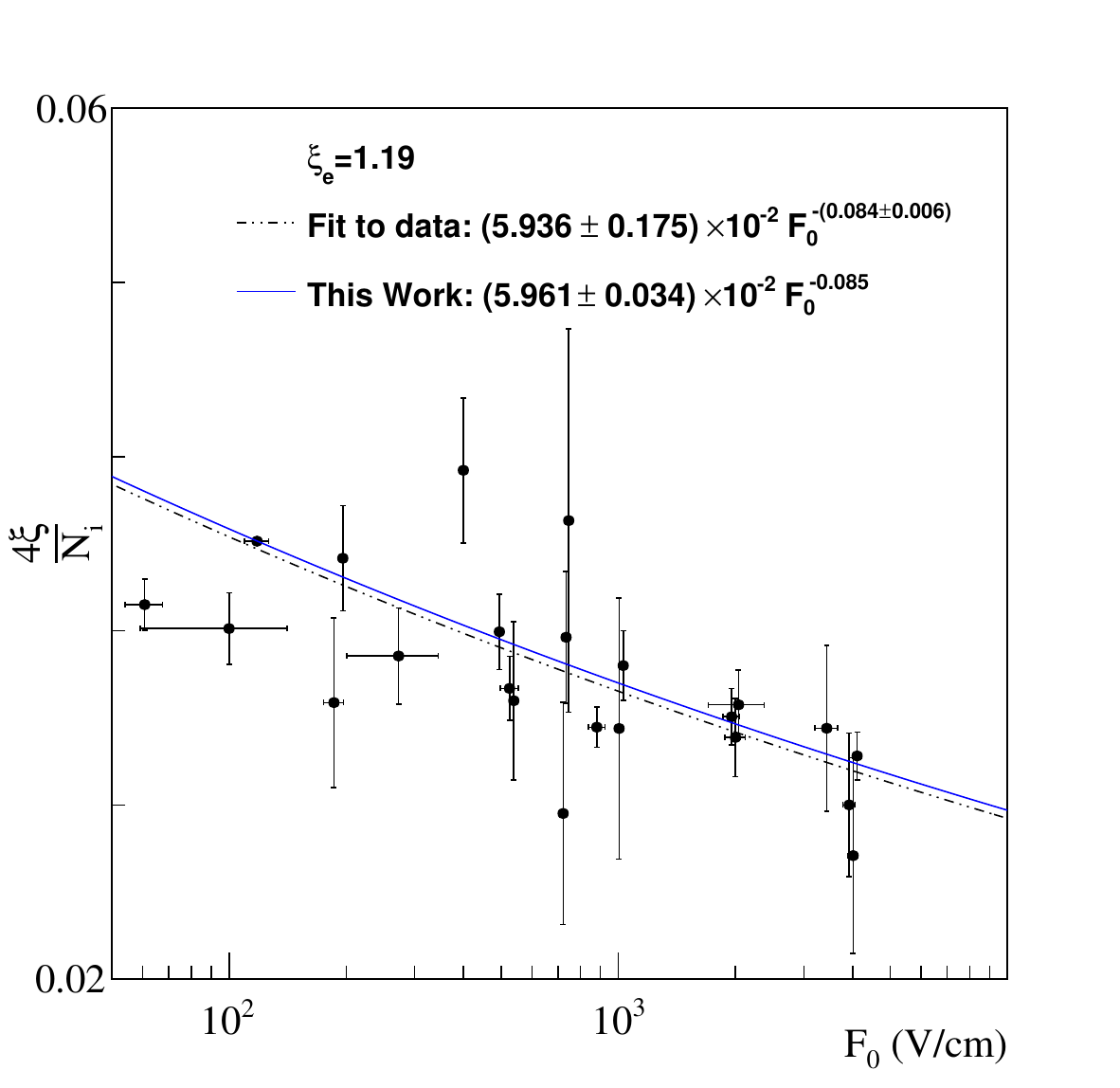}
    \textbf{a)}\\
    \includegraphics[width=0.99\linewidth]{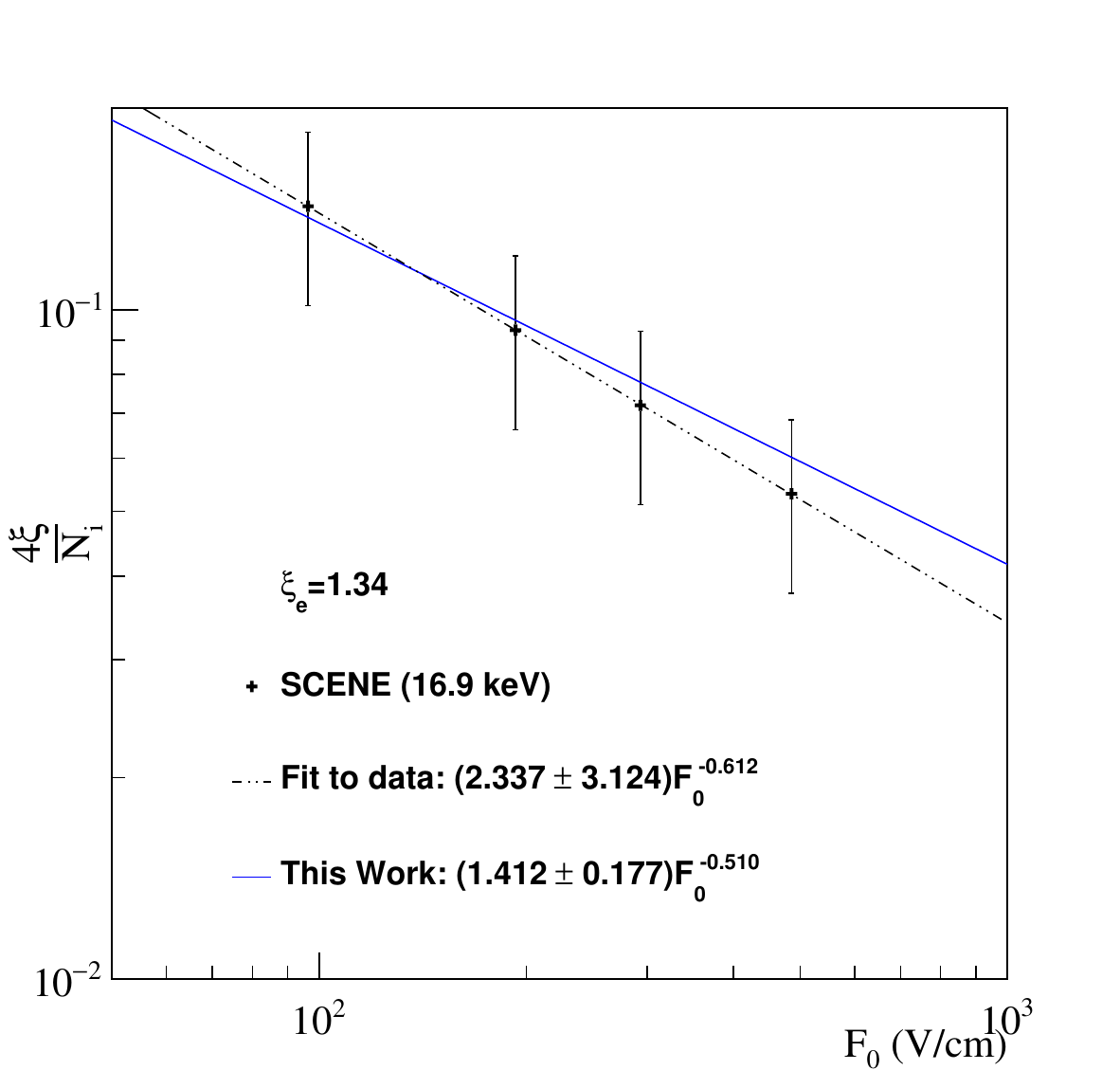}    
    \textbf{b)}\\
    \caption{ Measurements for $4\xi/N_i$ parameter from Eq. \eqref{ThomasImel} as function of electric field $F_0$ for (a) LXe and (b) LAr. (blue line) Model using the power law dependence computed from Eq. \eqref{BoxSizeEquation}, where we extract the one fit constant $C_0$ from LXe.}
    \label{fig:RecombinationPar}
\end{figure}

 The dimensionless parameter $4\xi/N_i$ as a function of the applied electric field is one of the main comparison tests to validate the physical consistency of our model. The dependence of this parameter on the electric field is not universal for the noble liquid elements. To determine the constant $C_0$ we fit the experimental data for LXe \cite{NESTModel} using a power-law function of the form $P_0/(F_0/(1\rm{V/cm}))^{1-2\delta}$, which yields $(4\xi/N_i)^{\text{fit}}_{\text{LXe}} = (0.05961 \pm 0.034 )/(F_0/(1\rm{V/cm}))^{0.082}$. This fit results in a value of $C_0 = 1.88 \times 10^3 \, \text{m}^3\text{s}^{-1}(\text{eV})^{-3/\Omega+1/2}(\text{V/cm})^{1-2\delta}$. Using this value in Eq. \eqref{ParameterFit} for LAr, the model predicts $(4\xi/N_i)_{\text{LAr}} = 1.36/(F_0/(1\rm{V/cm}))^{0.51}$, which is consistent with the direct experimental fit $(4\xi/N_i)^{\text{fit}}_{\text{LAr}} \cong (1.41\pm 0.18)/(F_0/(1\rm{V/cm}))^{0.51} $ as illustrated in Fig. \ref{fig:RecombinationPar}. 

\begin{figure}
    \centering
    \vspace*{-0.05cm}   \includegraphics[width=0.891\linewidth]{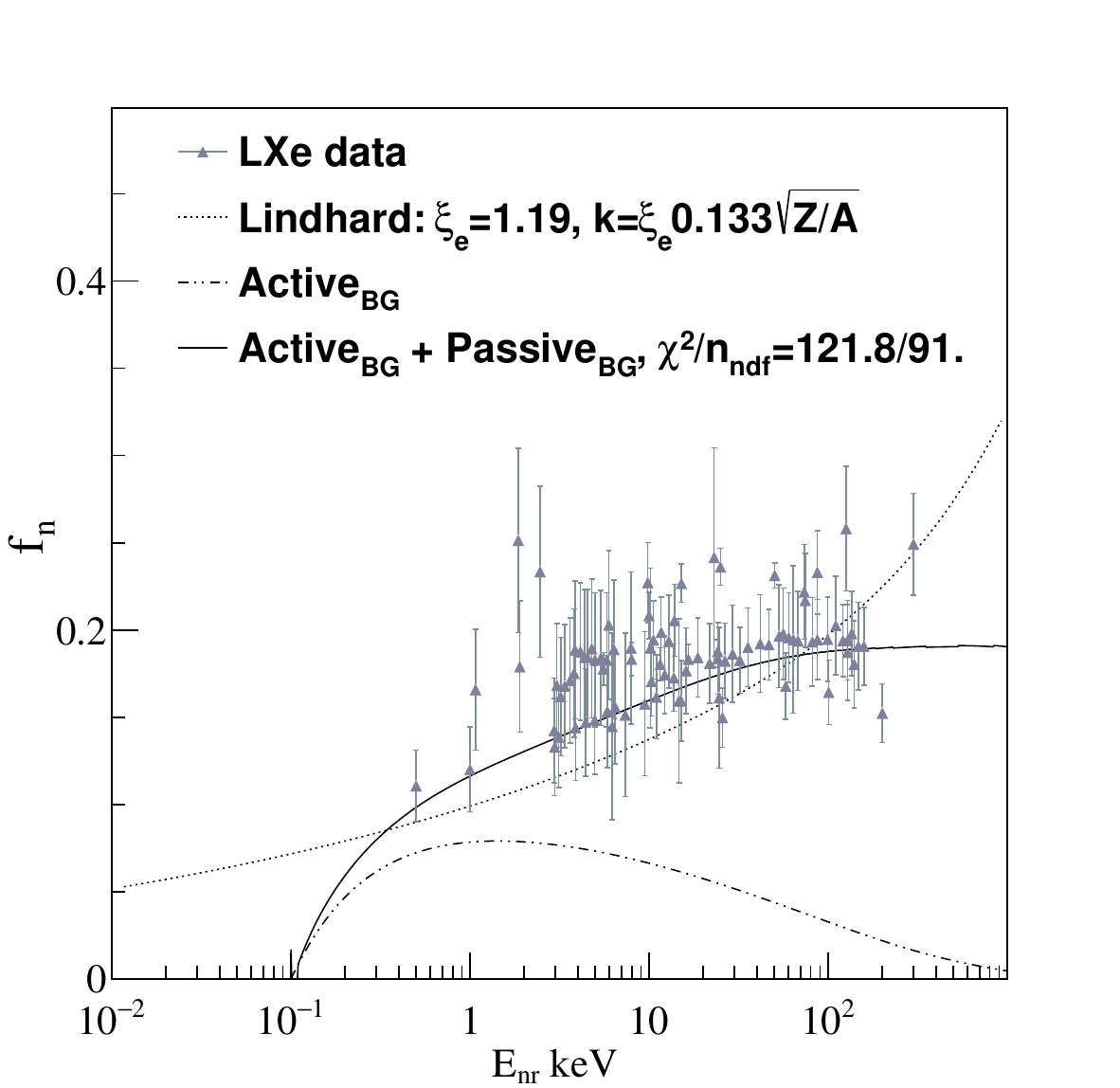}\\
 \textbf{a)}\\
   \vspace*{-0.05cm} \includegraphics[width=0.891\linewidth]{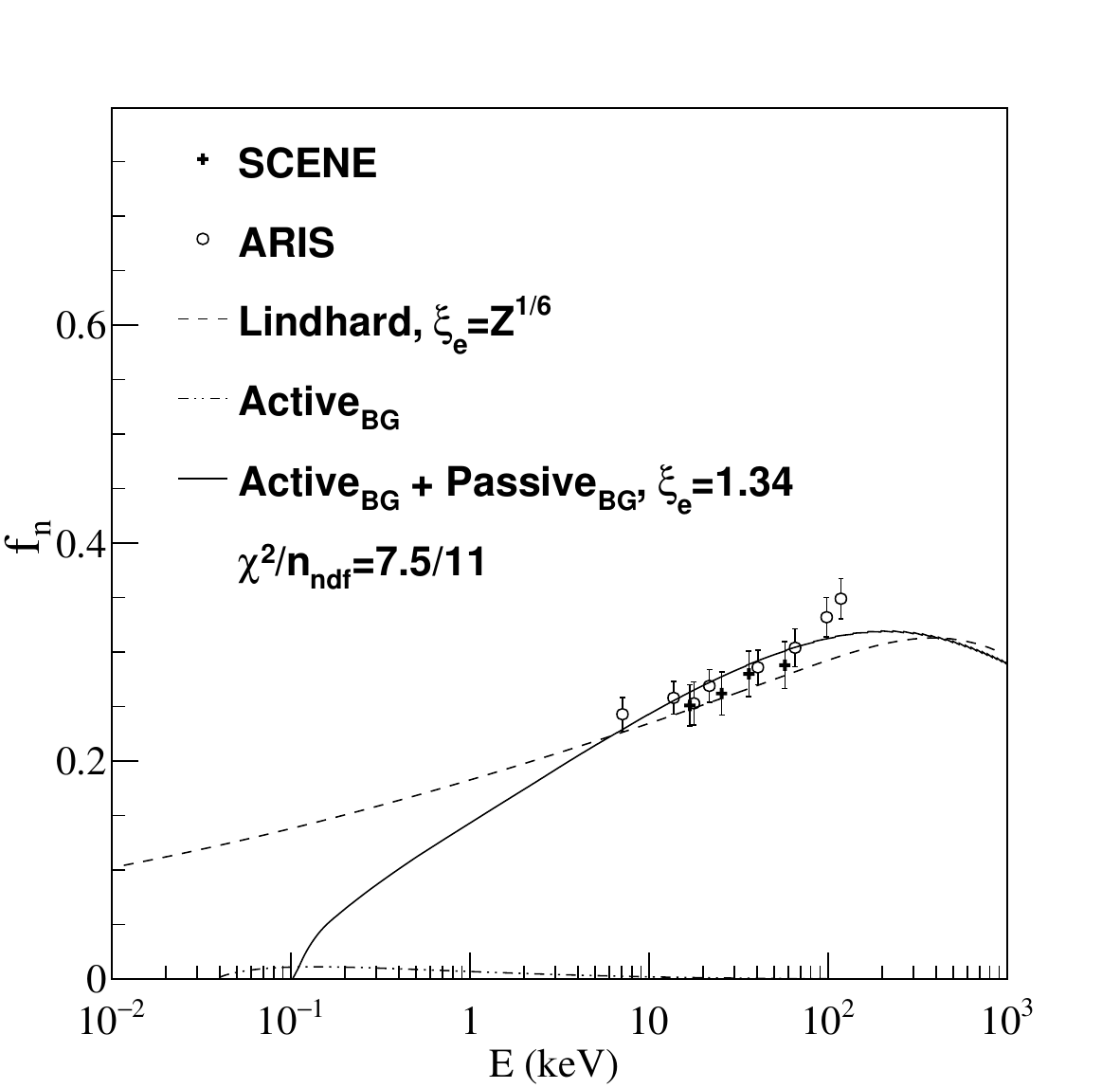}\\
     \textbf{b)}\\
      \vspace*{-0.05cm}   \includegraphics[width=0.891\linewidth]{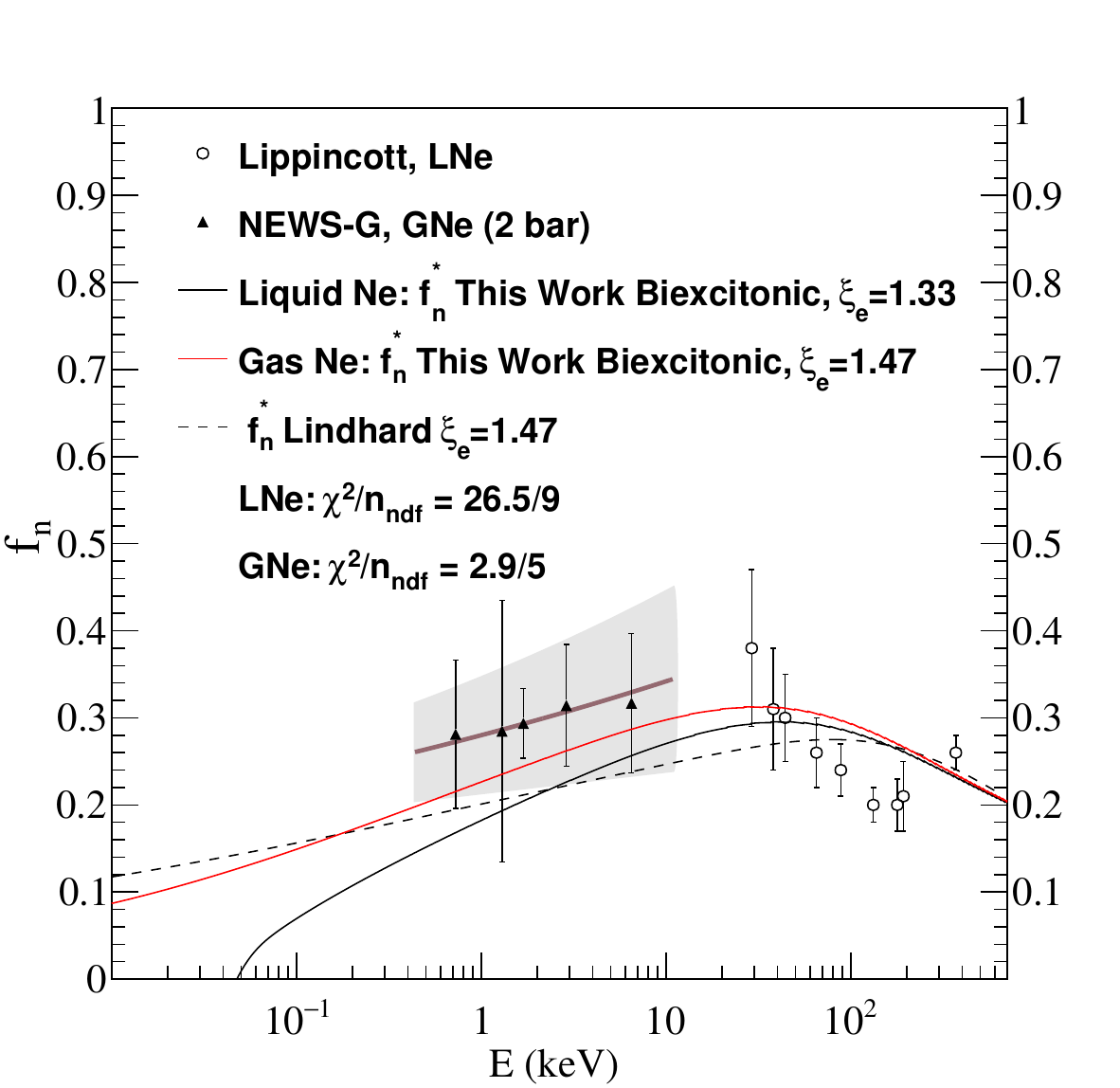}\\
     \textbf{c)}\\
  \vspace*{-0.3cm}   \caption{Total quanta prediction for (a) LXe, (b) LAr and (c) LNe-GNe compared with data.}
    \label{fig:QFLAr}
\end{figure}

The predictions of the model for LKr and LNe are $(4\xi/N_i)_{LKr}=(0.219 \, \rm{m^{3}s^{-1}})/(F_0/(1\rm{V/cm}))^{0.26}_0$ and $(4\xi/N_i)_{LNe}=(12.12 \, \rm{m^{3}s^{-1}})/(F_0/(1\rm{V/cm}))^{0.52}_0$. The general trend for any noble liquid with  $Z\geq 10$ is predicted to be $(4\xi/N_i)_Z= 2650[Z^{2.9 -1.5/Z^{1/2}} ]^{-1}/(F_0/(1\rm{V/cm}))^{1-2\delta}_0$. For the sake of completeness we present the results for total quanta using Eq. \eqref{eq:totQuantaInelastic} for LXe, LAr, GNe and LNe. The BG effect for neon is negligible and is not shown, see fig. \ref{fig:QFLAr}. 

\begin{figure}[ht]
    \centering
    \includegraphics[width=0.99\linewidth]{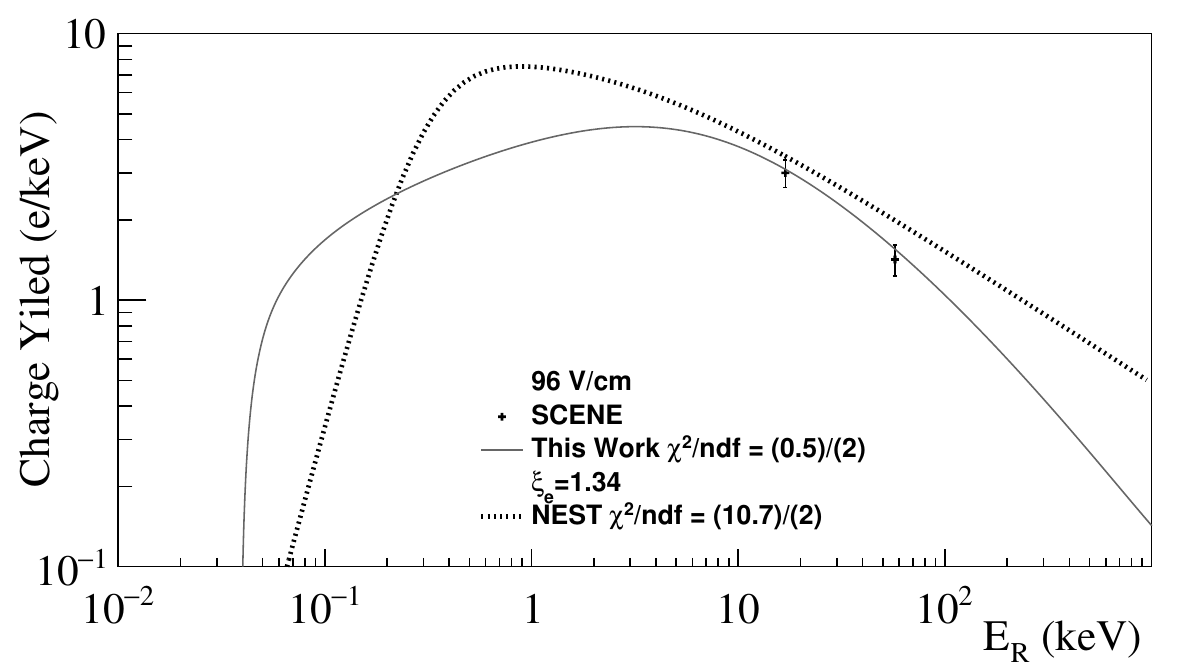}
        \includegraphics[width=0.99\linewidth]{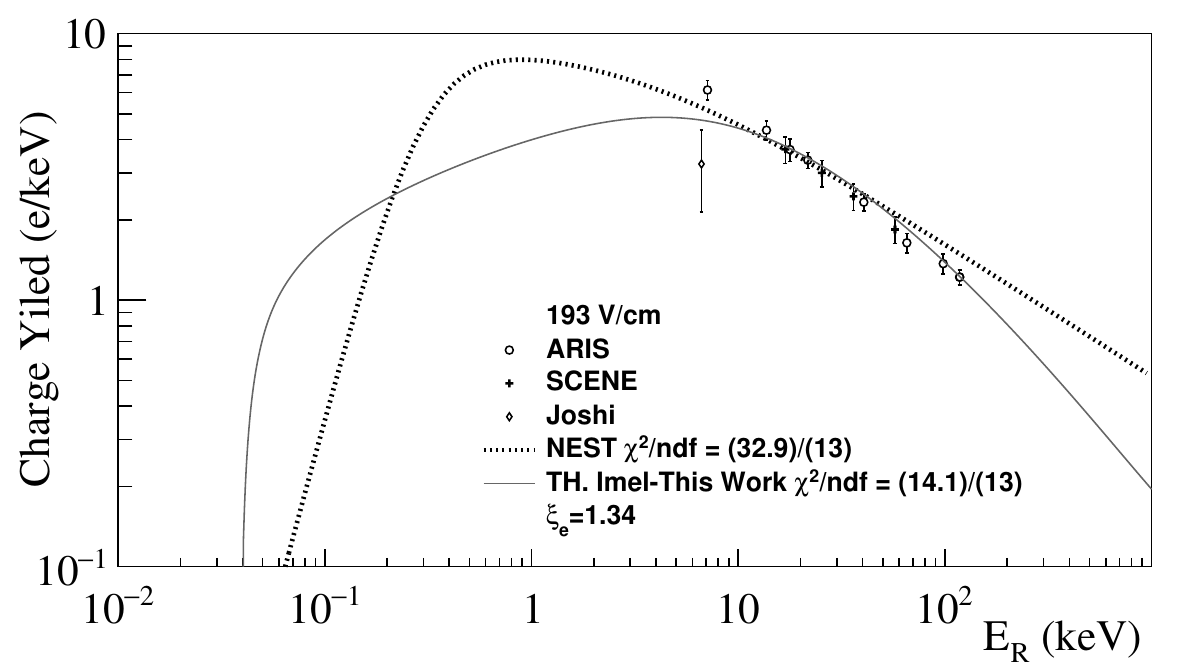}
           \includegraphics[width=0.99\linewidth]{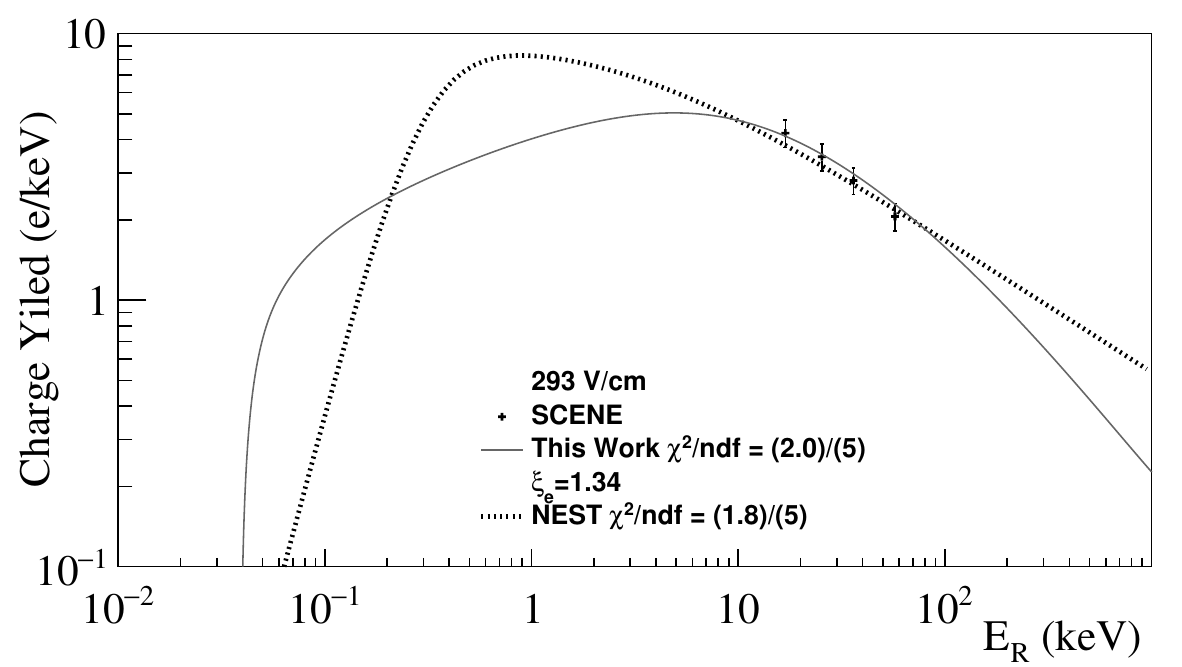}
        \includegraphics[width=0.99\linewidth]{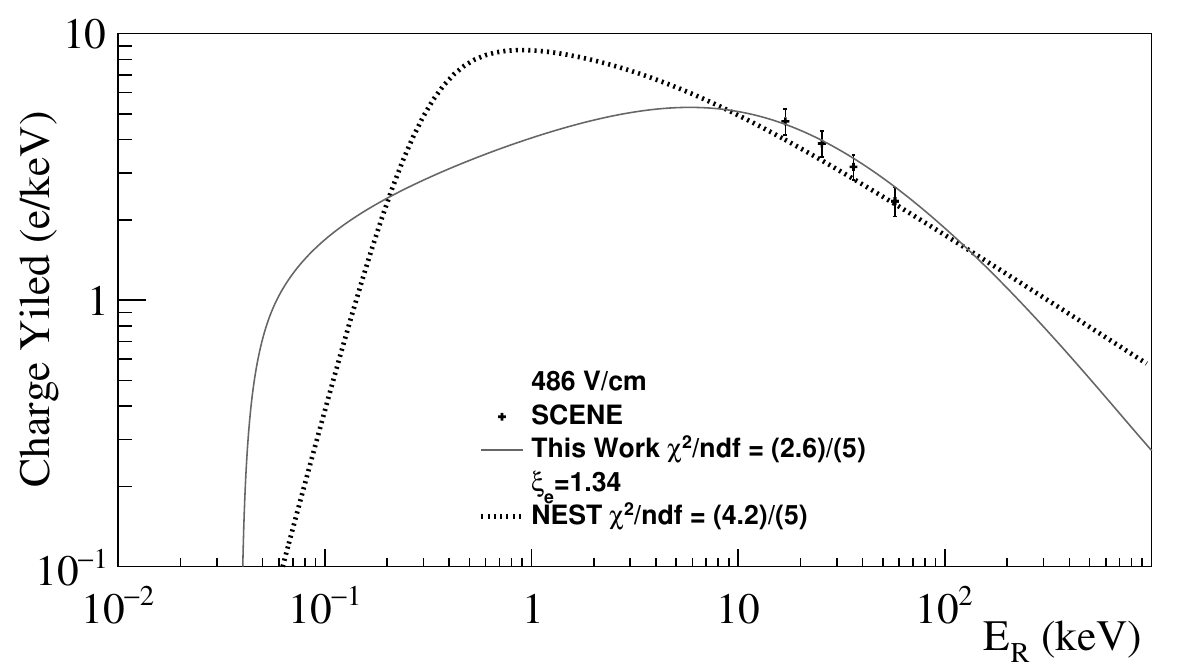}
        \includegraphics[width=0.99\linewidth]{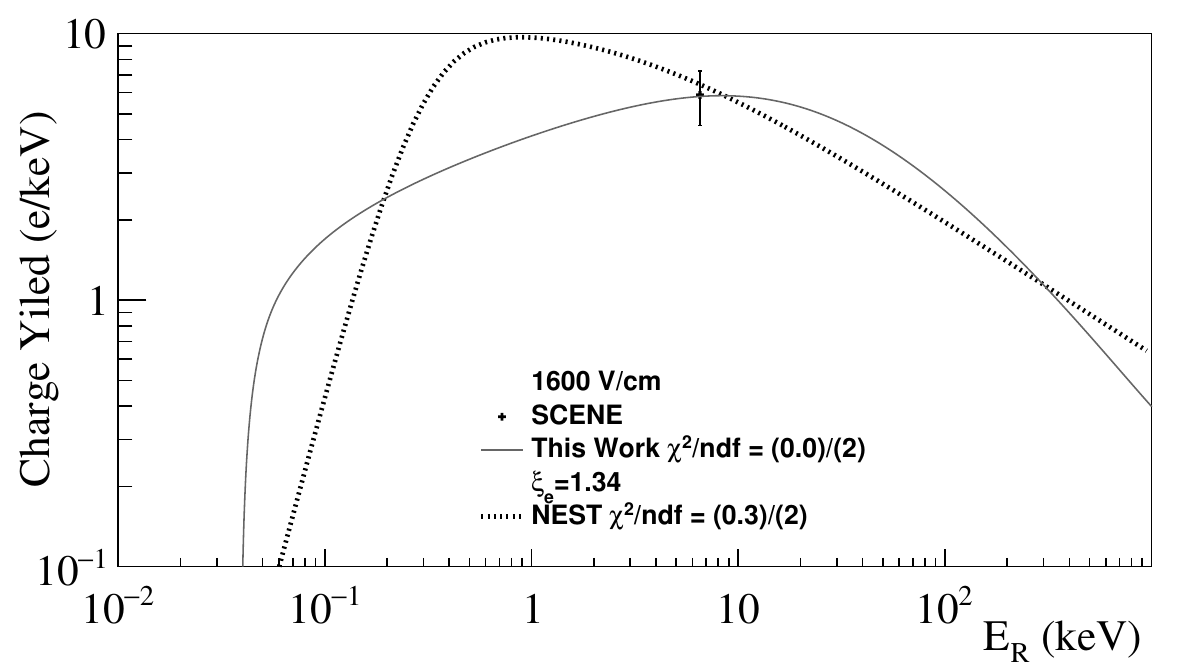}
    \caption{Charge yield for LAr for different electric fields, compared with data from \cite{NESTModel} and NEST curve.}
    \label{fig:ChargeYLAr}
\end{figure}

\begin{figure}[hb]
    \centering
    \includegraphics[width=0.99\linewidth]{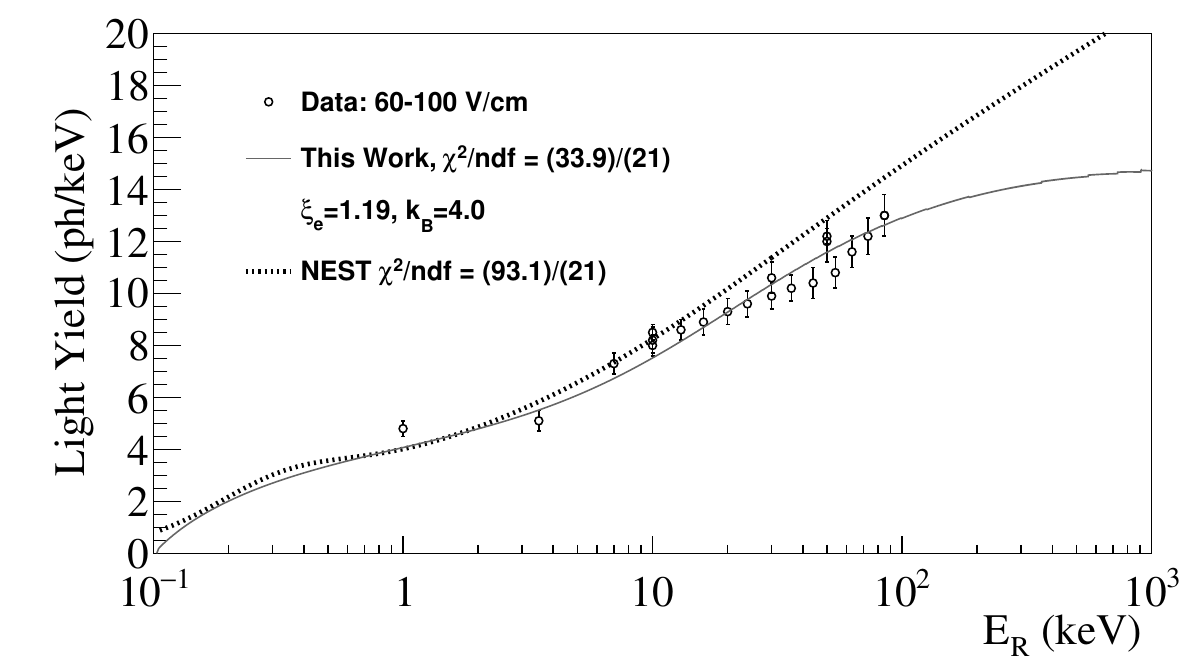}
        \includegraphics[width=0.99\linewidth]{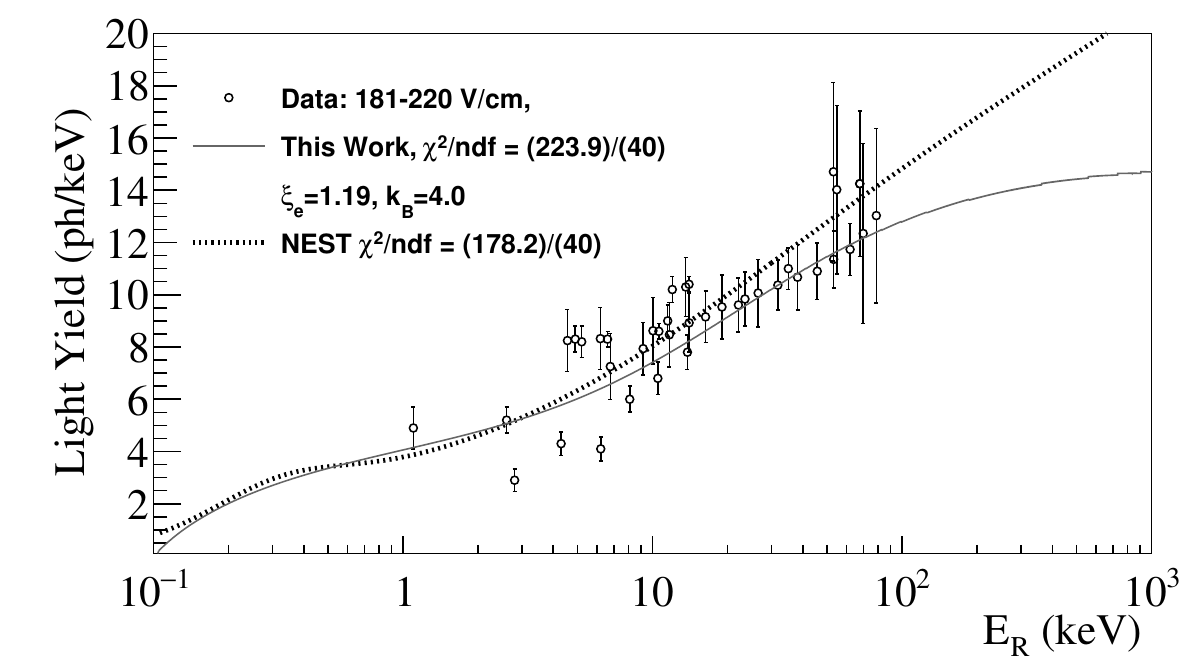}
    \includegraphics[width=0.99\linewidth]{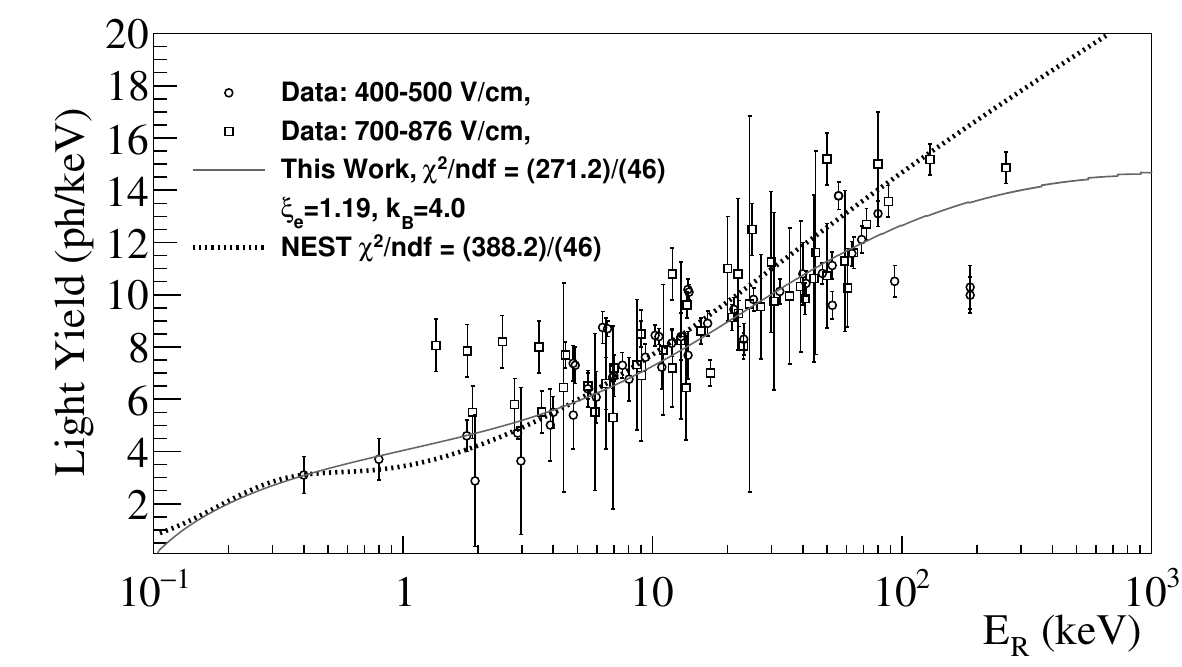}
         \includegraphics[width=0.99\linewidth]{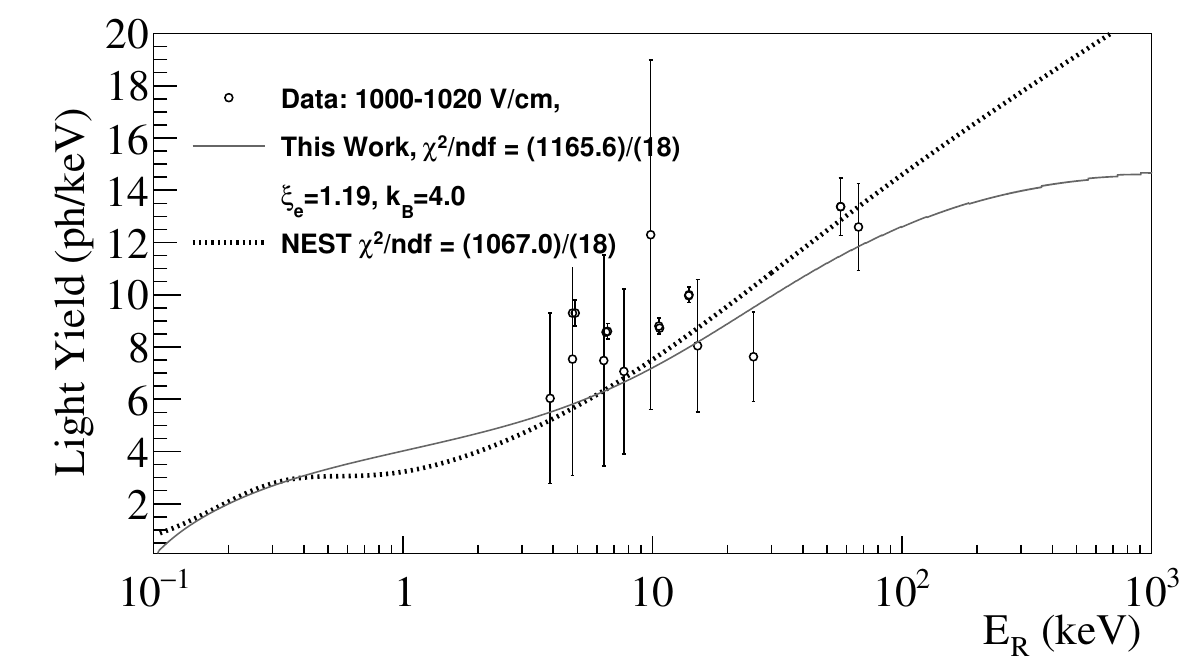}
    \includegraphics[width=0.99\linewidth]{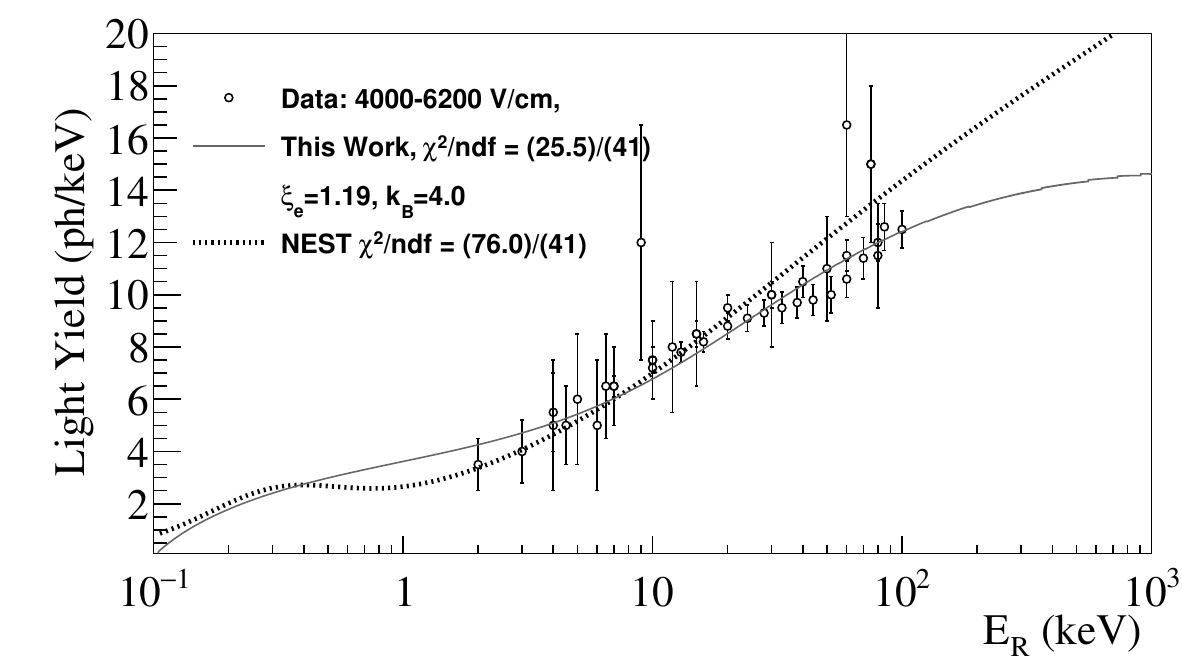}
    \caption{Light yield for LXe for different electric fields, compared with data from \cite{NESTModel} and NEST curve.}
    \label{fig:LightYLXe}
\end{figure}

\begin{figure}[htb]
    \centering
    \includegraphics[width=0.99\linewidth]{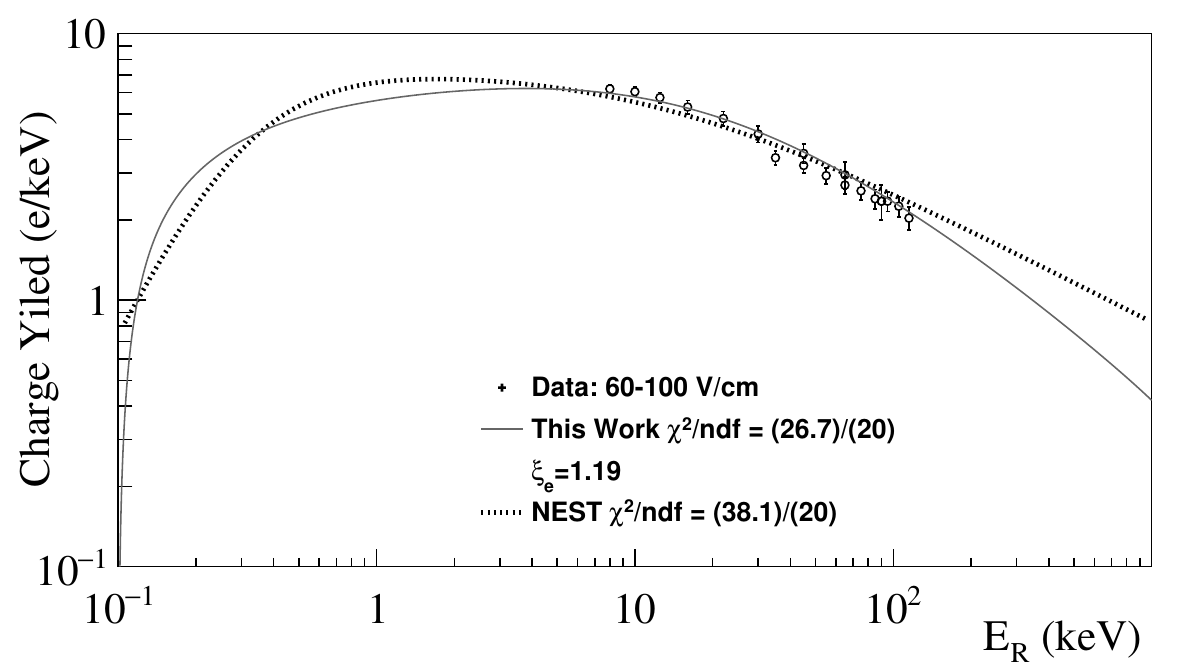}
        \includegraphics[width=0.99\linewidth]{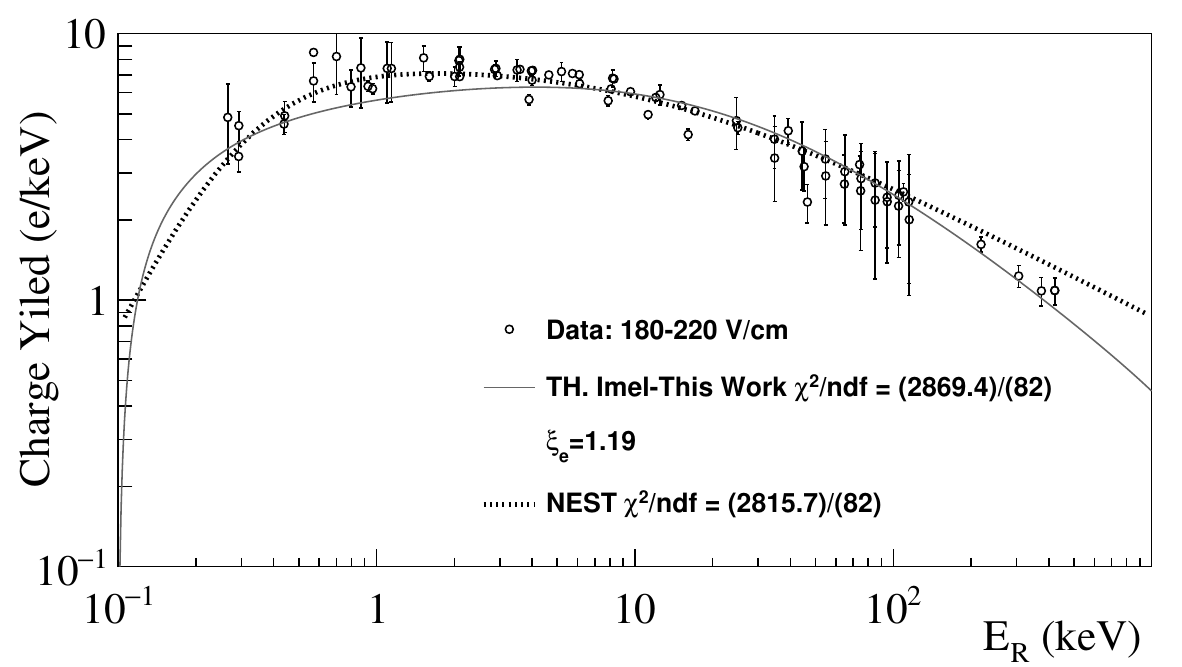}
        \includegraphics[width=0.99\linewidth]{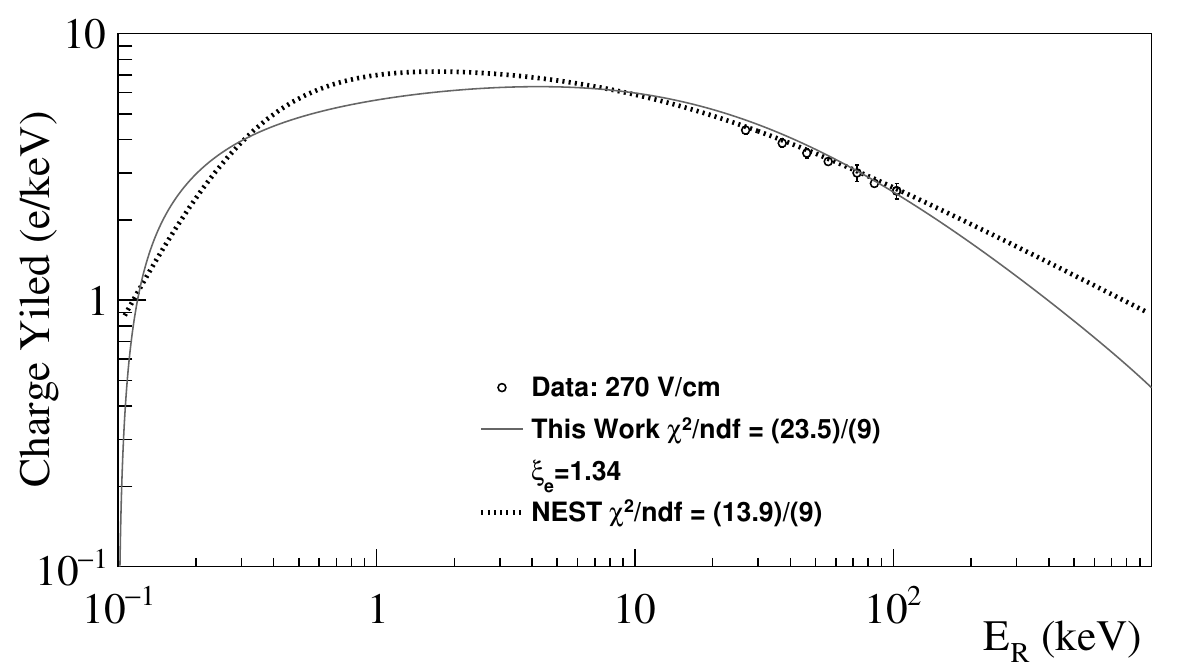}
        \includegraphics[width=0.99\linewidth]{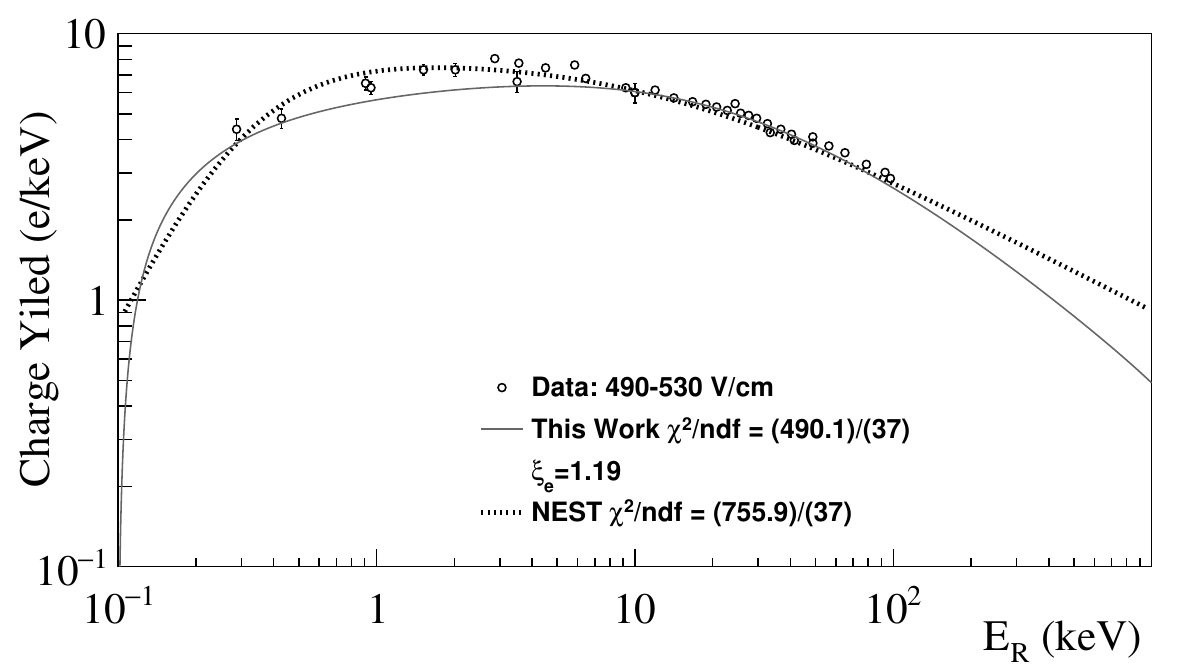}
    \caption{Charge yield for LXe for different electric fields, compared with data from \cite{NESTModel} and NEST curve.}
    \label{fig:ChargeYLXe}
\end{figure}

\begin{figure}[htb]
    \centering
    \includegraphics[width=0.99\linewidth]{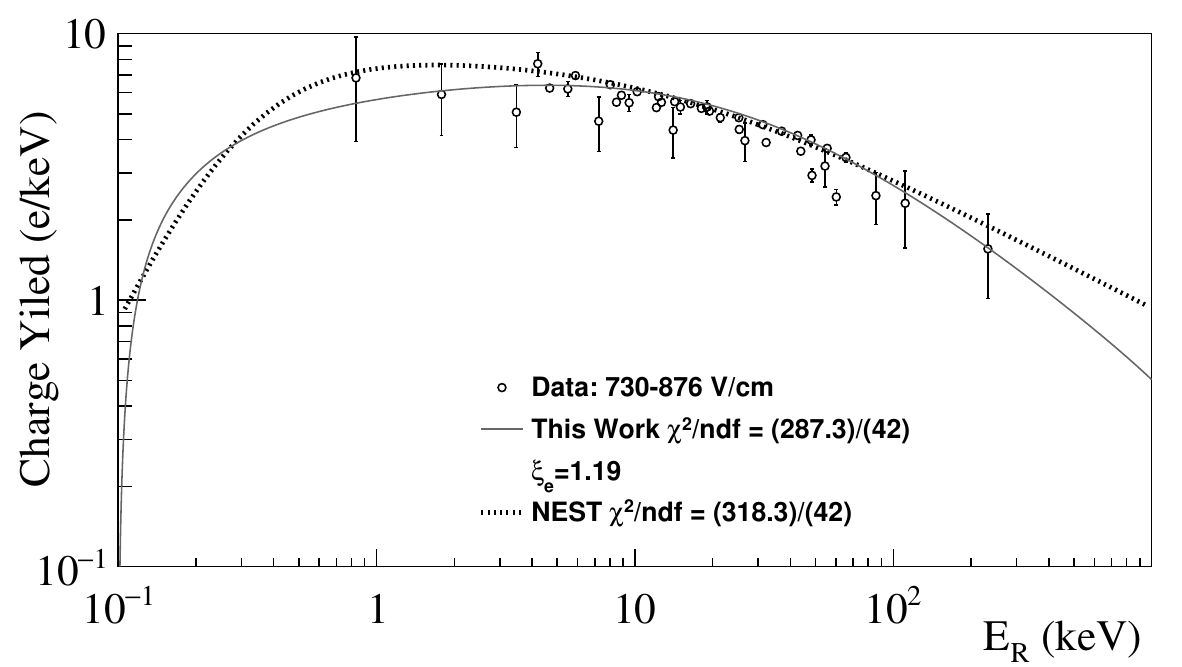}
        \includegraphics[width=0.99\linewidth]{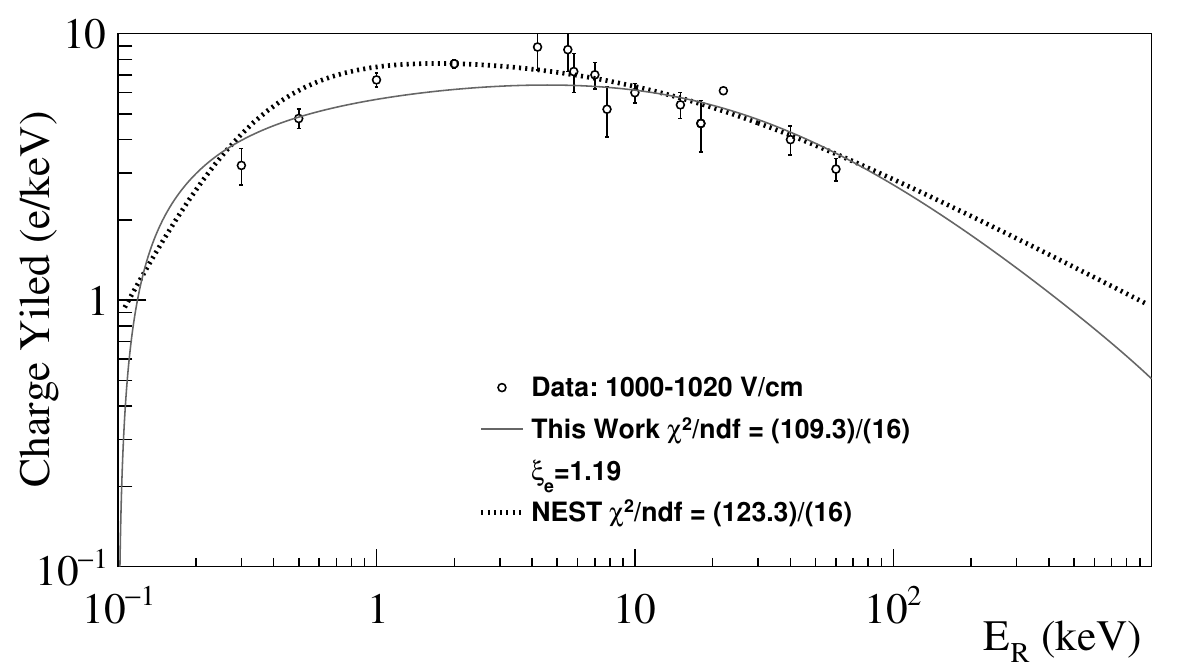}
        \includegraphics[width=0.99\linewidth]{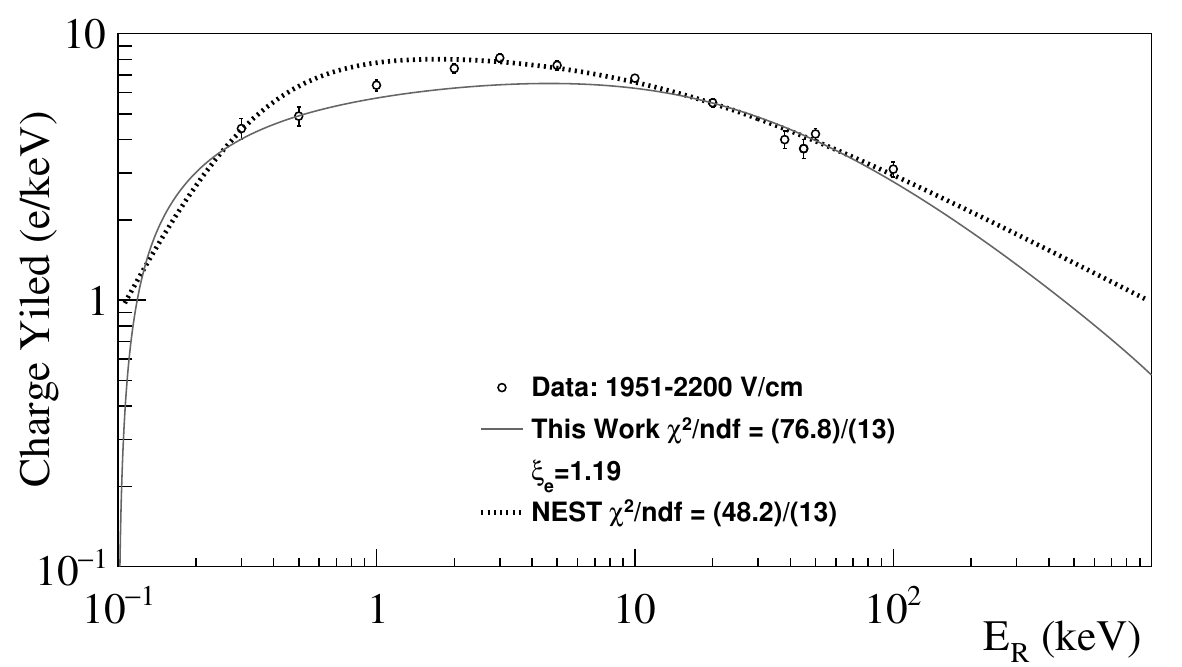}
        \includegraphics[width=0.99\linewidth]{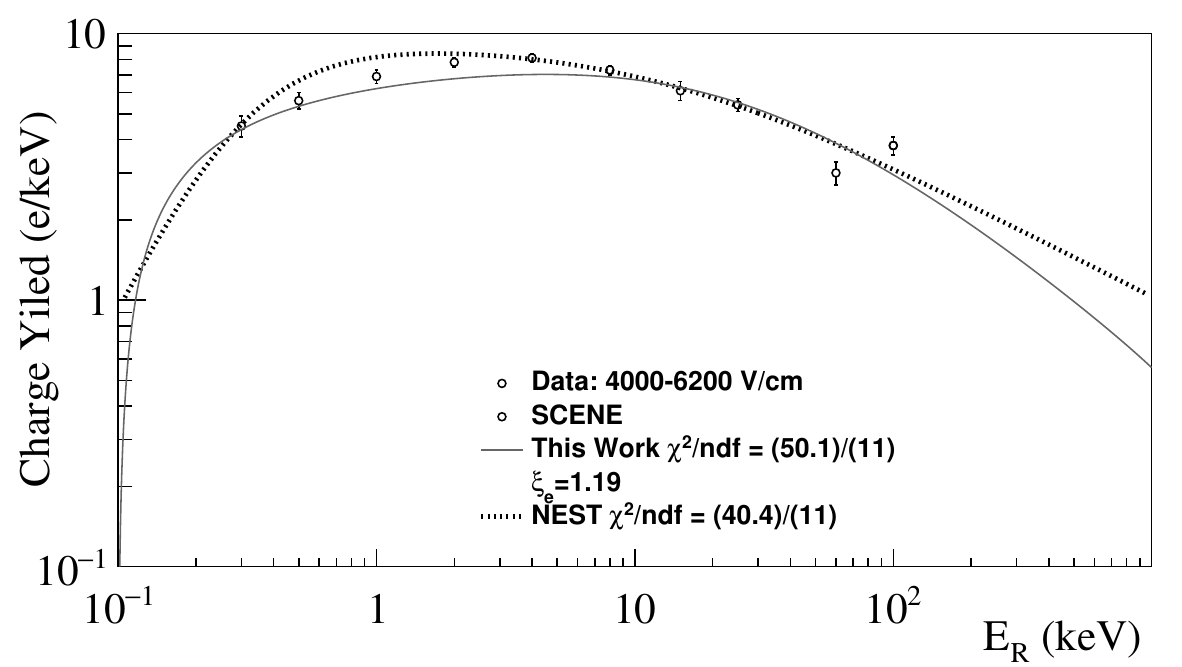}
    \caption{Charge yield for LXe for different electric fields, compared with data from \cite{NESTModel} and NEST curve.}
    \label{fig:ChargeYLXe2}
\end{figure}

The constant $C_0$ allows us to estimate the recombination probability constant, $\alpha_0$, providing deeper physical insight into the underlying modeling. From Eq. \eqref{ParameterFit}, we derive the ratio $\alpha_0/\tau_c = 2.2 \times 10^7 \, \text{s}^{-1}$, where $\tau_c$ is the mean collision time within the recombination box. This quantity is related to electron mobility by $\mu_e = e\tau_c/m$ and can be estimated in terms of the plasma frequency for LXe ($\omega^2_e =4\pi e^2 n/m\varepsilon_0 $), giving us $\tau_c\cong 2\pi/\omega_e = 3.4 \times 10^{-16} \, \text{s}$, which yields $\alpha_0 \cong 7.5 \times 10^{-9}$. By applying the Jaffe equations to the stationary regime for both ions and electrons \cite{ThomasImel}, the recombination constant can be approximated as $\alpha_0 \cong (\tau_r N_- r_0^2 v_F)^{-1}$, where $\tau_r$ is the recombination time for nuclear recoils and $N_-$ denotes the electron density, estimated as $N_- \simeq N_v n$. Proceeding in such a way, we obtain $\tau_r \cong 0.033 \, \text{ns}$, which is in good agreement with the results in the literature $\tau_r \ll 1 \, \text{ns}$ \cite{Dahl, NESTModel}. 
\begin{figure}[hbt!]
    \centering
    \includegraphics[width=0.99\linewidth]{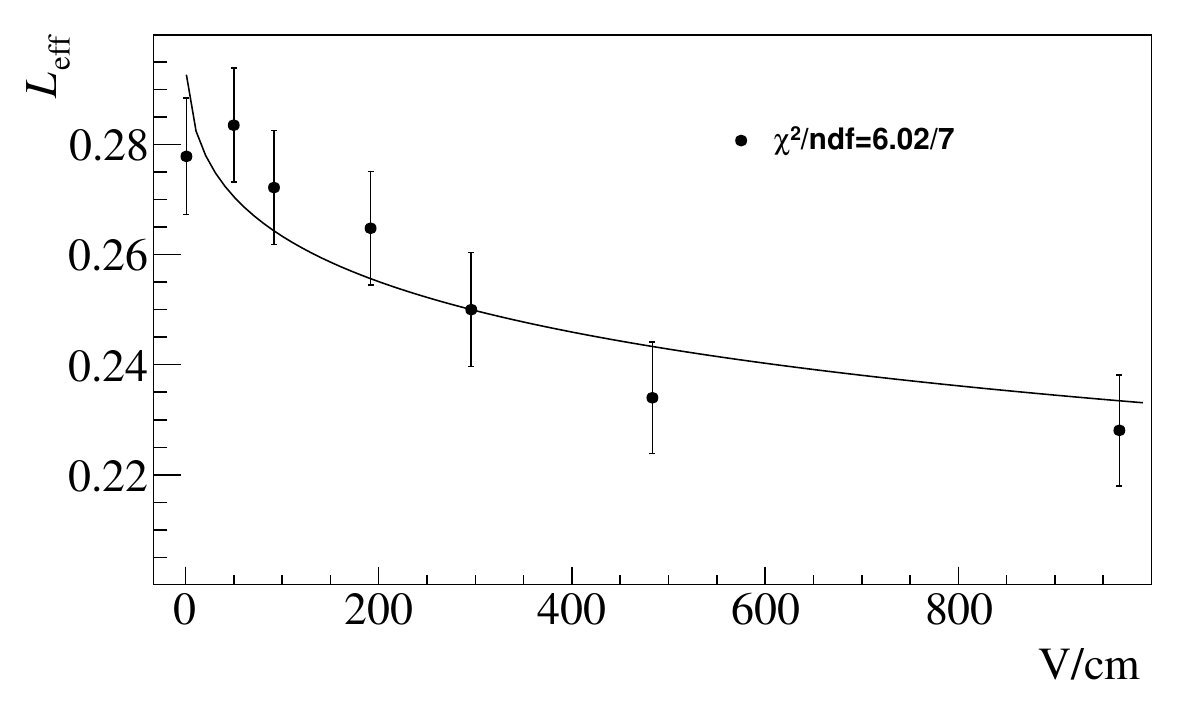}
    \caption{Scintillation efficiency in LAr for different electric fields, all points are at 36.1 keV energy. }
    \label{fig:Leff6.9keV}
\end{figure}
\begin{figure}[hbt]
    \centering
    \includegraphics[width=0.99\linewidth]{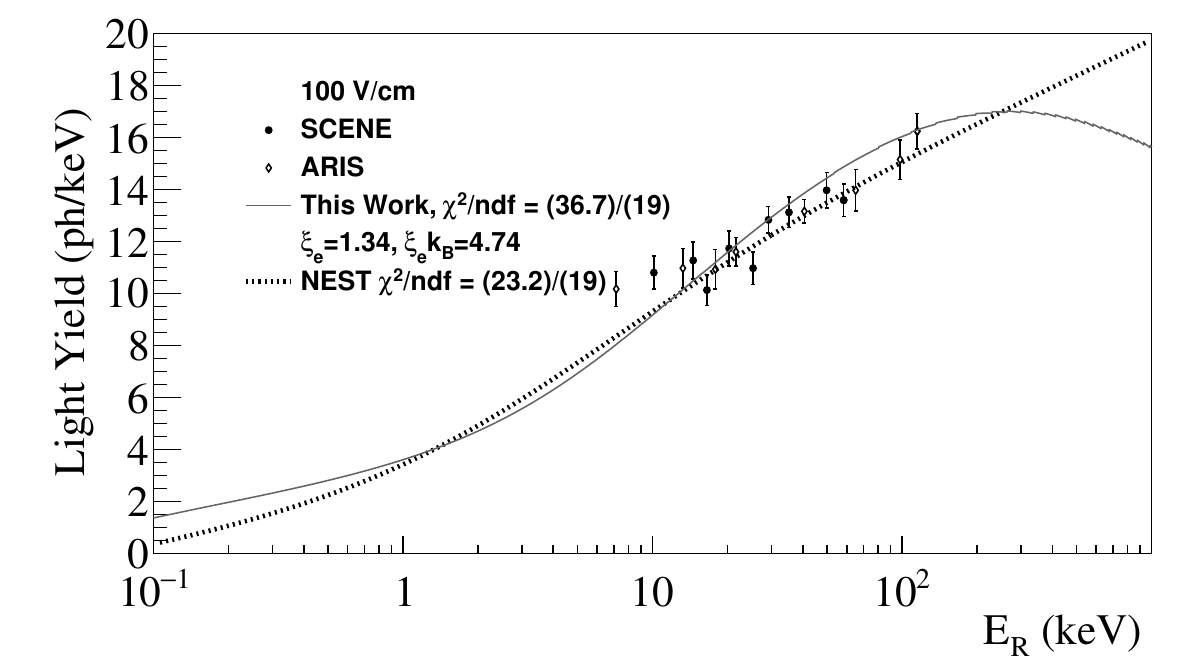}
        \includegraphics[width=0.99\linewidth]{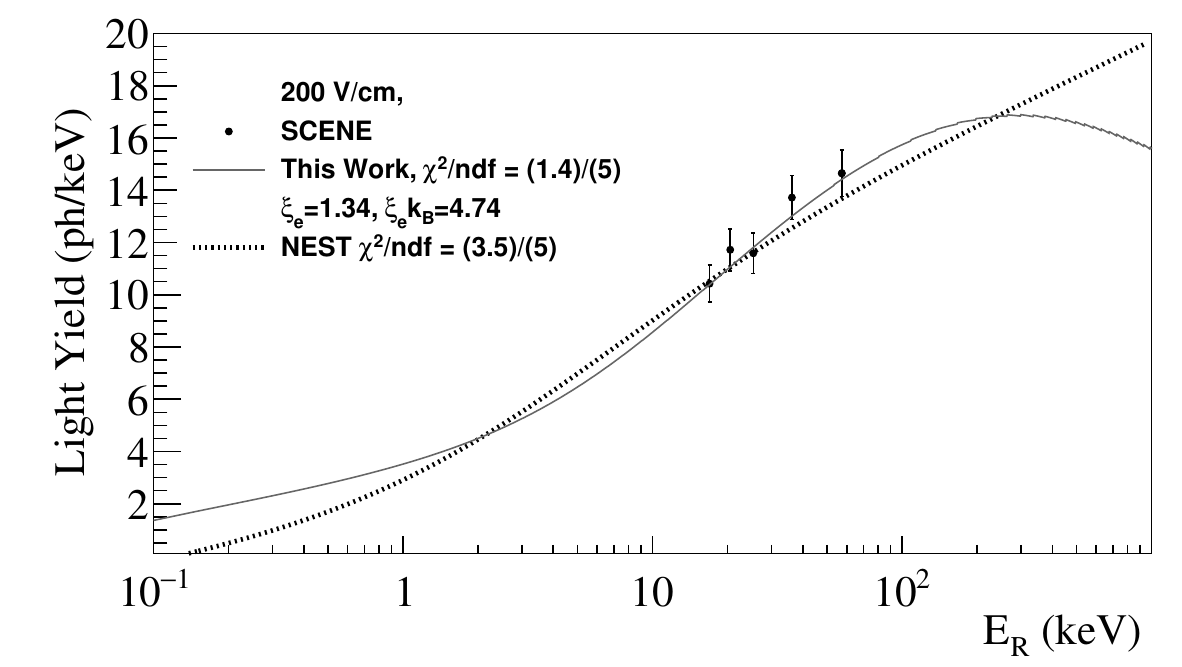}
    \includegraphics[width=0.99\linewidth]{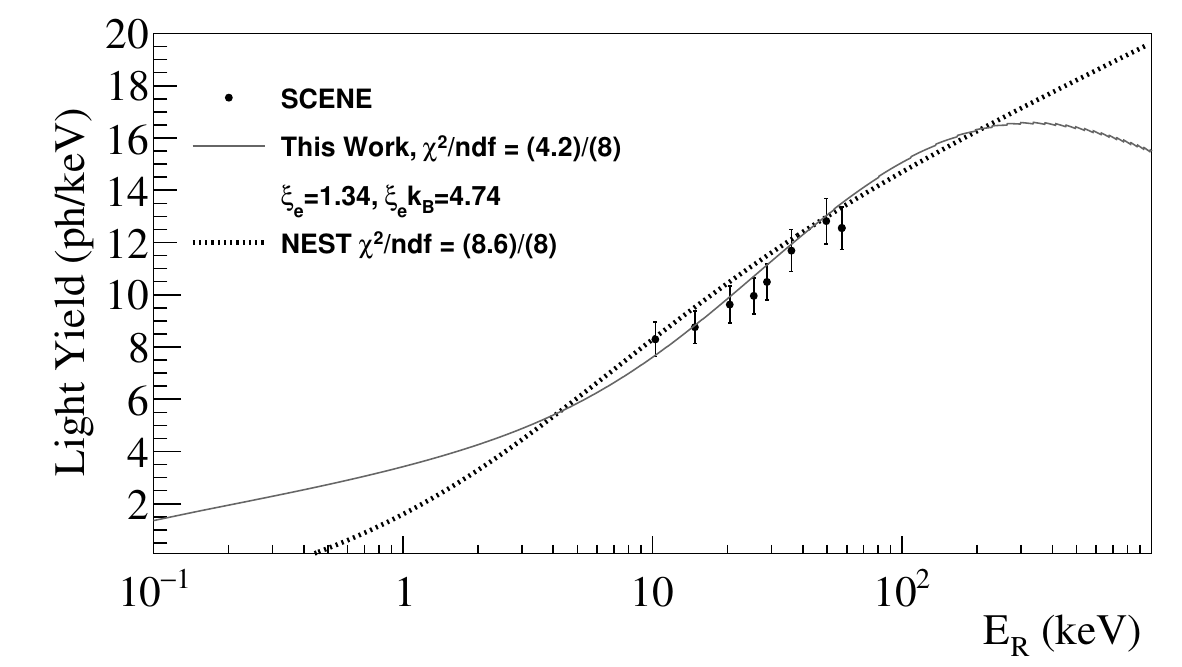}
    \caption{Light yield for LAr for different electric fields, compared with data from \cite{NESTModel} and NEST curve.}
    \label{fig:LightYLAr}
\end{figure}
Finally, in Figs. \ref{fig:LightYLAr}-\ref{fig:ChargeYLXe2},
we present the light and charge yields for LAr and LXe predicted by our model and compare them with the data and NEST predictions. We further compare our model predictions with the mean light-yield efficiency, $L_{\mathrm{eff}}$, in liquid argon (LAr) at 36.1 keV under varying applied electric-field strengths, see Fig.~\ref{fig:Leff6.9keV}.

\FloatBarrier

\section{Conclusions}
\label{Conclusions}
In this work, we have developed a first-principles theoretical formalism to describe total quanta production in noble liquids over an energy range spanning from the sub-keV regime to 1 MeV. The commonly used Lindhard formula is only valid for $\varepsilon > 1$ and assumes an electronic stopping power proportional to the velocity (only the parameter $k$). In contrast, our total quanta model is elaborated from the second-order integro-differential equation, which accounts for a general electronic stopping power, electronic straggling, and nuclear stopping power.  The electronic stopping is characterized using Tilinin's nonlinear analysis of the momentum transfer cross-section, incorporating recent experimental data for liquid argon and xenon, as well as low-energy Coulomb repulsion and high-energy Bohr stripping criteria.

 We demonstrated that elastic momentum transfer effects alone are insufficient to characterize ionization measurements in noble elements with $Z>20$. A significant contribution of this work is the inclusion of the Bates-Griffing dielectronic process to account for inelastic electron emission during the atomic cascade. The explicit mechanism  is formulated in terms of the electrostatic potential,  the spin configuration of the diatomic system, the exchange energy, and the kinetic energy of the projectile atom.

Furthermore, we  introduced the electronic scaling length, $\xi_e$, as a new fundamental physical parameter for nuclear recoil interactions. This parameter incorporates the Pauli exclusion principle within the Thomas-Fermi model, providing a more precise description of electronic orbitals during the nuclear cascade process. Within the recombination framework, $\xi_e$ gives a better characterization of the exciton to ion ratio, implemented as a function of the energy and the electric field, rather than as a constant value. The recombination process is described by an improved version of the Thomas-Imel box model, where the box size is derived from the electrostatic equilibrium between the external electric field and the field associated with the electrons and ions inside the box. This provides a physically grounded description of the field dependence, without the need for an empirical power-law fit.

 Finally, the model shows overall consistency for the noble liquids most commonly used in experiments: neon, argon, and xenon. All of them can be described by a single 
fitting parameter, while other relevant parameters are derived from fundamental atomic properties. The results presented  here show that our approach  outperforms empirical methods, such as NEST, particularly in the low-energy regime, predicting substantially higher light and charge yields. The model can provide an even more accurate description if we add an additional fitting parameter to describe an energy dependence of $\xi_e$. These findings highlight the qualities of the model as a robust tool for the next generation of ultra-low threshold experiments, including the search for dark matter and the detection of coherent elastic neutrino-nucleus scattering.

\section*{Acknowledgments}
This research was supported in part by SECIHTI grant CF-2023-I-1169, a SECIHTI postdoctoral grant, and DGAPA-UNAM grant PAPIIT-IN104026. The authors also acknowledge SNII support. \
\appendix
\section{Energy equilibrium for the cascade process}\label{A00}
Consider the collision between two atoms, with the target and projectile having $Z_1$ and $Z_2$ electrons and masses $M_1$ and $M_2$, respectively. 
The total kinetic energy of the target atom for each atomic collision in the CM frame is, $T_N=E'_R M_2/(M_1+M_2)$, where $E'_R$ is the nuclear recoil energy for each atomic collision in the lab frame. When the quasi-molecule is formed 
we assume that all electrons remain in a bound state, such that all the energy released to the system is absorbed by them, 
\begin{equation}\label{A00_eq1}
   T_N= T_e=\sum^{Z_1+Z_2}_{i=1} T_{ei},
\end{equation}
 with $T_{ei}$ being the energy of the electron $i$. We can define an average electron energy, $\bar T_e$, such that $T_e=(Z_1 +Z_2)\bar T_e$,
\begin{equation}\label{A00_eq2}
    \bar T_e=\frac{T_N}{Z_1+Z_2}.
\end{equation}
In the CM frame, $\bar T_e\cong \bar T^{lab}_e\equiv E_e$, since the electron mass is much lower than the atom mass; hence,
\begin{equation}\label{A00_eq3}
    E_e=\frac{E_R M_2}{(Z_1 + Z_2)(M_1 + M_2)}.
\end{equation}
The energy $E_e$, can be seen as the average energy of the  electron-atom interaction during atomic collision. For identical atoms, $E_e=E_R/4Z_1$, which coincides with the formula obtained using the Ziegler scale and can be compared with the Tilinin formula with the Lindhard scale, $E_e=E_R/2\sqrt{2}Z$ \cite{Tilinin}. 

For comparison, let us test the group IV of the periodic table. In particular  the minimum energy to excite an electron from the valence band to the Fermi level is $E_g/2$.  For an electron energy below this value, no ionization can occur, then $E_e\approx E_g/2$.   The minimal recoil energy, $E_{R_m}$, can be calculated from, $E_{R_m}=2Z_1E_g$. For energy gaps of $5.47, 1.17, 0.67$ eV, we obtain $E_{R_{m}}=65.7, 32.8, 42.9$ eV, for C, Si, and Ge, respectively. In the case of Si, the obtained value is close to those obtained in other studies \cite{SiliconTDDFT_Se}. According to the solution of the integral equation for the atomic motion (Eq.\ref{Ec:MainintegroDifStraggling}), the recoil threshold is approximately equal to $2U_F$, where $U_F$ is the Frenkel pair energy. From the previous results it can be directly obtained $U_F=E_{R_{m}}/2$, which compares  well with previous studies \cite{FrenkelGe,FrenkelSi,FrenkelDiamond}.

In noble liquids, the minimum energy to produce a metastable state is given by the  spin-exchange, where the energy range to form this state is about $0.1- 1.0$ eV \cite{MetaestableState,NIST_ASD}. Using eq. \eqref{A00_eq3} this gives, $4\,\rm{eV}<2U_c<216 \,\rm{eV}$, since the energy range under study is above 500 eV, as a first approximation we can take the average value, $U_c=50$ eV for all noble liquids.

\section{Free-electron model for the mean energy required to create an electron-hole pair}\label{AW}

To estimate $W$, we employ the free electron gas model in the limit $T \to 0$. In this framework, the change in internal energy, $\Delta \mathrm{U}$, of the valence band electrons due to an energy transfer ($\Delta E\propto k_B\Delta T$) from an ionizing particle is given by the following relation:
\begin{equation}
    \frac{\partial \mathrm{U} }{\partial E} = \frac{\pi^2 E}{2 E_F}
\end{equation}
where $E\ll E_F$ is the energy transferred to the Fermi gas and $E_F = E_V + E_g/2$, with $E_V$ denoting the valence-band energy. The energy $\Delta \mathrm{U}$ corresponds to the average energy required to excite a valence electron, $W_V$, multiplied by the total number of excitations. Thus, assuming that the derivative $\partial U/\partial E$ remains constant, we obtain $E/W_V = \pi^2 E / 2E_F$, where it follows that $W_V = 2E_F / \pi^2$. 
To account for the additional energy $E_g$ required to promote an electron from the valence band to the conduction band, the total average energy $W$ is given by:

\begin{equation}\label{Wtheory}
    W = \frac{2 E_F}{\pi^2} + E_g
\end{equation}

The width of the valence band can be estimated from the free-electron model as $E_V  \cong \hbar^2/2m(3\pi^2n_e)^{2/3}$, where $n_{e}$ is the number density of electrons ($N_{v}n$), see Tables  \ref{tab:material-properties} and \ref{tab:Eg_W_materials}.

\begin{table}[h]
\centering
\caption{Physical properties of noble liquids and semiconductors used to calculate the Fermi energy in the free-electron-gas model. The last column lists the band gap at $T\cong 100$ K.}
\label{tab:material-properties}
\begin{tabular}{|l|c|c|c|c|}
\toprule
\text{Material} & \text{$\rho$ (g/cm$^3$)} & \text{$n_e$ (m$^{-3}$)} & \text{$E_V$ (eV)} & \text{$E_g$ (eV)} \\
\hline
LNe     & 1.207 & $2.88 \times 10^{29}$ & 15.91 & 21.50 \\
LAr     & 1.395 & $1.68 \times 10^{29}$ & 11.11 & 14.30 \\
LKr     & 2.413 & $1.38 \times 10^{29}$ & 9.77  & 11.55 \\
LXe     & 2.940 & $1.08 \times 10^{29}$ & 8.26  & 9.30  \\
Si      & 2.330 & $1.99 \times 10^{29}$ & 12.47 & 1.17  \\
Ge      & 5.323 & $1.76 \times 10^{29}$ & 11.48 & 0.66  \\
Diamond & 3.510 & $7.04 \times 10^{29}$ & 28.86 & 5.48  \\
\hline

\end{tabular}
\end{table}

\begin{table}[ht]
\centering
\caption{Estimated mean energies $W$ for different materials, compared with measured values or reported ranges.}
\label{tab:Eg_W_materials}
\begin{tabular}{|l|c|c|}
\toprule
Material & $W_{exp}$ eV & $W$ eV Eq. \eqref{Wtheory} \\
\hline
$C_{Diamond}$ & 13.0-13.3&  11.88   \\
Si     & 3.66-3.75&   3.70  \\
Ge    & 2.8-3.0&   2.99  \\
LNe      & 25-27  &   26.9  \\    
LAr     & 19.5 & 18.0   \\
LKr     & 14.2-15.9 &  14.7  \\
LXe     & 11.5-13.7& 11.92    \\
\hline
\end{tabular}
\end{table}

\section{Free electron plasma dispersion}\label{A0}
Consider the average Fermi energy for a free electron gas, $\bar E_F=\frac{3}{5}(\hbar^2k^2_F/2m)$, where  $k_F$
is the maximum wave vector that an electron can have. The variation  of this energy,

\begin{equation}\label{eq:diff_Ef}
    \Delta  E_F=\frac{3\hbar^2 \bf{k_F}\cdot \Delta \bf{k_F}}{5m},
\end{equation}

  where $ \bf{k_F}\cdot \Delta \bf{k_F}$$=k_F\Delta k\cos\psi$, with $\psi$ the collision angle between the electrons in the laboratory frame. On the other hand, to estimate the maximum value of $\Delta E^{M}_F$, we can use Lindhard dielectric function \cite{lindhard1954properties}, to compare with \eqref{eq:diff_Ef},
\begin{equation}\label{eq:diff_Ef2}
    \Delta E^{M}_F\cong \frac{3\hbar v^2_F \Delta k_F^2}{10\omega_F}=\frac{3\hbar^2\Delta k_F^2}{5 m},
\end{equation}
where $\hbar \omega_F=mv^2_F/2$. By equating \eqref{eq:diff_Ef} and \eqref{eq:diff_Ef2} \cite{PlasmaDisprel}, we obtain the critical collision angle $\psi_c$,
\begin{equation}
    \cos\psi_c=\Delta k_F/k_F=\sin(\theta_c/2),
\end{equation}
where $\theta_c$  is the critical angle in the CM frame.

For angles $\theta<\theta_c$, then $\Delta E_F < \Delta E^{M}_F$ and there is no significant energy  transfer from free electrons to bonding electrons. In the opposite case,  the free-electron gas must be damped, due to direct transfer  of all its energy to a single inner orbital electron. This can be interpreted as meaning that for electron energies lower than $\Delta E_F$ we can approximate the electrons as free. In the context of TF theory, inner electrons with energies larger than $\Delta E_F$ do not contribute significantly to momentum transfer in atomic collisions.


\section{Effective number of electrons in atomic binary collisions}\label{A1}

From appendix \ref{A00} we see that the total electron energy given to the system of identical ions during a collision is $E_e=E_R/4Z$. Using the adiabatic criteria \cite{GumusH}, $v\leq\sqrt{3}v_F$, we have $E_F \approx E_e/3$. Then  calculate the effective electron distance for the electron to be stripped, $E_F=\hbar^2 (3\pi^3 n^{TF}_e)^{2/3}/2m$, where  $n^{TF}_e(x')=(1/3\pi^2)(2m \mid\phi^e_Z(x',\xi_e,Z)\mid/\hbar^2)^{3/2}$ is the TF electron density. Having the distance $x'$ we can compute the effective number of electrons in the binary collision by 
$$
Z-N=Z\left(\chi_Z -x'\frac{d\chi_Z}{dx'}\right),
$$
where $\chi_Z$ is the screen function associated with $\phi_Z(x',\xi_e)$.

\section{Amaldi model for ionized atoms}\label{A2}
According to the Thomas-Fermi model for the atom, the screen function $\chi_Z(x')$,  $x'=r/a$ with $a=a_00.885/Z^{1/3}$, obeys the equation
\begin{equation}\label{A2eq1}
    \chi_Z''=\frac{\chi_Z^{3/2}}{x'^{1/2}},
\end{equation}
with the boundary conditions $\chi_Z(0)=1$ and $\chi_Z(x'_0)=0$. 

For an ionized atom with $N$ electrons, the boundary condition at $x'_0$ becomes $\chi_Z(x'_0)=(Z-N)/Z$. Amaldi's approach \cite{AmaldiTF} approximates the ionized boundary condition, imposing  the \textit{ansatz}, 
\begin{equation}\label{A2eq2}
    \chi^{A}_Z (\kappa x')=\chi_Z(\kappa x')+\frac{(Z-N)}{Z}(1-\chi_Z(\kappa x')),
\end{equation}
with $(Z-N)/Z \ll 1$ and $\kappa$ a correction factor for screening length to be determined. Let us notice that this \textit{ansatz} satisfies the boundary conditions for an ionized atom mentioned above. 

Requiring that the expression  Eq.\eqref{A2eq2} satisfy \eqref{A2eq1}, we find
\begin{equation}
    \chi_Z''\cong \kappa^{-3/2}\left( \frac{N}{Z}\right)^{1/2} \frac{\chi_Z^{3/2}}{x'^{1/2}},
    \end{equation}
if we set $\kappa^{-3/2}(N/Z)^{1/2}=1$, we deduce Amaldi's screening length, $a_A=a_00.885/N^{1/3}$ screening length.

\section{Polarizability for free electron gas during recombination}\label{A3}
Here, we apply the Lindhard dielectric function theory \cite{lindhard1954properties} for a gas of free electrons in the presence of a static electromagnetic field. In his semiclassical model the dielectric function is, 
\begin{equation}\label{EqA3}
    \varepsilon(k_F,0)=1+\frac{3\omega^2_e}{(k_Fv_F)^2},
\end{equation}
where $\omega^2_e= Ze^2 \bar n_e /m\varepsilon_0$ is the electron plasma frequency (in standard units), with  $\bar n_e$  given by the Thomas-Fermi model:
$$
\bar n_e =\frac{(2mU_{TF})^{3/2}}{3\pi^2\hbar^3},
$$
where $U_{TF}=20.9\, Z^{4/3}/\xi_e^{2/3}$ eV. In Eq.~\eqref{EqA3} $v_F$ is the Fermi velocity associated with the average energy of the valence electrons  during recombination,  $k_Fv_F =(2/\hbar)(\Delta E_F/N_{v})$. 

The polarizability is given by
\begin{equation}\label{Eq2A3}
    \alpha_p=\frac{(\varepsilon(k_F,0)-1)\varepsilon_0}{\bar N_i},
\end{equation}
where $\bar N_i$ is the average ion density during recombination. Thus, 
$$
\omega^2_e=\frac{(270.25)(me^{7})^{1/2}\,Z^3}{3\pi^2\hbar^3\xi_e\varepsilon_0},
$$
from the previous result the polarizability is given by,

\begin{eqnarray}\label{Eq3A3}
    \alpha_p &=&\frac{(\varepsilon(k_F,0)-1)\varepsilon_0}{\bar N_i}\nonumber\\
    &=&\frac{N^2_{v}(270.25)(me^{7})^{1/2}\,Z^3}{4\pi^2\hbar\,\xi_e (\Delta E_F)^2 \bar N_i}.
\end{eqnarray}

The average ion density can be estimated as $gN$, where $N$ is the atomic number density, and  $g=(r_W/\bar r)^3$, with $\bar r= (r_W + r_0)/2$. For LXe $g=2.37$, $\alpha_p/(4\pi\varepsilon_0)=1.30\times 10^4 \r{A}^3$, and for LAr $g=6.07$, $\alpha_p/4\pi\varepsilon_0 = 7.77\times 10^3 \r{A}^3$. To test Eq. \eqref{Eq3A3} we estimate the polarizability of LXe and LAr in thermal equilibrium, in which case, instead of $U_{TF}$ we have to use the first ionization energy of the atom:  $12.8$ eV for LXe and $15.9$ eV for LAr.  Thus, we obtain $\alpha_p \cong 4.5 \r{A}^3$ for LXe  and $\alpha_p \cong  2.4\, \r{A}^3$ for LAr, which are similar to the measured values of $4.0 \,\r{A}^3$ and $1.7\,\r{A}^3$, respectively \cite{Polarizabilities}.

\section{Mean free path for electron ionization in atomic collisions}\label{AppendixE}
In Sec.~\ref{SubEnergyDistribution}, we  discussed the dielectronic emission process as a result of the collision between two atoms whose electrons interact through the electron-electron potential and  the screened  nucleus potential. At the threshold energy $W_i$, the electrons are emitted with an average energy $\bar \epsilon_D$. For energies larger than $W_i$ electrons feel the attractive potential of both  approaching atoms and, as a consequence, there is a probability of no  emission. The probability  of  electron pair creation is related to the mean free path of the sum of the elastic and inelastic  cross sections times the electron density in the binary collision.

For electrons with energies above $27.2\,  Z^{2/3}$ eV, the elastic  cross section is given by \cite{Tilinin}  
\begin{equation}\label{A5eq1}
    \sigma_{el}=\frac{4\pi a^{2}_0 Z^{4/3}E_0}{ E},\, 
\end{equation}
where $E_0=13.6$ eV. The inelastic channel   is given by the sum of all ionization (or excitation) processes \cite{IonizationCrossBethe}:  
\begin{eqnarray}
   \sigma_{i}&=&\sum_{i}\frac{S_i}{t_i+(u_i+1)} \\
  \nonumber &\times&\left[\frac{Q \ln t_i}{2}\left(1-\frac{1}{t_i^2}\right)+(2-Q)\left(1-\frac{1}{t_i}-\frac{\ln t_i}{t_i+1}\right)\right],
\end{eqnarray}
where $t_i=E/U_i$, $S_i=4\pi a^2_0n_i(E_0/U_i)^2$, $u_i\cong 1/2$, and $Q$ is the dipole constant defined in terms of the continuum dipole oscillator strength. Here, $U_i$ and $n_i$ are the corresponding binding energy and orbital occupation number.

The electron density, $N_e$, relevant for the BG process in atomic collisions, is related to the exchange energy.  Using the  TF model, we  estimate as  
\begin{equation}\label{A5eq2}
N_e=\frac{1}{3\pi^{2} \hbar^{3}}\left(2m\bar E_{ex}\right)^{3/2},
\end{equation}
where $\bar E_{ex}$ is the average exchange potential energy in eV, calculated in terms of the sum of all orbital contributions. According to the electron free gas model \cite{ExchangeEnergy}, we obtain 
\begin{equation}
\bar E_{ex}=-\frac{54.4a_0(6/\pi)^{1 / 3}}{Z} \sum_{i} F(\eta_i) (N_i/2)^{1 / 3} \,\rm{eV}.   
\end{equation}
Here,  $N_i$ is the electron density in orbital $i$ and
\begin{equation}
    F(\eta_i)=\frac{1}{2}+{\frac{1-\eta_i^2}{4 \eta_i}} \ln \left(\frac{1+\eta_i}{1-\eta_i}\right),
\end{equation}
with $\eta_i=k_i / \Delta k_F=\sqrt{U_i/2\Delta E_F}$, where by  the virial theorem ($T_i=U_i/2$). The exchange energy contributes only if $l\neq 0$ and  $\eta_i\leq1$, for $k_i>\Delta k_F$ electrons are above the Fermi level and, $F(\eta_i)=0$, consistent with the interpretation of $\Delta E_F$.  To compute $N_i$ we use the Thomas-Fermi model with the AVG potential \cite{WilsonF} evaluated at the corresponding radii, $x'_i=a_0n^2/bZ^i_{eff}$, where $n$  is the  principal number of the orbital  and $Z^i_{eff}$  effective charge of  orbital $i$, see Tables \ref{tab:ExchangeTableXe} and \ref{tab:ExchangeTableAr}.
\begin{table}[h]
    \centering
    \begin{tabular}{|c|c|c|c|}
        \hline 
        $i$ Shell &  $Z^i_{eff}$ &$U_i$ eV & $n_i$\\
         \hline
         1 ($5s$) & 14.22  & 23.3 & 2\\
         1 ($5p$) & 12.42 & 12.75 & 6 \\
         \hline
         2 ($4s$) & 26.17 & 213.2 & 2 \\
         2 ($4p$) & 24.96 & 146.1 & 6\\
         2 ($4d$) & 21.89 & 68.5 & 10\\
         \hline
         3 ($3s$) & 35.58 & 1148.7 & 2 \\
         3 ($3p$) & 35.67 & 971.4 & 6\\
         3 ($3d$) & 39.95 &  682.7 & 10\\
         \hline
         4 ($2s$) & 39.8 & 5453 & 2 \\
         4 ($2p$) & 49.83 & 4946.5 & 6\\
         \hline
         5 ($1s$) & 52.92 & 34561 & 2\\
         \hline
    \end{tabular}
    \caption{Effective charge, ionization energies and occupation number for Xenon atom \cite{NIST}.}
    \label{tab:ExchangeTableXe}
\end{table}

\begin{table}[hbt]
    \centering
    \begin{tabular}{|c|c|c|c|}
        \hline 
        $i$ Shell &  $Z^i_{eff}$ &$U_i$ eV & $n_i$\\
         \hline
         3 ($3s$) & 7.76 & 29.3& 2\\
         3 ($3p$) & 6.76 & 15.8  & 6\\
         \hline
         4 ($2s$) & 12.23 & 326.3 & 2 \\
         4 ($2p$) & 14.01 & 249.5 & 6\\
         \hline
         5 ($1s$) & 17.51 & 3206 & 2\\
         \hline
    \end{tabular}
    \caption{Effective charge, ionization energies and occupation number for Argon atom \cite{NIST}.}
    \label{tab:ExchangeTableAr}
\end{table}
For Ar, we need only consider the $3p$ orbital with electron density $8.68\times 10^{29} \,{e/\rm{m^3}}$, which gives $\bar E_{ex}=1.46$ eV.  In the case of Xe, the orbitals $5p$, $4p$, $4d$, $3p$, and $3d$ must be included. Their electron densities, in units of $e/\rm{m^3}$, are $3.86\times 10^{29}$, $2.94\times 10^{31}$, $2.90\times 10^{31}$, $5.91\times 10^{32}$, and $5.98\times 10^{32}$, respectively, and  $\bar E_{ex}=10.8$ eV.

 For the two-atom system, the mean free path is  $\lambda^{-1}=N_e\sigma_t(\Delta E_F)$, where $\sigma_t(\Delta E_F)=\sigma_{el}(\Delta E_F) + \sigma_{in}(\Delta E_F)$. By assuming a Poisson distribution, the probability of electron emission is $1-P$,  with $P=e^{-N_\lambda}$, where $N_\lambda=r/\lambda$ is the number of mean free paths for a separation $r$ of the approaching atoms.  In Ziegler units, $N_\lambda=  \xi^{2/3}_e\, 0.885\,q\,a_0 Z^{-0.23} N_e\sigma_t(\Delta E_F) x/2$, where we considered the dependence of the scaling length on the electron density. The probabilities for no emission are $P=e^{-0.068x}$ and $P=e^{-0.5x}$ for LAr and LXe, respectively. Note that the exponential parameters are  approximately proportional to $Z^{2}$, thus for light noble liquids this effect becomes negligible \footnote{We can set the ansatz $P=e^{-C_Zx}$, with $C_Z=(3.01\times 10^{-4})Z^2$}. 

\newpage
\bibliography{references}
\end{document}